\documentclass[12pt]{article}

\usepackage[utf8]{inputenc}
\usepackage[a4paper, margin=1in]{geometry}
\usepackage{amsmath,amssymb,bm, nccmath}
\usepackage{graphicx,float}
\usepackage{booktabs,adjustbox,capt-of,afterpage}
\usepackage{array,ragged2e,makecell,multirow,tabularx}
\usepackage{titlesec}
\usepackage{enumitem} 
\usepackage{listings}
\usepackage[normalem]{ulem}
\usepackage{accents,calc,tikz}
\usepackage{cite}
\usepackage{authblk}
\usepackage{caption, subcaption}
\usepackage{indentfirst}
\usepackage{longtable}
\usepackage[ruled,vlined]{algorithm2e}

\usepackage[colorlinks=true,
linkcolor=blue,
citecolor=blue,
urlcolor=blue]{hyperref}
\usepackage{cleveref}
\usepackage{amsthm}
\newtheorem{remark}{Remark}[section]

\numberwithin{equation}{section}
\numberwithin{figure}{section}
\numberwithin{table}{section}

\usepackage[titletoc]{appendix}
\usepackage{chngcntr}

\titleformat{\section}{\normalfont\normalsize\bfseries}{\thesection}{1em}{}
\titleformat{\subsection}{\normalfont\normalsize\bfseries}{\thesubsection}{1em}{}

 \title{\fontsize{14}{17}\selectfont\textbf{Bivariate Frank Copula: Some More Results on Point Estimation of the Association Parameter from a Bayesian Perspective and Revisiting the Goodness of Fit Tests with an Application to Model Groundwater Data from Dong Thap, Vietnam}}
 \date{}

\author[1,2,3]{\fontsize{11.5}{14}\selectfont Thi-Yen-Anh Pham\thanks{First author email: \href{mailto:ptyanh.sdh242@hcmut.edu.vn}{ptyanh.sdh242@hcmut.edu.vn}; \href{mailto:phamthiyenanh@tdtu.edu.vn}{phamthiyenanh@tdtu.edu.vn}}}
\author[1,2]{\fontsize{11.5}{14}\selectfont Dung T. Nguyen\thanks{Second author email: \href{mailto:dungnt@hcmut.edu.vn}{dungnt@hcmut.edu.vn}}}
\author[3]{\fontsize{11.5}{14}\selectfont Nabendu Pal\thanks{Third and corresponding author email: \href{mailto:nabendu.pal@tdtu.edu.vn}{nabendu.pal@tdtu.edu.vn}}}

\affil[1]{\fontsize{10.5}{13}\selectfont Faculty of Applied Science,
	Ho Chi Minh City University of Technology (HCMUT) \qquad \qquad \qquad 268 Ly Thuong Kiet Street, Dien Hong Ward, Ho Chi Minh City, Vietnam}  

\affil[2]{\fontsize{10.5}{14}\selectfont Vietnam National University Ho Chi Minh City, Linh Xuan Ward, Ho Chi Minh City, Vietnam}

\affil[3]{\fontsize{10.5}{14}\selectfont Faculty of Mathematics and Statistics, Ton Duc Thang University, Ho Chi Minh City, Vietnam}

\begin{document}
	\maketitle
	\vspace{-1.0cm}
	\begin{abstract}
		This work consists of two major parts. $(i)$ The first part of this work extends the recent comprehensive study of Pham et al.(2025)~\cite{Pham et al.(2025)} where three classical point estimators of the association parameter of a bivariate Frank Copula were compared and it was established that the maximum likelihood estimator (MLE) has the best overall performance (among those three) in terms of bias as well as the mean squared error (MSE). This current investigation of ours looks into the performance of two Bayes estimators under two natural priors, namely - the noninformative (generalized) flat prior, and the invariant Jeffreys prior; and finds a somewhat surprising result that the latter one dominates the former as well as the MLE uniformly in terms of MSE for small sample sizes, i.e., $n \leq 25$. For moderate to large sample sizes (i.e., $n > 25$), all these three estimators have almost identical performances both in terms of bias as well as MSE. We have also pointed out certain computational aspects while using the R package which have profound implications in computing the MLE, but may go unnoticed, and this impacts the MLE's bias and/or MSE computations for very small sample sizes. $(ii)$ The second part of this work uses a recent dataset from Vietnam and applies Frank Copula in analyzing the association between groundwater arsenic concentration and each of three other benign elements which are easy to monitor, and may help in understanding their inherent dynamics. In the process of this application of Frank Copula to model the groundwater data we revisit the two goodness of fit (GoF) tests  introduced by Genest et al.(2006)~\cite{GenestRemillard2006}. Not only have we explored some non-intuitive behavior of these
		two test statistics, but also provided extensive tables of critical values of those two test statistics using a comprehensive simulation study. This not only complements Genest et al.'s (2006)~\cite{GenestRemillard2006} seminal work, but also refines some of their limited computational results. With our groundwater data we have also demonstrated that usage of nonparametric marginals can have benefits in predicting the arsenic concentration (using a benign
		element) over the usage of parametric marginals using the Frank Copula based regression model.
		
	\end{abstract}
	
	
	
	
	\newpage
	\section{Introduction}
	\subsection{Preliminaries}
	
	In multivariate data analysis quite often a multivariate normal distribution (MND) is assumed either by omisson or by commission. However, marginal sample distributions of the data may point toward a non - MND, - a distribution which can be obtained by combining (possibly nonnormal) the marginal distributions through a suitable `copula'. In this work our focus is restricted to the bivariate case only.
	
	In a nutshell, a copula is a link function that combines two (or more) marginal distributions to construct a bivariate (or, multivariate) joint probability distribution. To be precise, let $X$ and $Y$ be two absolutely continuous random variables with marginal \textit{cdfs}, say, $F_X(x)$ and $F_Y(y)$, respectively. Then, according to the pathbreaking result due to Sklar(1959)~\cite{Sklar1959}, there exists a link function, called a `copula' $C$, such that the joint \textit{cdf} $F(x,y) \text{ of } (X, Y)$ can be expressed as
	
	\begin{equation}
		F(x,y) = C\big(F_X(x), F_Y(y)\big).
		\label{eq:sklar}
	\end{equation}
	
	By using the probability integral transforms, say 
	$U = F_X(X)$ and $V = F_Y(Y)$, 
	it is seen that both $U$ and $V$ are uniformly distributed over the interval $(0,1)$; with joint \textit{cdf} 
	
	\begin{equation}
		C(u,v) = P(U \leq u,\, V \leq v) 
		= P\!\big(X \leq F_X^{-1}(u),\, Y \leq F_Y^{-1}(v)\big).
		\label{eq:copula_definition}
	\end{equation}
	
	
	If we define $C^{(u,v)}(u,v)$ as
	\begin{equation}
		C^{(u,v)}(u,v) = 
		\frac{\partial}{\partial u} \frac{\partial}{\partial v} C(u,v) 
		= c(u,v) \quad \text{(say)},
		\label{eq:copula_density}
	\end{equation}
	
	\noindent then it is seen that vector $(U,V)^{\prime}$ has the \textit{pdf} $c(u,v)$ as given in \eqref{eq:copula_density}. Also, the joint \textit{pdf} of the original vector $(X,Y)^{\prime}$ can be expressed as 
	
	
	\begin{equation}
		f(x,y) = c \, \big(F_X(x), F_Y(y)\big)\, f_X(x)\, f_Y(y), 
		\label{eq:joint_density}
	\end{equation}
	
	\noindent where $f_X$ and $f_Y$ are the marginal \textit{pdfs} of $X$ and $Y$, respectively.
	\medskip
	
	In most of this work (except the last two sections) we are going to assume that the marginal \textit{cdfs} $F_X$ and $F_Y$ are completely known since our objective is to study the link function $C(u,v) (\text{or } c(u,v))$. This is because when the marginal \textit{cdfs} are unknown, they can be approximated fairly well by their empirical versions such as \[
	\hat{F}_X(s) = (1/n)\sum_{i=1}^n I(X_i \leq s)
	\quad \text{and} \quad
	\hat{F}_Y(t) = (1/n)\sum_{i=1}^n I(Y_i \leq t)
	\]
	based on $n$ \textit{iid} observations $(X_i, Y_i), \; 1 \leq i \leq n$, on $(X,Y)$. Sometimes a suitable adjustment is done  to avoid the extreme values $0$ and $1$ (at values smaller than the smallest entry, and larger than the largest entry, respectively) by using 
	
	\begin{equation}
		\hat{F}_X^{(\text{adj})}(s)
		=
		\{
		\textstyle{\sum_{i=1}^{n}} I(X_i \le s) + 0.5
		\}/(n+1),
		\label{eq:adjusted_ecdf}
	\end{equation}
	
	\noindent and likewise for $\hat{F}_Y^{(\text{adj})}(t)$ as well.
	\medskip
	
	More details about the general copula theory can be found in a recent work by Pham et al.(2025)~\cite{Pham et al.(2025)} or Genest et al.(1986)~\cite{Genest1986} and further references therein. To gauge the importance of general copula theory in banking and finance one may see the publications along with the further references therein: Bouye et al.(2000)~\cite{Bouye2000}, Cherubini et al.(2004)~\cite{Cherubini2004}, Embrechts et al.(2002)~\cite{Embrechts2002}, Huisman et al.(2001)~\cite{Huisman2001}, Meneguzzo et al.(2004)~\cite{Meneguzzo2004}, and Junker et al.(2006)~\cite{Junker2006}. 
	\medskip
	
	The objective of this paper is to focus on a specific copula, known as Frank Copula (FC), which has been found to have wide applications in finance and economic studies. The Standard Bivariate Frank Copula Distribution with parameter $\theta$ (or, $\mathbf{SBFCD}(\boldsymbol{\theta})$, in short) is obtained when $(U,V)^{\prime}$ has the joint \textit{pdf} on the unit square $(0,1) \otimes (0,1)$ given as 
	
	
	\begin{equation}
		c(u,v \mid \theta) \;=\;
		\frac{ \theta \bigl(1 - exp \,(-\theta)\bigr)\, \bigl(exp \,(-\theta (u+v))\bigr)}
		{ \left\{ exp \,(-\theta u) + exp \,(-\theta v) - exp \,(-\theta) - exp \,\bigl(-\theta (u+v)\bigr) \right\}^2 } \raisebox{0.5ex}{,}
		\label{eq:frank_density}
	\end{equation}
	
	\noindent	where the parameter $\theta \in \mathbb{R}$ indicates the association  (or, its strength) between  the components $U$ and $V$. The corresponding bivariate joint \textit{cdf} of $(U,V)$, i.e., the Frank Copula, is $C(u,v \mid \theta) 
	= \iint c(u,v \mid \theta)\,du\,dv $ with the expression 
	
	
	\begin{equation}
		C(u,v \mid \theta) 
		= -(1/\theta) \ln \!\left[ 1 + \{\,exp(-\theta u)-1\,\}\{\,exp(-\theta v)-1\,\}/ \{exp(-\theta)-1\} \right].
		\label{eq:frank_copula}
	\end{equation}

	The objective here is to estimate $\theta$ based on the \textit{iid} observations \((U_i, V_i)^\prime,\;1\leq i \leq n,
	\text{on } (U,V)^\prime \) following \eqref{eq:frank_density} (or \eqref{eq:frank_copula}).	
	\medskip
	
	Even though the literature mentions the parameter space (i.e., the permissible  range of $\theta$) as \(\Theta = \mathbb{R} \setminus \{0\}\), there is no harm in including the value $\{0\}$ in $\Theta$. However, the special case of $\theta = 0$ in \eqref{eq:frank_copula} requires a deft handling since a dicrect plug-in of $\theta = 0$ in \eqref{eq:frank_density} or \eqref{eq:frank_copula} yields $0/0$. Therefore, to see what happens at $\theta = 0$, one should take limit $\theta \to 0$ either by L'Hospital's rule or by using the approximations (when $\theta \approx 0$) as \(exp(-w) - 1 \;\approx\; - w\) and \(ln(1+w) \approx w\) as $w \to 0$. This yields \(c(u,v \mid \theta) = 1\) and \(C(u,v \mid \theta) = uv\), as $\theta \to 0$, thereby indicating the independence of the components $U$ and $V$.
	
	
	\subsection { Kendall's and Spearman's Rank Correlations as functions of \texorpdfstring{$\bm{\theta}$}{theta}}
	\medskip
	
	Note that the usual product--moment correlation, which is the Pearson's correlation ent, denoted by $\rho_P = \rho_P(X,Y)$, measures only the degree of linear association between two variables, say $X$ and $Y$. Further, the inferences on $\rho_P$, such as interval estimation and/or hypothesis testing, use the assumption of bivariate normality of $(X,Y)$	which may not be valid in many applications. Therefore, two other correlation measures which go beyond linear association
	and/or the normality assumption are -- Kendall's rank correlation
	(also known as ``Kendall's `tau' ") and Spearman's rank correlation,
	to be denoted by $\rho_K$ and $\rho_S$, respectively.
	\medskip
	
	Suppose $(X_*,Y_*)$ and $(X_{**},Y_{**})$ be two \text{iid} observations
	on the bivariate vector $(X,Y)$ having some continuous distribution
	(\textit{cdf}) $F(x,y)$. The pairs $(X_*,Y_*)$ and $(X_{**},Y_{**})$ are called
	concordant (or discordant) if
	$
	\Delta_0 \text{ (say)} =(X_* - X_{**})(Y_* - Y_{**}) \, $ $ > \text{(or } < ) \,0$. Then $\rho_K$ is defined as
	
	\begin{equation}
		\rho_K = \pi_C - \pi_D,
		\label{eq:kendall_tau1}
	\end{equation}
	
	\noindent where $ \pi_C = P(\Delta_0 > 0) $ and $ \pi_D = P(\Delta_0 < 0).$ Notice that $\rho_K \in [-1,1]$, and $\rho_K=-1$ (or $1$)
	in the case of a perfect negative (or positive) association. Also, the nonparametric estimate $\hat{\rho}_K$ of $\rho_K$ can be obtained from the data $(X_i,Y_i)$, $1 \le i \le n$, by using
	$
	\hat{\rho}_K=\hat{\pi}_C-\hat{\pi}_D,
	$
	where $\hat{\pi}_C$ and $\hat{\pi}_D$ are the sample estimates of
	$\pi_C$ and $\pi_D$, respectively. These estimates can be obtained
	as the proportion of concordant and discordant pairs out of $ N=\binom{n}{2} $ possible pairwise comparisons among the $n$ observations.
	\medskip
	
	On the other hand, Spearman's rank correlation, also known  as the ``grade correlation", is the Peason's correlation between $U = F_X(X)$ and $V = F_Y(Y)$, i.e., 
	
	\begin{equation}
		\rho_S = \rho_P(U, V).
		\label{eq:Spearman}
	\end{equation}
	
	In a sample setting, one can replace $U$ and $V$ by
	$U_i=\hat{F}_X(X_i),$ and $V_i=\hat{F}_Y(Y_i)$ (or by $\hat{F}_X^{adj}(X_i)$ and $\hat{F}_Y^{adj}(Y_i)$,
	or simply by $rank(X_i)$ and $rank(Y_i)$), respectively, and obtains the nonparametric estimate of $\rho_S$ as
	$
	\hat{\rho}_S=(1/n)\sum_{i=1}^{n}U_i^{*}V_i^{*}
	$
	where $U_i^{*}$ and $V_i^{*}$ are the standardized values of
	$U_i$ and $V_i$, respectively;
	$
	U_i^{*}= (U_i-\overline{U})/s_U, \,
	V_i^{*}=(V_i-\overline{V})/s_V,
	$
	with
	$
	s_U=(\sum_{i=1}^{n}(U_i-\overline{U})^2/n)^{1/2}
	$
	and
	$
	s_V=(\sum_{i=1}^{n}(V_i-\overline{V})^2/n)^{1/2}.
	$
	($\overline{U}$ and $\overline{V}$ are the means of $U_i$ and $V_i$ values, respectively.) Again, note that the range of $\rho_S$ (and that of $\hat{\rho}_S$)
	is $[-1,+1]$.
	\medskip
	
	In the context of the above-mentioned Frank Copula set-up, both $\rho_K$ and $\rho_S$ are functions of the copula association parameter $\theta$, given as (see Genest (1987))~\cite{Genest1987}
	
	
	\begin{equation}
		\rho_K
		= \rho_K(\theta)
		= 1 - (4/\theta)\{1 - D_1(\theta)\}, \, \text{and}
		\label{eq:kendall_tau}
	\end{equation}
	
	
	\begin{equation}
		\rho_{S}
		= \rho_{S}(\theta)
		= 1 - (12/\theta)\{D_1(\theta) - D_2(\theta)\},
		\label{eq:spearman_rho}
	\end{equation}
	
	\noindent where 
	$ D_1(\theta)
	= (1/\theta)
	\int_{0}^{\theta}
	t / \{\exp(t) - 1\}
	\,dt
	$
	and 	$ D_2(\theta)
	= (2/\theta^2)
	\int_{0}^{\theta}
	t^{2} / \{\exp(t) - 1\}
	\,dt.
	$
	\noindent
	Therefore, when a given bivariate dataset is assumed to obey the Frank Copula-based joint distribution, one can estimate $\rho_K$ and $\rho_S$ by their parametric estimates as
	$
	\hat{\hat{\rho}}_K = \rho_K(\hat{\theta}) \quad \text{and} \quad \hat{\hat{\rho}}_S = \rho_S(\hat{\theta}),
	$
	respectively, where $\hat{\theta}$ is a suitable estimatior of $\theta$ (and we will see more on this in section 5).
	\medskip
	
	The following Figure~\ref{fig:Ken_Spear} shows the plots of $\rho_K$ and $\rho_{S}$ as functions of $\theta$.
	
	
	\begin{figure}[H]
		\centering
		\includegraphics[width=0.6\textwidth]{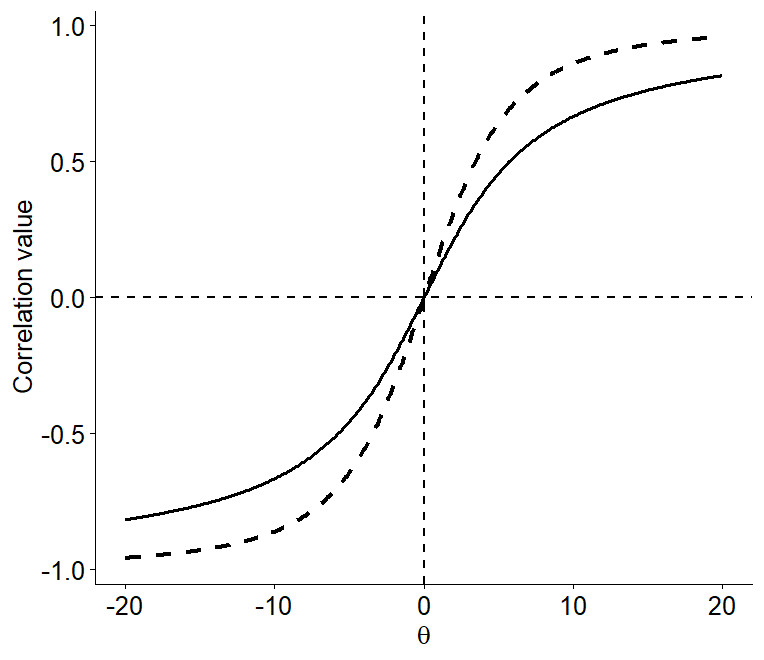}
		\caption{Plots of $\rho_K$ (solid line (\ref{eq:kendall_tau} )) and $\rho_S$ (dash line (\ref{eq:spearman_rho})) as functions of $\theta$}
		\label{fig:Ken_Spear}
	\end{figure}


	
	\subsection{A Motivating Example: Vietnam Groundwater Contamination Data}
	
	Severe arsenic concentration in groundwater has been a major problem
	in densely populated areas of South/Southeast Asian countries. One of
	the hotspots that has been severely affected by arsenic contamination
	in groundwater is the Mekong Delta Region (MDR) in the southern part of
	Vietnam. For an indepth and comprehensive history of arsenic contamination in MDR one may see Berg et al.(2007)~\cite{Berg2007}, Nguyen(2008)~\cite{Nguyen2008}, Pham(2015)~\cite{Pham2015}, and Pham et al.(2015)~\cite{Pham2015b}.
	\medskip
	
	Researchers have been carrying out surveys in MDR to assess arsenic prevalence and to have a better understanding of how this carcinogenic element is associated with other benign elements present in groundwater. Along with concentration of Arsenic (\textit{As}), which is well known to have adverse effects on human health when consumed beyond a safe limit, concentration of Chloride (\textit{Cl}), Redox level (\textit{Eh}), and pH level (\textit{pH}) of the groundwater were recorded based on samples collected from several water-wells scattered across Dong Thap province as reported by
	Merola et al.(2015)~\cite{Merola2015}. The study area covered two subregions, namely North and South, and the complete dataset pertaining to the aforementioned elements has been
	reproduced in Appendix \ref{app:groundwater_data}
	\medskip
	
	The fact that the two subregions (North and South) are vastly
	different can be seen from the following Table~\ref{tab:kst_pvalues}, which gives the p-values of the Kolmogorov--Smirnov Test (KST) comparing
	the empirical distributions of each element (or, say, variable)
	for the two subregions.
	
	
	\begin{table}[h]
		\centering
		\caption{The p-values of KST to test the equality of distributions in two subregions}
		\label{tab:kst_pvalues}
		\resizebox{0.95\textwidth}{!}{
			\begin{tabular*}{\textwidth}{@{\extracolsep{\fill}}c cccc}
				\hline
				North vs.\ South & \textit{As} & \textit{Cl} & \textit{Eh} & \textit{pH} \\
				\hline
				p-value & < 0.0001 & 0.0002 & < 0.0001 & < 0.0001 \\
				\hline
			\end{tabular*}
		}
	\end{table}
	
	Table \ref{tab:kst_pvalues} makes it clear that the probability distribution of each variable differs significantly from North to South. (More details on the dataset are provided in Section 5.) 
	\medskip	
	
	The relative frequency histograms of the four variables (\textit{As, Cl, Eh} and \textit{pH}) in two subregions appear to be mostly non-normal as the following Table \ref{tab:normality_tests} provides p-values of two formal normality tests, namely-- Anderson--Darling Test (ADT), and Shapiro--Wilk Test (SWT). The sample sizes of North and South are $\mathbf{n_N = 23}$ and $\mathbf{n_S = 44}$, respectively.
	
	
	\begin{table}[H]
		\centering
		\caption{ADT and SWT p-values to test normality in two subregions}
		\label{tab:normality_tests}
		
		\small
		\renewcommand{\arraystretch}{1.2}
		
		\begin{tabular*}{0.95\textwidth}{@{\extracolsep{\fill}}c ccccc}
			\hline
			Test & Subregion & \textit{As} & \textit{Cl} & \textit{Eh} & \textit{pH} \\
			\hline
			\multirow{2}{*}{ADT} & North & < 0.0001 & 0.0413 & 0.1404 & 0.0029 \\
			& South & 0.0001 & < 0.0001 & < 0.0001 & 0.4363 \\
			\hline
			\multirow{2}{*}{SWT} & North & < 0.0001 & 0.0286 & 0.0914 & 0.0012 \\
			& South & 0.0002 & < 0.0001 & < 0.0001 & 0.6933 \\
			\hline
		\end{tabular*}
		
	\end{table}
	
	Even though the four variables in two subregions appear to have distinct and mostly nonnormal distributions, it should be kept in mind that the two normality tests are asymptotic in nature and the p-values may be inaccurate, especially for North with a sample size of $23$ only.
	\medskip
	
	The three correlation coefficients, -- $\rho_P, \rho_K$ and $\rho_S$, have been estimated for four pairs of variables (which are relevant) in two subregions, and these have been shown in Table~\ref{tab:correlation_estimates}. The objective of this study is to explore how the carcinogenic element \textit{As} is associated with the other three benign elements (\textit{Cl, Eh} and \textit{pH}). Also, we have included the correlations between \textit{Eh} and \textit{pH} due to their association as explained in Remark~\ref{rem:remark1.2}.
	

	\begin{table}[H]
		\centering
		\caption{Three standard correlation estimates in two subregions}
		\label{tab:correlation_estimates}
		
		\small
		\renewcommand{\arraystretch}{1.2}
		\setlength{\tabcolsep}{4pt}
		
		\begin{tabular}{|c|cccc|cccc|}
			\hline
			& \multicolumn{4}{c|}{North ($n_N = 23)$} & \multicolumn{4}{c|}{South ($n_S = 44)$} \\
			\cline{2-9}
			
			\makecell{Estimated\\Correlation}
			& (\textit{Cl, As}) & (\textit{Eh, As}) & (\textit{pH, As}) & (\textit{Eh, pH})
			& (\textit{Cl, As}) & (\textit{Eh, As}) & (\textit{pH, As}) & (\textit{Eh, pH}) \\
			
			\hline
			
			$\hat{\rho}_P$ 
			& -0.183 & -0.525 & 0.754 & -0.503
			& -0.325 & -0.668 & 0.119 & -0.142 \\
			
			$\hat{\rho}_K$ 
			& 0.012 & -0.323 & 0.188 & -0.429
			& -0.230 & -0.577 & 0.101 & 0.062 \\
			
			$\hat{\rho}_S$ 
			& 0.018 & -0.414 & 0.260 & -0.545
			& -0.320 & -0.753 & 0.156 & 0.050 \\
			
			\hline
		\end{tabular}
		
	\end{table}
	
	
	\begin{remark}
		\label{rem:remark1.1}
		Even though Pearson's correlation estimates (i.e., $\hat{\rho}_P $) have been provided for the three pairs in two subregions, they should be discounted due to the non-normal nature of the data. In terms of $\hat{\rho}_K$ and $\hat{\rho}_S$, $(a)$ As and pH appear to be mildly correlated in both the subregions; $(b)$ As and Cl appear to be correlated only in South, but \underline{not} in North; and $(c)$ As and Eh appear to be correlated in both North and South (especially, strongly in South). By the way, the sampled water wells in North, in the area Tan Hong, is far from the Tien river, whereas those in South, in Thanh Binh, are adjacent to Tien river.
	\end{remark}

	
	\begin{remark}
		\label{rem:remark1.2}
		The pair $(Eh,pH)$ plays a fundamental role in hydro-geochemistry.
		The variable $Eh$, known as the redox potential, measures the
		oxidation--reduction condition of the aqueous environment, whereas
		$pH$ measures the acidity or alkalinity of the water. Many geochemical reactions occurring in groundwater involve both
		electron transfer and proton exchange. As a result, the chemical
		equilibrium of many dissolved elements depends simultaneously on the redox condition and the acidity of the medium. In general, $Eh$ controls the oxidation state of chemical elements, while $pH$
		influences their solubility and speciation in aqueous systems. 	
		For this reason, the joint analysis of $(Eh,pH)$ is essential in
		hydro-geochemical studies. The relationship between these two
		variables is often illustrated using the well-known $Eh$--$pH$
		diagram (also called the Pourbaix diagram), which identifies the
		stable chemical forms of elements under different environmental
		conditions. Therefore, investigating the dependence structure between $Eh$ and
		$pH$ can provide important insights into the geochemical processes
		governing groundwater systems.	
		
	\end{remark}
	
	
	\begin{remark}
		\label{rem:remark1.3}
		Going beyond $\rho_K$ and $\rho_S$, can we model the bivariate distributions of the aforementioned three pairs by Frank Copula? This will be explored in more details in Section 5. 
		
	\end{remark}
	
	
	\subsection{A synopsis of a most recent work and a summary of this work}
	
	As mentioned earlier, this work is a sequel to the most recent work of Pham et al.(2025)~\cite{Pham et al.(2025)} which complemented Genest's (1987)\cite{Genest1987} expository work on Frank Copula in a bivariate set-up using the model \eqref{eq:frank_density} which focused on estimating $\theta$ 
	using three classical estimators, namely -- the maximum likelihood estimator (MLE), the method of moment estimator of first kind (MME1) (based on Kendall's rank correlation), and the method of moment estimator of second kind (MME2) (based on Spearman's rank correlation). It was shown through a comprehensive simulation study, 
	among other things, that:
	\begin{enumerate}[label=(\alph*)]  
		\item In terms of bias, the MLE over-estimates $\theta$ when $\theta > 0$, 
		and under-estimates $\theta$ when $\theta < 0$. On the other hand, 
		the two MMEs act in completely opposite directions, i.e., 
		under (over) estimate $\theta$ when $\theta > (<) 0$. 
		At $\theta = 0$, all the three estimators have bias $= 0$. Also, in terms of absolute bias, the MLE out performs the MMEs by wide margins.
		
		\item In terms of MSE, again the MLE exhibits much superior performance than the MMEs, except for a small neighborhood of $0$ where the MMEs tend to perform better than the MLE. This neighborhood of $0$, given as \(\Theta_{\varepsilon(n)} = \{ \theta \mid |\theta| < \varepsilon \,\}\), where $\varepsilon = \varepsilon(n)$ is a function of $n$, and shrinks (i.e., $\varepsilon$ decreases) as $n$ increases.
		
		\item Further, the asymptotic behavior of the MLE was explored, and it was found that the asymptotic variance of the MLE 
		(i.e., the CramÃ©r--Rao lower bound) is attained fairly well with a minimum sample size of $75$.
	\end{enumerate}
	
	This current work extends the above point estimation results of $\theta$ further by considering two Bayes estimators, and to the best of our knowledge this was never done before for Frank Copula. (As such, there is a dearth of studies when it comes to copula parameter estimation from a Bayesian point of view, and hopefully this work will pique the interest of other researchers in this direction.)
	\medskip
	
	Two noninformative (generalized) priors have been considered to study the corresponding Bayes estimators of $\theta$ as follows.
	
	\begin{enumerate}[label=(\roman*)]
		\item The `flat prior' $\pi_{FP}(\theta) = d > 0,$ over $\Theta = \mathbb{R}$ has been used to construct the Bayes Flat Prior Estimator (BFPE) under the squared error loss, i.e., the posterior mean under the flat prior.
		
		\item Jeffreys prior \(\pi_J(\theta) = (I(\theta))^{1/2}\) over $\Theta = \mathbb{R}$ has been used to construct the Bayes Jeffreys Prior Estimator (BJPE), where $I(\theta)$ = Fisher information (per observation) (FIPO) with $(U,V) \sim c(u,v \mid \theta) $ in (\ref{eq:frank_density}). (Considerable care has been taken to compute BJPE numerically as it involves a far more complex structure than BFPE.)
		
		\item Both BFPE as well as BJPE have been compared against the MLE in terms of bias and MSE in a comprehensive way. Interestingly, the simulation results indicate that, in terms of MSE, BJPE beats the other two estimators uniformly for ``small" $n$ (i.e., $n \leq 25$). For ``large" $n$, i.e., $n > 25$, the dominance of BJPE is marginal, and all the three estimators become ``statistically equivalent". As a result, the take away message of this investigation is that for $n \leq 25$, one definitely should use BJPE over the MLE (provided the bivariate data can be modelled by Frank Copula).
		
		\item As discussed earlier, this work concludes with an interesting application of Frank Copula 
		to model several bivariate pairs of groundwater elements (or variables), including Arsenic.
		The estimated association parameter for each pair gives a better idea about the dynamics 
		between the pair components. But while undertaking this application we have turned our 
		attention to a more fundamental problem, that is  - the goodness of fit (GoF) test of Frank Copula 
		for a given bivariate dataset. We have revisited the seminal work of Genest et al.(2006)~\cite{GenestRemillard2006} and reexamined the behavior of two test statistics for our application purposes. In this process, we have provided expanded  tables of critical values for the two test statistics which are not only comprehensive and more accurate (due to the far larger number of replications used), but also complementary to the above work.
		
		\item In a real-life application, one needs to revert back to $(X,Y)$ from $(U,V) \; (\sim \mathrm{SBFCD}(\theta))$ in order to fit $Y$ from $X$. In this regard, one may use the nonparametric marginal transformations $X \to U$ and $Y \to V$ (and the reverse transformations) using (\ref{eq:adjusted_ecdf}),
		or may adopt suitable parametric marginal models. For our groundwater data it has been noticed that the nonparametric marginals tend to perform better (over the parametric approach) when the goal is to fit $Y$ using the Frank Copula-based regression model.
		
	\end{enumerate}
	
	The rest of the paper has been organized as follows. In section 2 we introduce the two noninformative generalized priors as well as the corresponding Bayes estimators. Section 3 has been devoted to a comprehensive simulation study comparing the two Bayes estimators against the MLE in terms of bias as well as MSE. In section 4 we revisit the two goodness of fit (GoF) tests for Frank Copula suggested by Genest et al.(2006)~\cite{GenestRemillard2006} and explore some of their properties. In Section 5 we demonstrate the applicability of bivariate Frank Copula to model the groundwater data, and obtain the estimated association parameter between the two components of various pairs of groundwater elements. 
	
	\medskip
	In Section 6 we explore the implications of using nonparametric marginals as well as parametric marginals when the objective is to fit $Y$ using the Frank Copula-based regression model in the context of our groundwater data. Since this work extends the in-depth study of Pham et al.(2025)~\cite{Pham et al.(2025)}, some of the details and lengthy derivations would be left out for brevity, and any interested reader can check the details provided in the aforementioned last paper. The details of our computational code for Bayes estimation and related matters have been provided in Pham et al. (2026).

	\section{The generalized Bayes estimators under two noninformative priors}
	
	It is assumed that we have \textit{iid} observations $(U_i, V_i), 1 \leq i \leq n$, following the \textit{pdf} \eqref{eq:frank_density}. The likelihood function of the `data' $ = \{(U_i, V_i) \mid 1 \leq i \leq n \}$ is 
	
	\begin{equation}
		L(\theta \mid \text{data}) = \textstyle{\prod_{i=1}^n} \, c \,(U_i, V_i \mid \theta),
		\label{eq:likelihood}
	\end{equation}
	where $c(u,v \mid \theta)$ is given in \eqref{eq:frank_density}.
	
	The MLE of $\theta$ is found by solving the normal equation, say $H(\theta) = 0$, where 
	
	\begin{equation}
		H(\theta) = (1/\theta) + \big( \mathit{exp}(\theta) - 1 \big)^{-1} 
		- (\overline{U} + \overline{V}) + (2/n) \, \textstyle{\sum_{i=1}^n} \left( A_{1i}/A_{2i} \right); 
		\label{eq:score_function}
	\end{equation}
	
	\begin{align*}
		A_{1i} &= A_{1i}(U_i, V_i) = \{ U_i  \mathit{exp}(-\theta U_i) 
		+ V_i  \mathit{exp}(-\theta V_i) - \mathit{exp}(-\theta) 
		- (U_i + V_i)  \mathit{exp}\big(-\theta (U_i+V_i)\big) \}, \\[6pt]
		A_{2i} &= A_{2i}(U_i, V_i) = \{ \mathit{exp}(-\theta U_i) 
		+ \mathit{exp}(-\theta V_i) - \mathit{exp}(-\theta) 
		- \mathit{exp}\big(-\theta (U_i+V_i)\big) \}.
	\end{align*}
	
	The existence of the MLE, henceforth denoted by $\hat{\theta}_{ML}$, and the behavior of the function $H(\theta)$, has been thoroughly discussed in Pham et al.(2025)~\cite{Pham et al.(2025)}.\\
	
	Suppose $\pi (\theta)$ is a prior on the parameter space $\Theta = \mathbb{R}$. Then the posterior distribution of $\theta$ given the `data' is 
	
	\begin{equation}
		\pi (\theta \mid \text{data}) = \pi(\theta) \, \textstyle{\prod_{i=1}^n} c(U_i, V_i \mid \theta) \Big/ \int_{-\infty}^{\infty} \pi(\theta) \, \textstyle{\prod_{i=1}^n} c(U_i, V_i \mid \theta) \, d\theta,
		\label{eq:posterior}
	\end{equation}
	
	Under the squared error loss function, the Bayes rule $\hat{\theta}_B$ is the posterior mean, i.e., 
	
	\begin{equation}
		\hat{\theta}_B = E(\theta \mid \theta \sim \pi(\theta \mid \text{data} ) ) = \textstyle{\int_{-\infty}^{\infty}} \theta \, \pi(\theta \mid \text{data}) \, d\theta,
		\label{eq:bayes_estimator}
	\end{equation}
	provided the above mean exists. 
	
	\medskip
	The first prior we are going to consider is the noninformative generalized `flat prior', given as $\pi_{FP}(\theta)$, where
	\begin{equation}
		\pi_{FP}(\theta) = d > 0, \quad \theta \in \Theta = \mathbb{R}. 
		\label{eq:flat_prior}		
	\end{equation}
	
	\medskip
	The resultant Bayes Flat Prior Estimator (BFPE), denoted by $\hat{\theta}_{BFPE}$, is (from \eqref{eq:bayes_estimator} and \eqref{eq:flat_prior})
	
	\begin{equation}
		\hat{\theta}_{BFP} = \textstyle{ \int_{-\infty}^{\infty}} \theta \, \prod_{i=1}^n c \,(U_i,V_i \mid \theta)  \, d\theta \Big/ \int_{-\infty}^{\infty} \textstyle {\prod_{i=1}^n} c \,(U_i,V_i \mid \theta)  \, d\theta.
		\label{eq:bayes_flat_estimator}
	\end{equation}
	
	The second prior we are going to consider is the noninformative invariant generalized `Jeffreys prior', given as $\pi_{JP} (\theta)$, where 
	
	
	\begin{equation}
		\pi_{JP}(\theta) = (I(\theta))^{1/2}, \quad \theta \in \Theta = \mathbb{R}, \label{eq:jeffreys_prior}
	\end{equation}
	\noindent	with $I(\theta)$ being the Fisher Information Per Observation (FIPO).
	\medskip
	
	The above $\pi_{JP}(\theta)$ has the unique feature that it is invariant under any reparameterization. Note that the flat uniform prior $\pi_{FP}(\theta)$ does not possess this property. (For more on invariant prior selection see Jeffreys (1946)~\cite{Jeffreys1946} or Jaynes (1968)~\cite{Jaynes1968}.)  The following Figure~\ref{fig:figure21} shows 
	the plots of the two aforementioned priors against the association parameter $\theta$.
	\medskip
	
	For the Frank Copula \textit{pdf} \eqref{eq:frank_density} the FIPO expression of $I(\theta)$ is a bit cumbersome, and does not have a closed form. It is given as (see Pham et al.(2025)~\cite{Pham et al.(2025)})
	
	\begin{equation}
		I(\theta) = \mathbb{E}\!\left[-\,(\partial^{2}/\partial\theta^{2}) \ln c(U,V \mid \theta)\right] = I_1(\theta) - I_2(\theta),
		\label{eq:fisher_information}
	\end{equation}
	
	{
		\begin{fleqn}
			\[
			\text{where } I_1(\theta) = \theta^{-2} + \mathit{exp}(\theta) / (\mathit{exp}(\theta) - 1)^2 ;
			\quad \text{and } I_2(\theta) = 2 \mathbb{E}[J(U,V \mid \theta)],
			\]
			\[
			\text{with } J(U,V \mid \theta) = \{J_1(U,V \mid \theta) / J_2(U,V \mid \theta)\}, 
			\quad (U,V) \sim c(u,v \mid \theta) \text{ (in \eqref{eq:frank_density}),}
			\]
			\[
			J_1(U,V \mid \theta) = \mathit{exp}(-\theta(U+V)) \Big\{
			-(U-V)^2 + U^2 \mathit{exp}(-\theta V) + V^2 \mathit{exp}(-\theta U)
			\]
			\[
			\hspace{2.5 cm} - (U+V-1)^2 \mathit{exp}(-\theta) \Big\} + (V-1)^2 \mathit{exp}(-\theta(V+1))
			+ (U-1)^2 \mathit{exp}(-\theta(U+1)) ,
			\]
			\[
			J_2(U,V \mid \theta) = \Big\{
			\mathit{exp}(-\theta U) + \mathit{exp}(-\theta V) - \mathit{exp}(-\theta) - \mathit{exp}(-\theta(U+V))
			\Big\}^2.
			\]
		\end{fleqn}
	}
	
	Note that while the term $I_1(\theta)$ is a simple expression, the term $I_2(\theta)$ has to be evaluated through numerical double integration
	
	
	\begin{equation}
		\mathbb{E}(J(U,V \mid \theta)) = 
		\int_{0}^{1} \!\! \int_{0}^{1} 
		J(u,v \mid \theta)\, c(u,v \mid \theta)\, du\, dv,
		\label{eq:expectation_J}
	\end{equation}
	for every $\theta$ in $\Theta = \mathbb{R}$, which is a bit time consuming.
	\medskip
	
	The Bayes rule under $\pi_{JP}(\theta)$, henceforth referred to as Bayes Jeffreys Prior Estimator (BJPE), and denoted by $\hat{\theta}_{BJPE}$, is 
	
	
	\begin{equation}
		\hat{\theta}_{BJPE} = \textstyle{ \int_{-\infty}^{\infty} } \theta \, (I(\theta))^{1/2} \, \textstyle{\prod_{i = 1}^n} c(U_i,V_i \mid \theta) \, d\theta \Big/ \textstyle{ \int_{-\infty}^{\infty} } \, (I(\theta))^{1/2} \, \textstyle{\prod_{i = 1}^n} c(U_i,V_i \mid \theta) \, d\theta.
		\label{eq:bayes_jeffreys_estimator}
	\end{equation}
	
	
	\begin{figure}[H]
		\centering
		\includegraphics[width=0.78\textwidth]{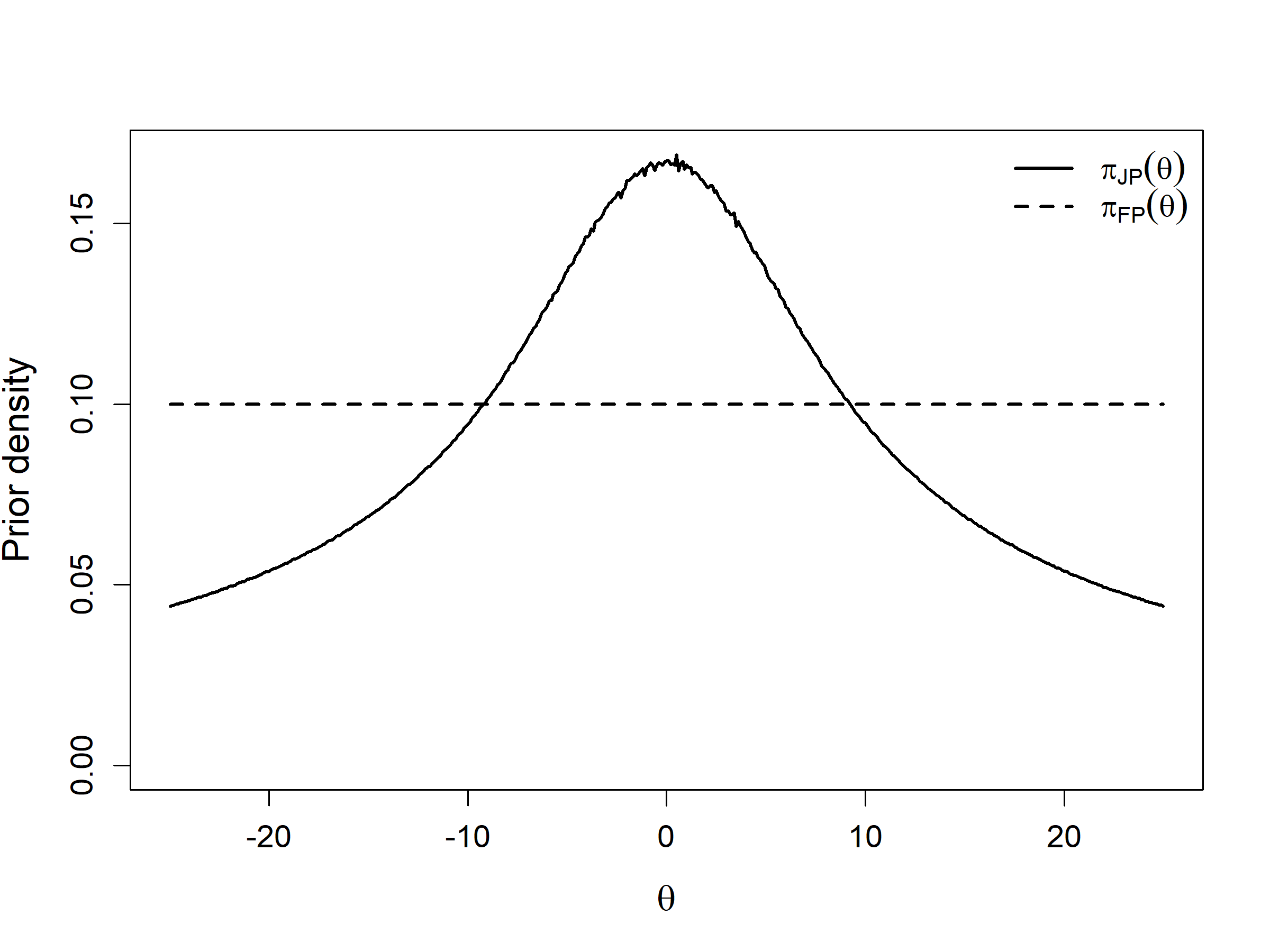}
		\caption{The priors $\pi_{FP}(\theta)$ and $\pi_{JP}(\theta)$ plotted against the association parameter $\theta$.}
		\label{fig:figure21}
	\end{figure}
	\vspace{-0.3cm}
	In Section 3 we have presented the results of our simulation study to compare the two Bayes estimators against the MLE in terms of bias and MSE. Computation of the BJPE requires triple integrations, and this has been done with utmost care to attain a high precision as discussed later. For this reason, each computed value of bias and MSE has been provided with its own standard error (SE) value. Also note that the bias and MSE of the MLE are consistent with those reported in Pham et al.(2025)~\cite{Pham et al.(2025)}.
	
	\section{A Comparison of MLE, BFPE and BJPE in terms of Bias and MSE}
	
	A comparision of the three estimators, i.e., MLE, BFPE and BJPE, has been undertaken through a comprehensive simulation study based on $M = 4 \times 10^4$ replications. For a generic estimator $\hat{\theta}$, let $\hat{\theta}^{(m)}$ be its computed value in the \( m^{\text{th}}\) replication, $1 \leq m \leq M$. Then the simulated Bias and MSE of $\hat{\theta}$ are computed as
	\[
	Bias(\hat{\theta}) \simeq \, \sum_{m=1}^M \, (\hat{\theta}^{(m)} - \theta)/M \text{ and } MSE(\hat{\theta}) \simeq \, \sum_{m=1}^M \, (\hat{\theta}^{(m)} - \theta)^2 / M.
	\]
	The computation of the MLE by solving \eqref{eq:score_function} is straightforward  as described  in Pham et al.(2025)\cite{Pham et al.(2025)}. However, there is a minor issue for very small sample sizes $(n \leq 10)$ which has been clarified in Remark~\ref{rem: 3.1}.
	\medskip
	
	The numerical integrations involved in the numerator and denominator of $\hat{\theta}_{BFPE}$ (in \eqref{eq:bayes_flat_estimator}) and $\hat{\theta}_{BJPE}$ (in \eqref{eq:bayes_jeffreys_estimator}) have been done through adaptive quadrature method (in R package), and this was a bit time consuming given the large number of replications (i.e., $M = 4 \times 10^4$) involved. Each numerical integration with respect to $\theta$ over the range $\Theta = \mathbb{R} = (-\infty, \infty )$ was approximated by the subrange $(\theta - A, \theta + A)$ with $A = 25$. (In fact, it was verified that such integrations do not change beyond $A = 15$).  The subrange $(\theta - A, \theta + A)$ was then subdivided into $L = 2,000$ subintervals (with each subinterval having a width of $0.025$) to approximate the integration over the subrange $(\theta - A, \theta + A)$ by the corresponding Riemann sum.
	\medskip
	
	For each given value of $\theta$ (i.e., the input parameter value), which ranged from $(-10)$ to $(+10)$, computations of Bias and MSE of each estimator (based on $M = 4 \times 10^4$ replications) were done in four smaller batches of $10^4$ replications each using Ton Duc Thang University's (TDTU's) High Performance Computing (HPC) facility. The simulations were executed using the R packages \texttt{foreach} and \texttt{doParallel} to perform the Monte Carlo replications in parallel, thereby substantially reducing the computational time. This was done for two reasons; -- (a) to obtain the results based on each  batch (of $10^4$ replications) in a reasonable amount of time (which ranged from 5 to 8 minutes for each combination of $n$ and $\theta$), and then use the set of four batch results to compute the standard error (SE) of the overall MSE value. The SE of each Bias value was computed by the overall MSE (obtained from $M = 4 \times 10^4$ replications) divided by $M$, and then taking its square-root. 
	
	\medskip
	The following tables (Table~\ref{tab:sim_bias_mse_a}~(a) - (b)) show the simulated Bias and MSE of the three estimators only for positive $\theta$ values ranging from $0.1$ to $10.0$ (due to the reasons explained in Remark~\ref{rem: 3.2}). The SE of each Bias and MSE value has been reported in brackets next to it.

	\begin{remark}
		\label{rem: 3.1}
		Obtaining the MLE of $\theta$ for a given dataset using R package requires some careful observations. One can use either the ``optim" command to maximize the log-likelihood function $\ell(\theta) = ln L (\theta)$ (in \eqref{eq:likelihood}) directly, or use the ``nleqslv" command to solve the normal equation `$H(\theta) = 0$' (in \eqref{eq:score_function}). Theoretically there is no difference between these two approaches, but the computational results do vary slightly, especially for very small sample sizes $(n \leq 10)$. For such small samples, the above second approach appears to be less accurate, thereby yielding larger Bias and MSE compared to those of the first approach. However, these two approaches yield almost identical  results for $n \geq 15$, and that is why we have used $n \geq 15$ wherever applicable. A summary of the above two approaches is given in Appendix~\ref{app:mle_comparison}.
	\end{remark}
	
	
	\begin{remark}
		\label{rem: 3.2}
		The simulation study has revealed the following trends:
	\begin{enumerate}[
		label=(\roman*),
		leftmargin=0pt,
		itemindent=3.0em,
		labelsep=0.3em,
		labelwidth=2.3em,
		align=left,
		topsep=6pt,
		itemsep=8pt
		]
			
		\item All the three estimators (MLE, BFPE and BJPE) tend to over (under) estimate $\theta$ when $\theta > (< ) \,0$. Further, the Bias, as a function of $\theta$, appears to be symmetric about $0$, i.e., Bias at $(-\theta) = - (\text{Bias at} \, (+\theta)).$ The Bias in estimation approaches to $0$ as $\theta \to 0$ (for any fixed $n$), and/or as $n \to \infty$ (for any fixed $\theta$). Also, in terms of Bias, both BJPE and MLE seem to be better than BFPE, and these two estimators have nearly identical behavior.
			\item In terms of MSE, the estimators are symmetric in $\theta$ about 0, and interestingly BJPE shows uniformly better performance over the other two estimators. However, this superiority of BJPE is more pronounced for small sample sizes, i.e., $n \leq 25$ (as shown in boldface). For larger $n \,(n > 25)$, the improvements provided by BJPE is marginal, especially over the MLE, and these are statistically insignificant.
		\end{enumerate}
		
	\end{remark}

	
	\begin{remark}
		\label{rem: 3.3}
		Our simulation study has used $\theta$ ranging from $0.1$ (near 0) to 10 only, and this is due to the symmetric behavior as noted before. Even though both the Bias (in absolute sense) and the MSE are increasing with respect to $|\theta|$ (for fixed $n$), the rate of increment is actually decreasing. This can be seen from the standardized Bias (SB) and standardized MSE (SMSE) defined as $SB = (Bias/|\theta|)$, and $SMSE =  (MSE/\theta^2)$. It is easy to see (but not reported here explicitly) that for all the three estimators both $SB$ and $SMSE$ are exploding near $|\theta| = 0$, and are of order $o(|\theta|)$ as $|\theta| \to \infty$.
	\end{remark}
	
	
	\begin{table}[H]
		\centering
		\caption{(a). Simulated Bias and MSE of three estimators of $\theta$ ($n=15,20,25$).}
		\label{tab:sim_bias_mse_a}
		
		\renewcommand{\arraystretch}{1.15}
		\setlength{\tabcolsep}{6.0pt}
		\resizebox{0.92\textwidth}{!}{%
			\begin{tabular}{|c|c|c|c|c|c|c|c|}
				\hline
				\multirow{2}{*}{$n$} & \multirow{2}{*}{$\theta$} &
				\multicolumn{2}{c|}{MLE} &
				\multicolumn{2}{c|}{BFPE} &
				\multicolumn{2}{c|}{BJPE} \\ \cline{3-8}
				&  & Bias(SE) & MSE(SE) & Bias(SE) & MSE(SE) & Bias(SE) & MSE(SE) \\
				\hline
				
				\multirow{14}{*}{15} & 10   & 0.644(0.013) & 9.777(0.039) & 1.102(0.013) & 11.237(0.028) & 0.658(0.013) & \textbf{9.558} (0.049) \\
				& 9    & 0.605(0.007) & 8.630(0.033) & 1.010(0.006) & 10.041(0.044)  & 0.620(0.005) & \textbf{8.479}(0.023) \\
				& 8    & 0.528(0.008) & 7.268(0.036) & 0.875(0.009) & 8.465(0.041)  & 0.543(0.009) & \textbf{7.219}(0.034) \\
				& 7    & 0.485(0.003) & 6.333(0.027) & 0.774(0.002) & 7.328(0.038)  & 0.499(0.002) & \textbf{6.297}(0.035) \\
				& 6    & 0.439(0.007) & 5.360(0.043) & 0.668(0.007) & 6.178(0.044)  & 0.448(0.007) & \textbf{5.362}(0.038) \\
				& 5    & 0.362(0.007) & 4.694(0.013) & 0.532(0.009) & 5.329(0.017)  & 0.364(0.008) & \textbf{4.665}(0.048) \\
				& 4    & 0.302(0.007) & 4.060(0.038) & 0.417(0.006) & 4.527(0.041)  & 0.297(0.006) & \textbf{4.088}(0.038) \\
				& 3    & 0.220(0.005) & 3.663(0.040) & 0.288(0.005) & 3.912(0.039)  & 0.208(0.005) & \textbf{3.630}(0.038) \\
				& 2    & 0.122(0.009) & 3.303(0.031) & 0.159(0.009) & 3.446(0.034)  & 0.111(0.009) & \textbf{3.282}(0.032) \\
				& 1.5  & 0.103(0.008) & 3.260(0.020) & 0.130(0.008) & 3.350(0.022)  & 0.094(0.008) & \textbf{3.221}(0.020) \\
				& 1    & 0.061(0.006) & 3.186(0.021) & 0.077(0.007) & 3.233(0.021)  & 0.055(0.007) & \textbf{3.131}(0.022) \\
				& 0.75 & 0.052(0.004) & 3.172(0.020) & 0.065(0.004) & 3.205(0.020)  & 0.047(0.004) & \textbf{3.116}(0.019) \\
				& 0.5  & 0.033(0.006) & 3.155(0.010) & 0.040(0.006) & 3.171(0.010)  & 0.030(0.006) & \textbf{3.091}(0.010) \\
				& 0.1  & 0.016(0.007) & 3.153(0.019) & 0.018(0.007) & 3.156(0.020)  & 0.015(0.007) & \textbf{3.087}(0.018) \\
				\hline
				
				\multirow{14}{*}{20} & 10   & 0.479(0.010) & 6.940(0.035) & 0.821(0.011) & 7.846(0.037) & 0.488(0.011) & \textbf{6.882}(0.031) \\
				& 9    & 0.457(0.007) & 6.007(0.026) & 0.756(0.007) & 6.817(0.034) & 0.465(0.007) & \textbf{5.985}(0.030) \\
				& 8    & 0.408(0.010) & 5.160(0.026) & 0.664(0.011) & 5.840(0.029) & 0.415(0.010) & \textbf{5.151}(0.022) \\
				& 7    & 0.353(0.012) & 4.331(0.040) & 0.565(0.013) & 4.882(0.048) & 0.360(0.012) & \textbf{4.327}(0.040) \\
				& 6    & 0.318(0.005) & 3.722(0.043) & 0.485(0.006) & 4.172(0.047) & 0.322(0.005) & \textbf{3.731}(0.042) \\
				& 5    & 0.263(0.006) & 3.198(0.032) & 0.384(0.006) & 3.552(0.034) & 0.263(0.006) & \textbf{3.219}(0.032) \\
				& 4    & 0.204(0.006) & 2.810(0.004) & 0.285(0.006) & 3.061(0.003) & 0.199(0.006) & \textbf{2.829}(0.004) \\
				& 3    & 0.169(0.005) & 2.529(0.015) & 0.216(0.006) & 2.693(0.016) & 0.161(0.005) & \textbf{2.538}(0.015) \\
				& 2    & 0.108(0.003) & 2.364(0.009) & 0.132(0.002) & 2.443(0.009) & 0.100(0.002) & \textbf{2.357}(0.009) \\
				& 1.5  & 0.080(0.003) & 2.294(0.017) & 0.096(0.003) & 2.342(0.016) & 0.073(0.003) & \textbf{2.278}(0.017) \\
				& 1    & 0.052(0.007) & 2.241(0.009) & 0.061(0.007) & 2.261(0.010) & 0.047(0.007) & \textbf{2.215}(0.009) \\
				& 0.75 & 0.035(0.008) & 2.229(0.006) & 0.041(0.008) & 2.241(0.006) & 0.031(0.008) & \textbf{2.200}(0.006) \\
				& 0.5  & 0.024(0.006) & 2.225(0.017) & 0.029(0.006) & 2.227(0.017) & 0.022(0.006) & \textbf{2.193}(0.016) \\
				& 0.1  & 0.010(0.002) & 2.212(0.015) & 0.010(0.002) & 2.212(0.015) & 0.010(0.002) & \textbf{2.179}(0.015) \\
				\hline
				
				\multirow{14}{*}{25} & 10   & 0.379(0.015) & 5.231(0.033) & 0.649(0.015) & 5.824(0.044) & 0.383(0.015) & \textbf{5.225}(0.035) \\
				& 9    & 0.349(0.007) & 4.537(0.027) & 0.586(0.007) & 5.048(0.032) & 0.353(0.007) & \textbf{4.536}(0.028) \\
				& 8    & 0.317(0.004) & 3.917(0.030) & 0.520(0.005) & 4.342(0.033) & 0.321(0.004) & \textbf{3.915}(0.028) \\
				& 7    & 0.280(0.004) & 3.367(0.024) & 0.448(0.004) & 3.715(0.027) & 0.279(0.002) & \textbf{3.364}(0.027) \\
				& 6    & 0.258(0.009) & 2.904(0.019) & 0.388(0.007) & 3.175(0.014) & 0.257(0.007) & \textbf{2.895}(0.011) \\
				& 5    & 0.215(0.009) & 2.482(0.022) & 0.310(0.008) & 2.694(0.021) & 0.214(0.007) & \textbf{2.480}(0.018) \\
				& 4    & 0.167(0.009) & 2.136(0.013) & 0.230(0.009) & 2.293(0.014) & 0.164(0.009) & \textbf{2.147}(0.012) \\
				& 3    & 0.135(0.006) & 1.937(0.008) & 0.171(0.006) & 2.039(0.008) & 0.130(0.006) & \textbf{1.943}(0.007) \\
				& 2    & 0.083(0.004) & 1.794(0.009) & 0.100(0.004) & 1.843(0.009) & 0.077(0.004) & \textbf{1.791}(0.009) \\
				& 1.5  & 0.065(0.007) & 1.751(0.009) & 0.075(0.007) & 1.778(0.010) & 0.060(0.007) & \textbf{1.742}(0.010) \\
				& 1    & 0.048(0.005) & 1.699(0.015) & 0.053(0.005) & 1.716(0.014) & 0.044(0.005) & \textbf{1.701}(0.012) \\
				& 0.75 & 0.025(0.004) & 1.711(0.014) & 0.028(0.004) & 1.716(0.014) & 0.021(0.004) & \textbf{1.695}(0.014) \\
				& 0.5  & 0.025(0.005) & 1.697(0.008) & 0.027(0.005) & 1.696(0.008) & 0.023(0.005) & \textbf{1.679}(0.008) \\
				& 0.1  & 0.003(0.004) & 1.682(0.013) & 0.004(0.004) & 1.679(0.013) & 0.002(0.004) & \textbf{1.663}(0.012) \\
				\hline
				
			\end{tabular}
		}
	\end{table}
	

	\newpage
	\addtocounter{table}{-1}
	\begin{table}[H]
		\centering
		\caption{(b). Simulated Bias and MSE of three estimators of $\theta$ ($n=50,75,100$).}
		\label{tab:sim_bias_mse_b}
		
		\setlength{\tabcolsep}{6.0pt}
		\renewcommand{\arraystretch}{1.35}
		
		\resizebox{0.9\textwidth}{!}{%
			\begin{tabular}{|c|c|c|c|c|c|c|c|}
				\hline
				\multirow{2}{*}{$n$} & \multirow{2}{*}{$\theta$} &
				\multicolumn{2}{c|}{MLE} &
				\multicolumn{2}{c|}{BFPE} &
				\multicolumn{2}{c|}{BJPE} \\ \cline{3-8}
				&  & Bias(SE) & MSE(SE) &
				Bias(SE) & MSE(SE) &
				Bias(SE) & MSE(SE)\\
				\hline
				
				\multirow{14}{*}{50}
				& 10   & 0.194(0.006) & \textbf{2.439}(0.008) & 0.325(0.007) & 2.578(0.006) & 0.192(0.007) & \textbf{2.434}(0.006) \\
				& 9    & 0.184(0.011) & \textbf{2.070}(0.019) & 0.299(0.009) & 2.191(0.019) & 0.183(0.009) & \textbf{2.068}(0.019) \\
				& 8    & 0.145(0.006) & \textbf{1.783}(0.011) & 0.243(0.006) & 1.880(0.012) & 0.144(0.006) & \textbf{1.782}(0.012) \\
				& 7    & 0.139(0.004) & \textbf{1.546}(0.008) & 0.220(0.004) & 1.627(0.007) & 0.138(0.004) & \textbf{1.544}(0.007) \\
				& 6    & 0.130(0.004) & \textbf{1.347}(0.010) & 0.193(0.004) & 1.408(0.011) & 0.128(0.005) & \textbf{1.341}(0.011) \\
				& 5    & 0.104(0.006) & \textbf{1.148}(0.009) & 0.152(0.006) & 1.199(0.008) & 0.104(0.006) & \textbf{1.148}(0.008) \\
				& 4    & 0.082(0.002) & \textbf{1.007}(0.005) & 0.113(0.002) & 1.045(0.006) & 0.082(0.002) & \textbf{1.007}(0.005) \\
				& 3    & 0.066(0.002) & \textbf{0.905}(0.006) & 0.082(0.002) & 0.936(0.008) & 0.064(0.002) & \textbf{0.905}(0.006) \\
				& 2    & 0.048(0.005) & \textbf{0.830}(0.003) & 0.055(0.005) & 0.842(0.004) & 0.046(0.005) & \textbf{0.830}(0.003) \\
				& 1.5  & 0.026(0.002) & \textbf{0.809}(0.005) & 0.029(0.002) & 0.816(0.005) & 0.024(0.002) & \textbf{0.808}(0.005) \\
				& 1    & 0.018(0.002) & \textbf{0.802}(0.003) & 0.020(0.002) & 0.805(0.003) & 0.017(0.002) & \textbf{0.800}(0.003) \\
				& 0.75 & 0.014(0.003) & \textbf{0.786}(0.006) & 0.015(0.003) & 0.787(0.006) & 0.013(0.003) & \textbf{0.784}(0.006) \\
				& 0.5  & 0.012(0.002) & \textbf{0.786}(0.002) & 0.015(0.002) & 0.786(0.002) & 0.013(0.002) & \textbf{0.783}(0.002) \\
				& 0.1  & 0.007(0.003) & \textbf{0.786}(0.005) & 0.007(0.003) & 0.785(0.005) & 0.006(0.003) & \textbf{0.783}(0.005) \\
				\hline
				
				\multirow{14}{*}{75}
				& 10   & 0.127(0.007) & \textbf{1.580}(0.008) & 0.215(0.007) & 1.641(0.010) & 0.127(0.007) & \textbf{1.578}(0.009) \\
				& 9    & 0.109(0.006) & \textbf{1.355}(0.004) & 0.187(0.006) & 1.406(0.005) & 0.110(0.006) & \textbf{1.353}(0.004) \\
				& 8    & 0.104(0.004) & \textbf{1.173}(0.010) & 0.170(0.004) & 1.218(0.010) & 0.104(0.004) & \textbf{1.174}(0.010) \\
				& 7    & 0.087(0.004) & \textbf{1.014}(0.003) & 0.142(0.004) & 1.046(0.006) & 0.087(0.004) & \textbf{1.010}(0.006) \\
				& 6    & 0.076(0.005) & \textbf{0.872}(0.004) & 0.119(0.006) & 0.900(0.006) & 0.076(0.005) & \textbf{0.871}(0.005) \\
				& 5    & 0.066(0.003) & \textbf{0.756}(0.006) & 0.097(0.006) & 0.776(0.006) & 0.066(0.003) & \textbf{0.754}(0.006) \\
				& 4    & 0.055(0.003) & \textbf{0.652}(0.002) & 0.076(0.003) & 0.668(0.002) & 0.055(0.003) & \textbf{0.653}(0.002) \\
				& 3    & 0.042(0.003) & \textbf{0.584}(0.004) & 0.053(0.003) & 0.594(0.004) & 0.042(0.003) & \textbf{0.584}(0.004) \\
				& 2    & 0.022(0.005) & \textbf{0.543}(0.002) & 0.026(0.005) & 0.548(0.002) & 0.021(0.005) & \textbf{0.543}(0.002) \\
				& 1.5  & 0.016(0.004) & \textbf{0.528}(0.002) & 0.019(0.004) & 0.531(0.002) & 0.016(0.004) & \textbf{0.528}(0.002) \\
				& 1    & 0.016(0.002) & \textbf{0.518}(0.004) & 0.017(0.002) & 0.525(0.004) & 0.016(0.002) & \textbf{0.523}(0.003) \\
				& 0.75 & 0.014(0.004) & \textbf{0.522}(0.002) & 0.014(0.004) & 0.523(0.002) & 0.013(0.004) & \textbf{0.522}(0.003) \\
				& 0.5  & 0.011(0.003) & \textbf{0.511}(0.003) & 0.011(0.003) & 0.511(0.003) & 0.010(0.003) & \textbf{0.510}(0.003) \\
				& 0.1  & 0.007(0.002) & \textbf{0.502}(0.005) & 0.007(0.002) & 0.502(0.006) & 0.007(0.002) & \textbf{0.501}(0.005) \\
				\hline
				
				\multirow{14}{*}{100}
				& 10   & 0.092(0.002) & \textbf{1.158}(0.004) & 0.153(0.002) & 1.189(0.005) & 0.091(0.002) & \textbf{1.155}(0.005) \\
				& 9    & 0.092(0.002) & \textbf{1.019}(0.006) & 0.147(0.003) & 1.045(0.006) & 0.089(0.003) & \textbf{1.015}(0.006) \\
				& 8    & 0.081(0.002) & \textbf{0.866}(0.004) & 0.131(0.002) & 0.892(0.004) & 0.081(0.002) & \textbf{0.866}(0.004) \\
				& 7    & 0.065(0.003) & \textbf{0.739}(0.005) & 0.104(0.003) & 0.757(0.006) & 0.063(0.003) & \textbf{0.737}(0.005) \\
				& 6    & 0.053(0.005) & \textbf{0.637}(0.003) & 0.086(0.005) & 0.653(0.003) & 0.054(0.005) & \textbf{0.637}(0.003) \\
				& 5    & 0.048(0.002) & \textbf{0.553}(0.005) & 0.071(0.002) & 0.565(0.005) & 0.048(0.002) & \textbf{0.553}(0.005) \\
				& 4    & 0.044(0.002) & \textbf{0.479}(0.002) & 0.059(0.002) & 0.488(0.002) & 0.044(0.002) & \textbf{0.479}(0.002) \\
				& 3    & 0.033(0.005) & \textbf{0.434}(0.002) & 0.042(0.004) & 0.440(0.002) & 0.033(0.004) & \textbf{0.434}(0.002) \\
				& 2    & 0.031(0.004) & \textbf{0.402}(0.003) & 0.034(0.004) & 0.405(0.003) & 0.031(0.004) & \textbf{0.402}(0.003) \\
				& 1.5  & 0.014(0.004) & \textbf{0.388}(0.002) & 0.015(0.004) & 0.390(0.002) & 0.014(0.004) & \textbf{0.388}(0.002) \\
				& 1    & 0.013(0.004) & \textbf{0.381}(0.003) & 0.015(0.004) & 0.382(0.003) & 0.012(0.004) & \textbf{0.381}(0.003) \\
				& 0.75 & 0.009(0.001) & \textbf{0.379}(0.002) & 0.011(0.003) & 0.380(0.003) & 0.009(0.003) & \textbf{0.379}(0.003) \\
				& 0.5  & 0.008(0.001) & \textbf{0.378}(0.002) & 0.007(0.001) & 0.378(0.002) & 0.004(0.002) & \textbf{0.378}(0.002) \\
				& 0.1  & 0.005(0.001) & \textbf{0.375}(0.001) & 0.004(0.001) & 0.375(0.001) & 0.004(0.001) & \textbf{0.375}(0.001) \\
				\hline
				
			\end{tabular}
		}
	\end{table}
	
	
	\section{Goodness of Fit (GoF) Tests for  Frank Copula}
	
	Based on \textit{iid} observations $(X_i,Y_i)$, $1 \le i \le n$, on $(X,Y)$,
	if our goal is to test whether $(X,Y)$ follows a bivariate Frank Copula
	with association parameter $\theta$, then it is equivalent to test that
	$(U,V)$ follows $\mathrm{SBFCD(\theta)}$ with \textit{pdf}
	$c(u,v \mid \theta)$ as given in \eqref{eq:frank_density}.
	\medskip
	
	Genest et al.(2006)~\cite{GenestRemillard2006} discussed two test statistics in a more general
	setting which are being discussed in the following at great length for
	our $\mathrm{SBFCD(\theta)}$. For details on the asymptotic behaviors of
	these two test statistics, one may see the aforementioned paper.
	\medskip
	
	Define the pseudo-observations $W_j$, $1 \le j \le n$, as follows:
	\begin{equation}
		\begin{aligned}
			W_j
			&= (1/n)\sum_{k=1}^{n} I(X_k \le X_j,\; Y_k \le Y_j) \\
			&= (1/n)\sum_{k=1}^{n} I(R_k^{X} \le R_j^{X},\; R_k^{Y} \le R_j^{Y}),
		\end{aligned}
		\label{eq:Wj}
	\end{equation}
	
	\noindent where $R_j^{X}$ is the rank of $X_j$ among $\{X_1, X_2, \ldots, X_n\}$,
	and $R_j^{Y}$ is the rank of $Y_j$ among $\{Y_1, Y_2, \ldots, Y_n\}$.
	\medskip
	
	Define $K_n(t)$, the empirical \textit{cdf} based on the pseudo-observations $W_j$, $1 \le j \le n$, as
	\begin{equation}
		K_n(t) = (1/n)\sum_{j=1}^{n} I(W_j \le t), \qquad t \in [0,1].
		\label{eq:Kn}
	\end{equation}
	\medskip		
	On the other hand, let $K(t,\theta)$ be the \textit{cdf} of $F(X,Y)$ at $t$
	under the assumed Frank Copula with the association parameter $\theta$,
	i.e.,
	
	\begin{equation}
		P(F(X,Y) \le t) = K(t,\theta),
		\label{eq:Ktheta}
	\end{equation}
	
	\noindent	(see Table 1, page 344 in Genest et al.(2006) \cite{GenestRemillard2006}).
	\medskip
	
	Let $k(t,\theta)$ be the density obtained from $K(t,\theta)$, i.e., 
	$	k(t,\theta) = (\partial/\partial t) K(t,\theta).$
	Note that for Frank Copula,
	\begin{equation}
		K(t,\theta)
		=
		t
		-
		\left\{(1-\mathit{exp}(\theta t))/\theta \right\}
		\mathit{ln} \left\{
		(1-\mathit{exp}(-\theta))/(1-\mathit{exp}(-\theta t))
		\right\}.
		\label{eq:K_frank}
	\end{equation}
	
	
	Hence, after some simplifications, we have
	\begin{equation}
		k(t,\theta)
		=
		\mathit{exp}(\theta t)
		\mathit{ln} \left\{
		(1-\mathit{exp}(-\theta))/(1-\mathit{exp}(-\theta t))
		\right\}.
		\label{eq:k_frank}
	\end{equation}
	
	The two GoF test statistics of interest are:
	\medskip
	
	\noindent	\textbf{(a) CramÃ©r--von Mises statistic} $\bm{S_n}$, given as
	\begin{equation}
		S_n
		=
		\int_{0}^{1}
		n \left\{ K_n(t) - K(t,\hat{\theta}) \right\}^{2}
		k(t,\hat{\theta}) \, dt , \text{ and}
		\label{eq:Sn}
	\end{equation}
	
	
	\noindent	\textbf{(b) Kolmogorov--Smirnov statistic} $\bm{T_n}$, given by
	\begin{equation}
		T_n
		=
		\mathit{sup}_{0 \le t \le 1}
		\left[
		\sqrt{n}\, | K_n(t) - K(t,\hat{\theta}) |
		\right],
		\label{eq:Tn}
	\end{equation}
	
	\noindent	where $\hat{\theta}$ is a consistent estimator of $\theta$.
	\medskip
	
	However, Genest et al.(2006)~\cite{GenestRemillard2006} (see their page 346) provided simplified
	expressions of $S_n$ and $T_n$ as follows.
	
	\begin{equation}
		\begin{aligned}
			S_n
			&=
			n/3
			+
			n \sum_{j=1}^{n-1}
			K_n^{2}(j/n)
			\left\{
			K((j+1)/n,\hat{\theta})
			-
			K(j/n,\hat{\theta})
			\right\}
			\\
			&\quad
			-
			n \sum_{j=1}^{n}
			K_n(j/n)
			\left\{
			K^{2}((j+1)/n,\hat{\theta})
			-
			K^{2}(j/n,\hat{\theta})
			\right\}, \text{ and}
		\end{aligned}
		\label{eq:Sn_simplified}
	\end{equation}
	
	\begin{equation}
		T_n
		=
		\sqrt{n}
		\operatorname*{\mathit{max}}_{\substack{i = 0,1; \\ 0 \le j \le n-1}}
		\left\{
		\left|
		K_n(j/n)
		-
		K((j+i)/n,\hat{\theta})
		\right|
		\right\}.
		\label{eq:Tn_simplified}
	\end{equation}

	Under some regularity conditions, which are satisfied by Frank Copula,
	it was shown that the functional
	$
	K_n^{*}(t)
	=
	\sqrt{n}\left\{K_n(t)-K(t,\hat{\theta})\right\}
	$
	converges weakly to a suitable Gaussian distribution.
	\medskip
	
	Genest et al.(2006)~\cite{GenestRemillard2006} carried out a very limited simulation to obtain the
	$90^{\text{th}}$ and $95^{\text{th}}$ percentile values of $S_n$ and $T_n$
	using $M = 1000$ replications, and $\hat{\theta} = \hat{\theta}_{MLE}$. The null distributions of $S_n$ and $T_n$  do
	depend on the unknown association parameter $\theta$ which, in turn,
	dicitates the value of $\rho_K$ (see (\ref{eq:kendall_tau})).	
	Genest et al.(2006)~\cite{GenestRemillard2006} provided the values of the two above mentioned
	percentiles for $n = 100, 250,$ and $1000$, and
	$\rho_K(\theta) = 0.20, 0.40, 0.60,$ and $0.80$.
	\medskip
	
	In this comprehensive work of ours, we provide extended percentile tables for the GoF test statistics of $S_n$ and $T_n$ based on $M = 10,000$ replications for each combination of $(n,\theta)$ and undertake an in-depth investigation about some distrubutional properties of $S_n$ and $T_n$. (For large sample sizes (i.e., $n > 25$), one may use  $\hat{\theta} = \hat{\theta}_{MLE}$ as shown in Section 4, but for smaller sample sizes $(i.e., n \leq 25)$ it is suggested that one should use  $\hat{\theta} = \hat{\theta}_{BJPE}$ as shown in the Appendix \ref{app:A4}).
	\medskip
	
	Let $S_n(1-\alpha \mid \theta)$ and $T_n(1-\alpha \mid\theta)$ denote the $100(1-\alpha)^{\text{th}}$ percentile values of $S_n$ and $T_n$ (when the dataset is assumed to have come from a joint distribution under Frank Copula). Therefore, ideally one should reject the Frank Copula model at level $\alpha$ if $S_n$ (and/or $T_n$) exceeds $S_n(1-\alpha \mid \theta)$ (and/or $T_n(1-\alpha \mid\theta)$.
	\medskip
	
	The following two tables (Table~\ref{tab:Sn_MLE_combined} - \ref{tab:Tn_MLE_combined}) tabulate the values of $S_n(1-\alpha \mid\theta)$ and $T_n(1-\alpha \mid\theta)$ with $(1-\alpha) = 0.90, 0.95;$ $n$ ranging from $25$ to $1,000$; and $\theta$ going from $(-10)$ to $(+10)$, and $\hat{\theta} = \hat{\theta}_{MLE}$. Also, these tables do reveal some unexpected trends as disscussed in the following remarks.
	\medskip

	%
	
	\begin{table}[H]
		\centering
		
		\caption{Values of $S_n(1-\alpha\mid\theta)$ for various $n$, $\theta$,
			and ($1-\alpha$)
			$\bigl(\hat{\theta}=\widehat{\theta}_{MLE}\bigr)$}
		\label{tab:Sn_MLE_combined}
		
		
		\scriptsize
		\renewcommand{\arraystretch}{0.92}
		\setlength{\tabcolsep}{2.6pt}
		
		\resizebox{0.9\textwidth}{!}{%
			\begin{tabular}{|c|c|cccccccccc|}
				\hline
				
				\multirow{2}{*}{$1-\alpha$}
				& \multirow{2}{*}{$\theta$}
				& \multicolumn{10}{c|}{$n$} \\
				\cline{3-12}
				
				&
				& $25$ & $50$ & $75$ & $100$ & $150$
				& $200$ & $250$ & $500$ & $750$ & $1000$ \\
				
				\hline
				
				
				\multirow{28}{*}{$0.90$}
				& -10.00 & 1.389 & 0.892 & 0.691 & 0.583 & 0.467
				& 0.404 & 0.362 & 0.266 & 0.235 & 0.219 \\
				
				& -9.00  & 1.243 & 0.798 & 0.630 & 0.529 & 0.428
				& 0.370 & 0.333 & 0.255 & 0.228 & 0.209 \\
				
				& -8.00  & 1.089 & 0.719 & 0.553 & 0.480 & 0.390
				& 0.337 & 0.304 & 0.237 & 0.214 & 0.197 \\
				
				& -7.00  & 0.969 & 0.623 & 0.497 & 0.427 & 0.354
				& 0.303 & 0.281 & 0.218 & 0.199 & 0.186 \\
				
				& -6.00  & 0.842 & 0.552 & 0.438 & 0.384 & 0.313
				& 0.283 & 0.257 & 0.210 & 0.192 & 0.179 \\
				
				& -5.00  & 0.723 & 0.472 & 0.391 & 0.334 & 0.276
				& 0.253 & 0.230 & 0.192 & 0.179 & 0.173 \\
				
				& -4.00  & 0.610 & 0.410 & 0.330 & 0.292 & 0.246
				& 0.222 & 0.206 & 0.178 & 0.168 & 0.158 \\
				
				& -3.00  & 0.505 & 0.342 & 0.286 & 0.251 & 0.218
				& 0.199 & 0.185 & 0.160 & 0.152 & 0.150 \\
				
				& -2.00  & 0.417 & 0.288 & 0.243 & 0.213 & 0.188
				& 0.171 & 0.166 & 0.146 & 0.138 & 0.134 \\
				
				& -1.50  & 0.375 & 0.267 & 0.224 & 0.194 & 0.175
				& 0.160 & 0.154 & 0.140 & 0.133 & 0.133 \\
				
				& -1.00  & 0.341 & 0.241 & 0.200 & 0.181 & 0.161
				& 0.152 & 0.146 & 0.134 & 0.129 & 0.127 \\
				
				& -0.75  & 0.324 & 0.223 & 0.195 & 0.174 & 0.155
				& 0.147 & 0.141 & 0.130 & 0.129 & 0.124 \\
				
				& -0.50  & 0.308 & 0.217 & 0.186 & 0.166 & 0.152
				& 0.145 & 0.139 & 0.129 & 0.125 & 0.122 \\
				
				& -0.10  & 0.285 & 0.205 & 0.172 & 0.159 & 0.146
				& 0.138 & 0.134 & 0.123 & 0.120 & 0.119 \\
				
				& 0.10   & 0.272 & 0.200 & 0.171 & 0.155 & 0.142
				& 0.136 & 0.131 & 0.123 & 0.119 & 0.117 \\
				
				& 0.50   & 0.254 & 0.182 & 0.162 & 0.148 & 0.138
				& 0.130 & 0.127 & 0.119 & 0.114 & 0.116 \\
				
				& 0.75   & 0.243 & 0.178 & 0.156 & 0.145 & 0.134
				& 0.130 & 0.126 & 0.116 & 0.115 & 0.113 \\
				
				& 1.00   & 0.229 & 0.171 & 0.151 & 0.141 & 0.130
				& 0.127 & 0.123 & 0.118 & 0.114 & 0.114 \\
				
				& 1.50   & 0.213 & 0.160 & 0.145 & 0.136 & 0.125
				& 0.122 & 0.119 & 0.112 & 0.112 & 0.110 \\
				
				& 2.00   & 0.196 & 0.149 & 0.137 & 0.129 & 0.124
				& 0.116 & 0.115 & 0.110 & 0.109 & 0.107 \\
				
				& 3.00   & 0.171 & 0.137 & 0.124 & 0.121 & 0.115
				& 0.111 & 0.111 & 0.106 & 0.106 & 0.104 \\
				
				& 4.00   & 0.153 & 0.126 & 0.116 & 0.113 & 0.110
				& 0.106 & 0.105 & 0.101 & 0.102 & 0.101 \\
				
				& 5.00   & 0.138 & 0.117 & 0.110 & 0.108 & 0.102
				& 0.101 & 0.099 & 0.097 & 0.096 & 0.095 \\
				
				& 6.00   & 0.128 & 0.109 & 0.102 & 0.099 & 0.096
				& 0.095 & 0.094 & 0.092 & 0.092 & 0.090 \\
				
				& 7.00   & 0.119 & 0.099 & 0.095 & 0.094 & 0.091
				& 0.089 & 0.088 & 0.087 & 0.087 & 0.086 \\
				
				& 8.00   & 0.110 & 0.095 & 0.091 & 0.089 & 0.085
				& 0.083 & 0.084 & 0.081 & 0.080 & 0.081 \\
				
				& 9.00   & 0.103 & 0.090 & 0.086 & 0.082 & 0.080
				& 0.080 & 0.078 & 0.077 & 0.077 & 0.077 \\
				
				& 10.00  & 0.098 & 0.084 & 0.080 & 0.078 & 0.075
				& 0.074 & 0.074 & 0.073 & 0.073 & 0.070 \\
				
				\hline
				
				
				\multirow{28}{*}{$0.95$}
				& -10.00 & 1.567 & 1.017 & 0.790 & 0.679 & 0.551
				& 0.488 & 0.442 & 0.336 & 0.296 & 0.282 \\
				
				& -9.00  & 1.401 & 0.915 & 0.730 & 0.628 & 0.509
				& 0.447 & 0.407 & 0.313 & 0.285 & 0.260 \\
				
				& -8.00  & 1.239 & 0.825 & 0.652 & 0.570 & 0.466
				& 0.409 & 0.379 & 0.302 & 0.270 & 0.249 \\
				
				& -7.00  & 1.105 & 0.723 & 0.597 & 0.505 & 0.430
				& 0.379 & 0.355 & 0.276 & 0.250 & 0.235 \\
				
				& -6.00  & 0.978 & 0.636 & 0.518 & 0.460 & 0.383
				& 0.353 & 0.321 & 0.264 & 0.242 & 0.225 \\
				
				& -5.00  & 0.851 & 0.559 & 0.465 & 0.405 & 0.338
				& 0.314 & 0.282 & 0.241 & 0.229 & 0.216 \\
				
				& -4.00  & 0.711 & 0.489 & 0.395 & 0.353 & 0.302
				& 0.277 & 0.250 & 0.223 & 0.211 & 0.205 \\
				
				& -3.00  & 0.597 & 0.409 & 0.348 & 0.304 & 0.268
				& 0.246 & 0.230 & 0.203 & 0.189 & 0.185 \\
				
				& -2.00  & 0.495 & 0.348 & 0.295 & 0.264 & 0.230
				& 0.212 & 0.206 & 0.181 & 0.175 & 0.167 \\
				
				& -1.50  & 0.447 & 0.320 & 0.272 & 0.241 & 0.214
				& 0.196 & 0.190 & 0.176 & 0.168 & 0.166 \\
				
				& -1.00  & 0.407 & 0.291 & 0.240 & 0.221 & 0.199
				& 0.186 & 0.182 & 0.165 & 0.157 & 0.158 \\
				
				& -0.75  & 0.385 & 0.272 & 0.239 & 0.213 & 0.189
				& 0.182 & 0.173 & 0.159 & 0.161 & 0.155 \\
				
				& -0.50  & 0.374 & 0.259 & 0.225 & 0.204 & 0.188
				& 0.176 & 0.173 & 0.159 & 0.155 & 0.148 \\
				
				& -0.10  & 0.343 & 0.248 & 0.209 & 0.197 & 0.180
				& 0.169 & 0.163 & 0.150 & 0.149 & 0.147 \\
				
				& 0.10   & 0.325 & 0.243 & 0.208 & 0.189 & 0.174
				& 0.165 & 0.159 & 0.151 & 0.148 & 0.146 \\
				
				& 0.50   & 0.306 & 0.217 & 0.199 & 0.180 & 0.167
				& 0.159 & 0.155 & 0.144 & 0.139 & 0.143 \\
				
				& 0.75   & 0.290 & 0.210 & 0.187 & 0.176 & 0.162
				& 0.157 & 0.154 & 0.141 & 0.139 & 0.138 \\
				
				& 1.00   & 0.274 & 0.204 & 0.182 & 0.168 & 0.157
				& 0.152 & 0.150 & 0.146 & 0.138 & 0.138 \\
				
				& 1.50   & 0.253 & 0.190 & 0.173 & 0.165 & 0.152
				& 0.146 & 0.143 & 0.136 & 0.137 & 0.134 \\
				
				& 2.00   & 0.231 & 0.175 & 0.162 & 0.156 & 0.149
				& 0.140 & 0.141 & 0.133 & 0.132 & 0.131 \\
				
				& 3.00   & 0.199 & 0.159 & 0.146 & 0.144 & 0.139
				& 0.134 & 0.133 & 0.128 & 0.129 & 0.125 \\
				
				& 4.00   & 0.177 & 0.147 & 0.137 & 0.132 & 0.129
				& 0.126 & 0.125 & 0.121 & 0.121 & 0.121 \\
				
				& 5.00   & 0.159 & 0.137 & 0.130 & 0.126 & 0.122
				& 0.118 & 0.118 & 0.117 & 0.111 & 0.116 \\
				
				& 6.00   & 0.147 & 0.125 & 0.119 & 0.115 & 0.115
				& 0.115 & 0.111 & 0.109 & 0.110 & 0.107 \\
				
				& 7.00   & 0.135 & 0.114 & 0.111 & 0.112 & 0.108
				& 0.105 & 0.104 & 0.103 & 0.104 & 0.102 \\
				
				& 8.00   & 0.125 & 0.110 & 0.106 & 0.104 & 0.099
				& 0.098 & 0.099 & 0.095 & 0.095 & 0.096 \\
				
				& 9.00   & 0.117 & 0.104 & 0.100 & 0.095 & 0.095
				& 0.094 & 0.092 & 0.092 & 0.091 & 0.091 \\
				
				& 10.00  & 0.110 & 0.097 & 0.093 & 0.090 & 0.088
				& 0.088 & 0.087 & 0.085 & 0.085 & 0.083 \\
				
				\hline
			\end{tabular}%
		}
		
		\vspace{-2mm}
	\end{table}

	
	\begin{table}[H]
		\centering
		
		\caption{Values of $T_n(1-\alpha\mid\theta)$ for various $n$, $\theta$,
			and ($1-\alpha$)
			$\bigl(\hat{\theta}=\widehat{\theta}_{MLE}\bigr)$}
		\label{tab:Tn_MLE_combined}
		
		
		\tiny
		\renewcommand{\arraystretch}{0.92}
		\setlength{\tabcolsep}{2.6pt}
		
		\resizebox{0.95\textwidth}{!}{%
			\begin{tabular}{|c|c|cccccccccc|}
				\hline
				
				\multirow{2}{*}{$1-\alpha$}
				& \multirow{2}{*}{$\theta$}
				& \multicolumn{10}{c|}{$n$} \\
				\cline{3-12}
				
				&
				& $25$ & $50$ & $75$ & $100$ & $150$
				& $200$ & $250$ & $500$ & $750$ & $1000$ \\
				
				\hline
				
				
				\multirow{28}{*}{$0.90$}
				& -10.00 & 1.951 & 1.858 & 1.717 & 1.614 & 1.489
				& 1.411 & 1.347 & 1.202 & 1.146 & 1.117 \\
				
				& -9.00  & 1.897 & 1.769 & 1.641 & 1.548 & 1.427
				& 1.361 & 1.304 & 1.181 & 1.123 & 1.093 \\
				
				& -8.00  & 1.815 & 1.696 & 1.565 & 1.484 & 1.376
				& 1.309 & 1.258 & 1.143 & 1.100 & 1.069 \\
				
				& -7.00  & 1.763 & 1.608 & 1.499 & 1.417 & 1.315
				& 1.250 & 1.213 & 1.106 & 1.071 & 1.040 \\
				
				& -6.00  & 1.686 & 1.531 & 1.419 & 1.363 & 1.252
				& 1.210 & 1.169 & 1.086 & 1.047 & 1.024 \\
				
				& -5.00  & 1.603 & 1.443 & 1.345 & 1.282 & 1.206
				& 1.162 & 1.121 & 1.049 & 1.015 & 1.012 \\
				
				& -4.00  & 1.506 & 1.368 & 1.273 & 1.215 & 1.140
				& 1.103 & 1.067 & 1.015 & 0.996 & 0.980 \\
				
				& -3.00  & 1.416 & 1.275 & 1.208 & 1.147 & 1.096
				& 1.056 & 1.030 & 0.979 & 0.959 & 0.954 \\
				
				& -2.00  & 1.320 & 1.200 & 1.134 & 1.082 & 1.038
				& 1.000 & 0.993 & 0.942 & 0.931 & 0.922 \\
				
				& -1.50  & 1.272 & 1.162 & 1.100 & 1.047 & 1.008
				& 0.981 & 0.964 & 0.937 & 0.920 & 0.914 \\
				
				& -1.00  & 1.230 & 1.117 & 1.054 & 1.022 & 0.881
				& 0.962 & 0.942 & 0.910 & 0.900 & 0.896 \\
				
				& -0.75  & 1.207 & 1.087 & 1.044 & 1.005 & 0.662
				& 0.947 & 0.932 & 0.903 & 0.899 & 0.896 \\
				
				& -0.50  & 1.183 & 1.070 & 1.025 & 0.983 & 0.956
				& 0.940 & 0.933 & 0.908 & 0.887 & 0.881 \\
				
				& -0.10  & 1.154 & 1.047 & 0.994 & 0.971 & 0.641
				& 0.921 & 0.911 & 0.886 & 0.877 & 0.872 \\
				
				& 0.10   & 1.126 & 1.035 & 0.988 & 0.959 & 0.930
				& 0.917 & 0.906 & 0.885 & 0.880 & 0.870 \\
				
				& 0.50   & 1.105 & 1.007 & 0.968 & 0.943 & 0.918
				& 0.898 & 0.893 & 0.875 & 0.863 & 0.862 \\
				
				& 0.75   & 1.082 & 0.994 & 0.948 & 0.935 & 0.909
				& 0.897 & 0.888 & 0.862 & 0.858 & 0.851 \\
				
				& 1.00   & 1.059 & 0.979 & 0.938 & 0.915 & 0.892
				& 0.888 & 0.879 & 0.865 & 0.852 & 0.847 \\
				
				& 1.50   & 1.032 & 0.953 & 0.922 & 0.908 & 0.879
				& 0.869 & 0.864 & 0.850 & 0.849 & 0.842 \\
				
				& 2.00   & 0.998 & 0.926 & 0.900 & 0.882 & 0.869
				& 0.856 & 0.850 & 0.844 & 0.839 & 0.834 \\
				
				& 3.00   & 0.945 & 0.884 & 0.867 & 0.864 & 0.847
				& 0.841 & 0.836 & 0.821 & 0.827 & 0.822 \\
				
				& 4.00   & 0.902 & 0.856 & 0.845 & 0.833 & 0.833
				& 0.823 & 0.818 & 0.816 & 0.816 & 0.812 \\
				
				& 5.00   & 0.859 & 0.834 & 0.821 & 0.819 & 0.808
				& 0.805 & 0.804 & 0.801 & 0.795 & 0.797 \\
				
				& 6.00   & 0.836 & 0.806 & 0.798 & 0.796 & 0.787
				& 0.792 & 0.787 & 0.784 & 0.786 & 0.781 \\
				
				& 7.00   & 0.802 & 0.779 & 0.778 & 0.777 & 0.775
				& 0.770 & 0.767 & 0.765 & 0.770 & 0.768 \\
				
				& 8.00   & 0.777 & 0.766 & 0.761 & 0.757 & 0.755
				& 0.752 & 0.755 & 0.751 & 0.746 & 0.753 \\
				
				& 9.00   & 0.750 & 0.749 & 0.744 & 0.737 & 0.738
				& 0.739 & 0.737 & 0.737 & 0.736 & 0.738 \\
				
				& 10.00  & 0.728 & 0.725 & 0.723 & 0.719 & 0.721
				& 0.720 & 0.721 & 0.723 & 0.725 & 0.720 \\
				
				\hline
				
				
				\multirow{28}{*}{$0.95$}
				& -10.00 & 2.088 & 1.975 & 1.821 & 1.717 & 1.590
				& 1.511 & 1.452 & 1.302 & 1.249 & 1.219 \\
				
				& -9.00  & 2.031 & 1.890 & 1.743 & 1.650 & 1.524
				& 1.463 & 1.410 & 1.278 & 1.227 & 1.192 \\
				
				& -8.00  & 1.954 & 1.821 & 1.670 & 1.591 & 1.482
				& 1.411 & 1.365 & 1.242 & 1.204 & 1.169 \\
				
				& -7.00  & 1.889 & 1.714 & 1.604 & 1.512 & 1.422
				& 1.360 & 1.317 & 1.205 & 1.164 & 1.138 \\
				
				& -6.00  & 1.805 & 1.637 & 1.519 & 1.463 & 1.352
				& 1.318 & 1.273 & 1.187 & 1.146 & 1.121 \\
				
				& -5.00  & 1.732 & 1.548 & 1.450 & 1.378 & 1.310
				& 1.265 & 1.229 & 1.150 & 1.117 & 1.107 \\
				
				& -4.00  & 1.623 & 1.471 & 1.371 & 1.308 & 1.238
				& 1.203 & 1.160 & 1.120 & 1.092 & 1.076 \\
				
				& -3.00  & 1.527 & 1.377 & 1.291 & 1.246 & 1.181
				& 1.157 & 1.128 & 1.073 & 1.050 & 1.043 \\
				
				& -2.00  & 1.427 & 1.293 & 1.227 & 1.176 & 1.131
				& 1.086 & 1.086 & 1.030 & 1.024 & 1.009 \\
				
				& -1.50  & 1.375 & 1.251 & 1.195 & 1.130 & 1.098
				& 1.066 & 1.050 & 1.022 & 1.000 & 0.994 \\
				
				& -1.00  & 1.327 & 1.206 & 1.138 & 1.105 & 1.071
				& 1.046 & 1.025 & 0.997 & 0.975 & 0.988 \\
				
				& -0.75  & 1.305 & 1.171 & 1.128 & 1.084 & 1.043
				& 1.035 & 1.020 & 0.977 & 0.985 & 0.973 \\
				
				& -0.50  & 1.281 & 1.158 & 1.107 & 1.065 & 1.037
				& 1.017 & 1.010 & 0.987 & 0.975 & 0.960 \\
				
				& -0.10  & 1.250 & 1.131 & 1.079 & 1.049 & 1.017
				& 1.002 & 0.992 & 0.962 & 0.951 & 0.948 \\
				
				& 0.10   & 1.220 & 1.122 & 1.075 & 1.040 & 1.011
				& 0.994 & 0.982 & 0.964 & 0.956 & 0.943 \\
				
				& 0.50   & 1.194 & 1.089 & 1.048 & 1.021 & 0.992
				& 0.975 & 0.970 & 0.949 & 0.936 & 0.936 \\
				
				& 0.75   & 1.176 & 1.069 & 1.031 & 1.006 & 0.981
				& 0.974 & 0.964 & 0.933 & 0.934 & 0.926 \\
				
				& 1.00   & 1.146 & 1.056 & 1.014 & 0.993 & 0.964
				& 0.961 & 0.952 & 0.947 & 0.928 & 0.926 \\
				
				& 1.50   & 1.111 & 1.023 & 0.995 & 0.977 & 0.954
				& 0.941 & 0.932 & 0.924 & 0.922 & 0.903 \\
				
				& 2.00   & 1.074 & 0.992 & 0.971 & 0.953 & 0.939
				& 0.925 & 0.924 & 0.915 & 0.912 & 0.908 \\
				
				& 3.00   & 1.013 & 0.949 & 0.928 & 0.927 & 0.918
				& 0.906 & 0.907 & 0.890 & 0.899 & 0.889 \\
				
				& 4.00   & 0.963 & 0.917 & 0.908 & 0.898 & 0.895
				& 0.886 & 0.882 & 0.881 & 0.880 & 0.879 \\
				
				& 5.00   & 0.915 & 0.895 & 0.883 & 0.882 & 0.869
				& 0.864 & 0.866 & 0.861 & 0.859 & 0.859 \\
				
				& 6.00   & 0.888 & 0.860 & 0.859 & 0.851 & 0.847
				& 0.854 & 0.844 & 0.844 & 0.840 & 0.844 \\
				
				& 7.00   & 0.854 & 0.832 & 0.832 & 0.832 & 0.829
				& 0.827 & 0.827 & 0.823 & 0.829 & 0.830 \\
				
				& 8.00   & 0.819 & 0.817 & 0.810 & 0.810 & 0.811
				& 0.808 & 0.808 & 0.810 & 0.802 & 0.810 \\
				
				& 9.00   & 0.794 & 0.800 & 0.795 & 0.792 & 0.793
				& 0.789 & 0.793 & 0.795 & 0.791 & 0.796 \\
				
				& 10.00  & 0.768 & 0.772 & 0.770 & 0.768 & 0.771
				& 0.776 & 0.774 & 0.775 & 0.777 & 0.771 \\
				
				\hline
			\end{tabular}%
		}
		
		\vspace{-2mm}
	\end{table}

	\begin{remark}
	\label{rem: 4.1}
		(a) The most striking feature of Tables~\ref{tab:Sn_MLE_combined}--\ref{tab:Tn_MLE_combined} is the lack of
		symmetry in the critical values $S_n(1-\alpha \mid \theta)$ and
		$T_n(1-\alpha \mid \theta)$ with respect to the association parameter $\theta$. Intuitively, one would expect that these critical values ought to be the same at $\theta$ and $(-\theta)$ for any $\theta \in \mathbb{R}$, but that is not what is being observed here. After all, the problem of goodness of fit does not change whether the assumed Frank Copula model has positive association or negative association. This is because the distribution function $K(t, \theta)$ (see \eqref{eq:K_frank}) lacks symmetry in the sense that $K(t, -\theta) \neq K(t, \theta)$, and as a result, the distribution of $S_n$ at $\theta$ is not same as that of $S_n$ at $(-\theta)$. The same holds true for the statistic  $T_n$ as well. This is the reason why $S_n(1-\alpha \mid -\theta) \neq S_n(1-\alpha \mid \theta)$ and $T_n(1-\alpha \mid -\theta) \neq T_n(1-\alpha \mid \theta)$.
		(b) An interesting relationship in $K(t, \theta)$ is observed, that is $K(t,-\theta) = 1- K(1-t, \theta)$ 		
		which follows from the fact that a random pair $(U,V)$ follows SBFCD $(\theta)$ if and only if the pairs $(1-U, V )$ and $(U, 1-V)$ have SBFCD $(\theta)$. In other words, the Frank Copula family is closed under reflection.
		This identity may explain why the distributions of the statistics $S_n$ and $T_n$ may appear somewhat approximately symmetric in simulation studies for some combinations of n and $\theta$, even though the statistics themselves are not symmetric in general. The following Figures \ref{fig:Sn_n_25_plot} -- \ref{fig:Tn_n_25_plot} show the relative frequency histograms of $10,000$ simulated values of $S_n$ and $T_n$ at $\theta = -7, -3, -1$ and $\theta = +1, +3, +7$, respectively, for $n = 25$, just as a demonstration. Exactly the same features have been observed for n = 50 and 100 (but not reported here for brevity).
	\end{remark}
	\vspace{-0.5cm}
	\begin{figure}[H]
		\centering
		\includegraphics[width=0.95\textwidth]{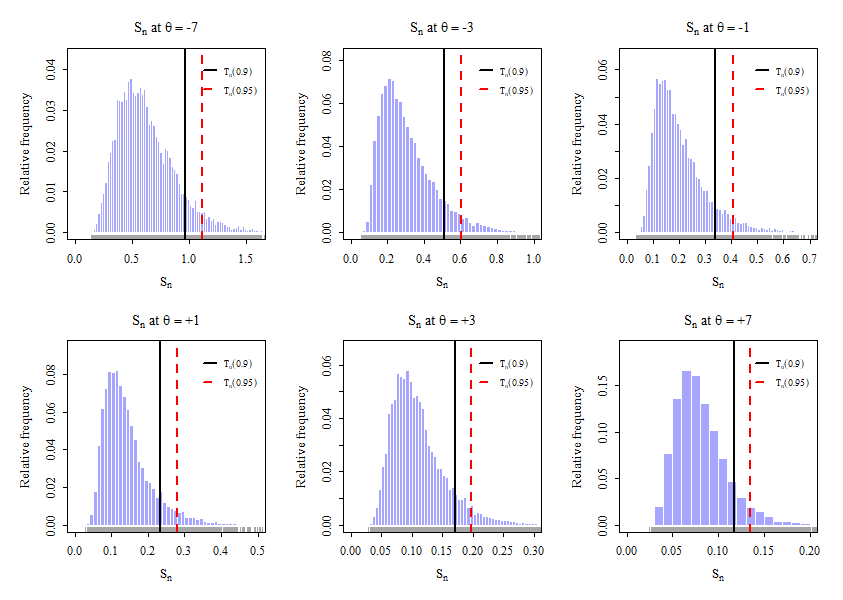}
		
		\caption{Relative frequency histogram of $S_n$ with $n = 25 $ at $ \theta = -7, -3, -1, +1, +3, +7$}
		\label{fig:Sn_n_25_plot}
	\end{figure}
	
	\vspace{-0.7cm}
	\begin{figure}[H]
		\centering
		\includegraphics[width=0.95\textwidth]{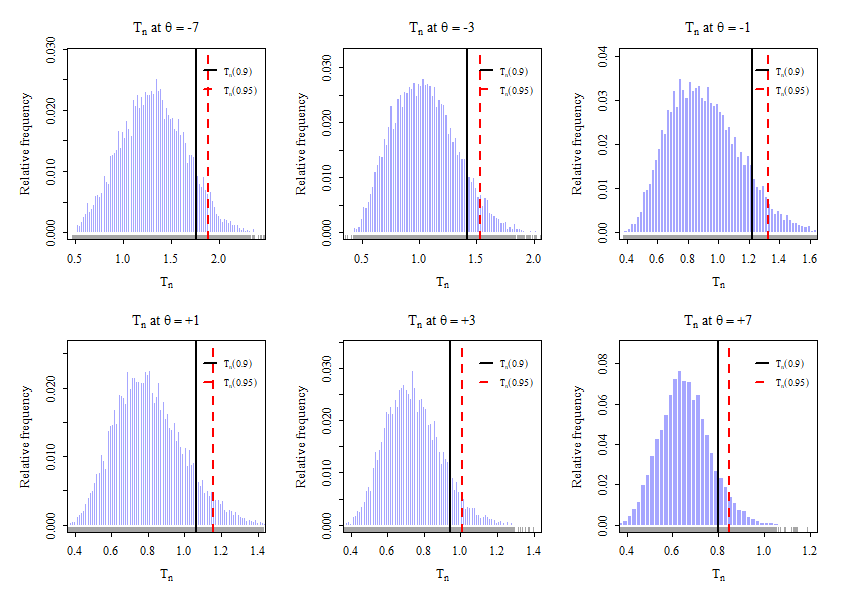}
		
		\caption{Relative frequency histogram of $T_n$ with $n = 25 $ at $ \theta =  -7, -3, -1, +1, +3, +7$}
		\label{fig:Tn_n_25_plot}
	\end{figure}
	
\begin{figure}[H]
		\centering
		\includegraphics[width= 0.95 \textwidth]{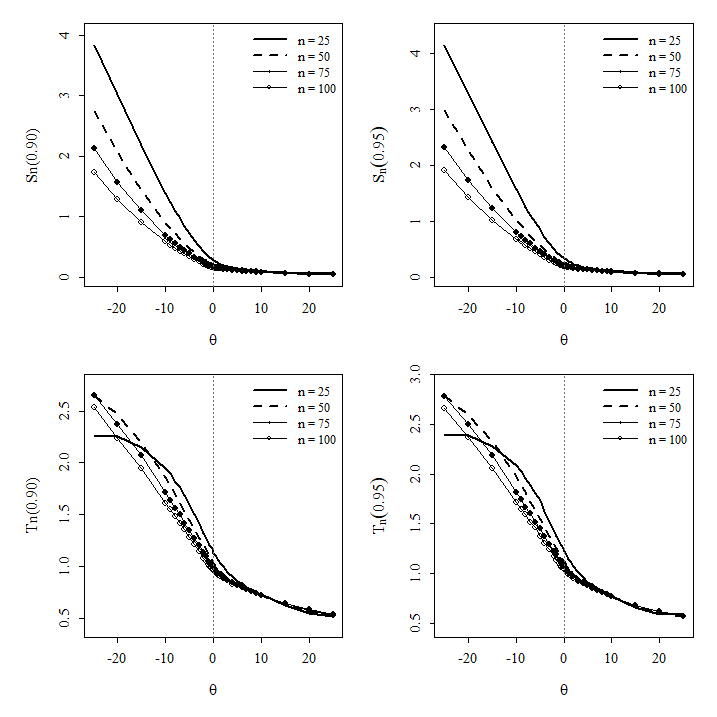}
		\caption{Plots of the critical values with $\theta \in (-25, 25)$ and $n = 25,50,75,100$}
		\label{fig:Sn_Tn_plot}
	\end{figure}

	\vspace{-0.5cm}
	\begin{remark}
		\label{rem: 4.2}
		An interesting feature  that has been observed in our comprehensive simulation study is that the critical values of the statistics $S_n$ and $T_n$ tend to decrease as the association parameter $\theta$ increases. This behavior can be explained by examining the distribution function $K(t, \theta)$ of the random variable $W = C(U,V \mid \theta)$ (see \eqref{eq:sklar} and \eqref{eq:frank_copula}) under the assumed Frank Copula model. (i) When $\theta > 0$, the variables $U$ and $V$ tend to move in the same direction, i.e., the data point $(u,v)$ on $(U,V)$ tend to concentrate around the main diagonal line $u = v$, and this concentration becomes more prominent when $\theta$ moves away more from $0$. In other words, the variable $W = C(U,V \mid \theta)$ shifts more toward right, i.e., $W$ increases stochastically in $\theta$. As a result, for any fixed $t \in (0,1), P(W \leq t)$ becomes smaller, and hence $K(t, \theta)$ takes smaller values for larger $\theta$ values. (ii) On the other hand, when $\theta < 0$, the variables $U$ and $V$ tend to move in opposite  directions, and the data points $(u,v)$ on $(U, V)$ tend to concentrate around the minor diagonal line $u = 1 - v$. In this case, the variable $W = C(U,V \mid \theta)$ tends to be smaller, and hence the distribution of $W$ shifts to the left. With $\theta$ negative, and moving away from $0$, the variable $W$ becomes stochastically smaller. Therefore, for any fixed $t \in (0, 1), K(t, \theta)$ takes large values with more extreme values of $\theta$ in the negative direction. The above justification thus explains why $K(t, \theta)$ should be decreasing, as a whole, as $\theta$ goes from $(-10)$ to $+ 10$, as used in Tables~\ref{tab:Sn_MLE_combined}--\ref{tab:Tn_MLE_combined}. This also explains the decreasing trends in $S_n(1-\alpha \mid \theta)$ and $T_n(1-\alpha \mid \theta)$, as functions of $\theta$, as shown in Figure \ref{fig:Sn_Tn_plot}.
	\end{remark}
	\vspace{-0.6cm}
	
	
	\begin{remark}\label{rem:gof_implementation}
		So, the big question is: ``How to implement the GoF tests for Frank
		Copula?'' Since $\theta$ is unknown, a lot depends on its estimated
		value $\hat{\theta}$, and it is suggested that one should use the BJPE
		for $n \leq 25$, and either the MLE or the BJPE for $n > 25$.		
		One way to apply the GoF tests is to read Tables \ref{tab:Sn_MLE_combined} and/or \ref{tab:Tn_MLE_combined} using
		$\theta \approx \hat{\theta}$, and check whether the observed value of
		$S_n$ and/or $T_n$ exceed(s) the corresponding critical value(s)
		$S_n(1-\alpha \mid \hat{\theta})$ and/or
		$T_n(1-\alpha \mid \hat{\theta})$, where $\alpha$ is the desired level.
		However, this approach can be erroneous due to the variability in
		$\hat{\theta}$. Therefore, a more nuanced approach would be to consider
		the range of $\theta$ as
		\(
		(
		\hat{\theta}-2\sqrt{\mathit{MSE}},
		\;
		\hat{\theta}+2\sqrt{\mathit{MSE}}
		),
		\)
		where the MSE of $\hat{\theta}$ can be seen from Tables \ref{tab:sim_bias_mse_a} (a)--(b).		
		Also, since the critical values of the test statistics $S_n$ and $T_n$
		are decreasing in $\theta$ (see Tables \ref{tab:Sn_MLE_combined} and \ref{tab:Tn_MLE_combined}, as well as
		Figures \ref{fig:Sn_n_25_plot} and \ref{fig:Tn_n_25_plot}), it is enough to look at the critical value at
		\(
		\theta=\hat{\theta}-2\sqrt{\mathit{MSE}}.
		\)
		As an example, think of a situation with $n=50$ and $\theta$ is
		estimated to be close to $3.0$, i.e.,
		$\hat{\theta}_{\mathit{MLE}}=3.0$. From Table \ref{tab:sim_bias_mse_b} (b), MSE
		(at $\theta=3.0$) is $0.905$, and hence
		\(
		\hat{\theta}-2\sqrt{\mathit{MSE}}
		\approx 1.1 \approx 1.
		\)
		Note that
		\(
		S_n(0.95\mid\theta=1)=0.204
		\)
		and
		\(
		T_n(0.95\mid\theta=1)=1.056.
		\)
		Therefore, the FC model should be retained if the observed $S_n$ and
		$T_n$ do not exceed these cut-off points. This approach of using
		$\theta\approx\hat{\theta}-2\sqrt{\mathit{MSE}}$ does make the overall
		method somewhat conservative, and the exact size and the powers of
		$S_n$ and $T_n$ based on this approach should be studied further.
		A more practical bootstrap based p-value approach has been discussed
		in Section 5.
	\end{remark}

	\begin{remark}
			\label{rem: 4.4}
		We would like to point out that our simulated critical values (based on $M = 10,000$ replications) are in line with those limited ones as reported by Genest at al.(2006)~\cite{GenestRemillard2006} (based on their $M = 1,000$ replications only) as shown in the following Table \ref{tab: comparison_summary}. However, we belive that our values are more accurate and much refined as they are based on a higher number of replications.
	\end{remark}
	\vspace{-0.4cm}
	\begin{table}[htbp]
		
		\centering
		
		\caption{Comparison of our simulated percentile values of $S_n$ and $T_n$ (based on $M = 10,000$) with those of Genest et al.(2006) \cite{GenestRemillard2006} given in [ ] (based on their $M = 1,000$)}
		\label{tab: comparison_summary}
		
		\setlength{\tabcolsep}{22pt}   
		\renewcommand{\arraystretch}{1.0} 
		\resizebox{0.9\textwidth}{!}{      
			
			\begin{tabular}{|c||c||c||c|}
				\hline
				$n$ & $\theta \,[ = \rho_K (\theta)]$ & $S_n(0.95\mid \theta)$ & $T_n(0.95 \mid \theta)$ \\
				\hline
				\multirow{4}{*}{100}
				& 1.861 [0.20]  & 0.157 [0.1515] & 0.961 [0.9743] \\
				& 4.161 [0.40]  & 0.133 [0.1254] & 0.895 [0.8883] \\
				& 7.930 [0.60]  & 0.102 [0.1017] & 0.813 [0.7945] \\
				& 18.192 [0.80] & 0.058 [0.0591] & 0.645 [0.6410] \\
				\hline
				\multirow{4}{*}{250}
				& 1.861 [0.20]  & 0.140 [0.1328] & 0.935 [0.9294] \\
				& 4.161 [0.40]  & 0.123 [0.1186] & 0.884 [0.8614] \\
				& 7.930 [0.60]  & 0.097 [0.0979] & 0.802 [0.7982] \\
				& 18.192 [0.80] & 0.054 [0.0536] & 0.652 [0.6472] \\
				\hline
				\multirow{4}{*}{1000}
				& 1.861 [0.20]  & 0.134 [0.1216] & 0.915 [0.9210] \\
				& 4.161 [0.40]  & 0.120 [0.1150] & 0.870 [0.8439] \\
				& 7.930 [0.60]  & 0.098 [0.0975] & 0.814 [0.7875] \\
				& 18.192 [0.80] & 0.052 [0.0492] & 0.658 [0.6496] \\
				\hline
			\end{tabular}
		}
	\end{table}
	\vspace{-0.4cm}
	\section{Application of Frank Copula Goodness of Fit Test for Groundwater Data}
	
	We now turn our attention to the groundwater data mentioned in Section 1 (shown in Appendix \ref{app:groundwater_data}). Our first objective is to see if 
	each pairwise bivariate data could be modeled by a suitable Frank Copula-based bivariate joint distribution, 
	and if so, how strong the association is between the components. Since the North and South subregions 
	have sample sizes $n_N = 23$ and $n_S = 44$, respectively, we first present the 90th and 95th percentile 
	values for the two test statistics $S_n$ and $T_n$ for these two sample sizes, as shown in Table~\ref{tab:SnTn}. Also, to be consistent with our tables in Section 4, the following Table \ref{tab:SnTn}  has been made with $\hat{\theta} = \hat{\theta}_{MLE}.$
	\begin{table}[H]
		\centering
		\caption{Simulated percentile values for $n=23$ and $n=44 \,(\hat{\theta} = \hat{\theta}_{MLE}) $}
		\label{tab:SnTn}
		\setlength{\tabcolsep}{3pt}   
		\renewcommand{\arraystretch}{1.0}
		\resizebox{1.0\textwidth}{!}{
			\begin{tabular}{|c| c |cccccccccccc|}
				\hline
				&  & \multicolumn{12}{c|}{$\theta$} \\ \cline{3-14}
				
				Percentile Value & $n$ & -8 & -5 & -3 & -1.5 & -1 & -0.5 & 0.5 & 1 & 1.5 & 3 & 5 & 8 \\
				
				\hline
				
				\multirow{2}{*}{$S_n(0.90)$}
				& 23 & 1.149 & 0.766 & 0.531 & 0.398 & 0.358 & 0.321 & 0.263 & 0.243 & 0.220 & 0.176 & 0.142 & 0.113 \\
				& 44 & 0.773 & 0.508 & 0.364 & 0.279 & 0.254 & 0.233 & 0.192 & 0.178 & 0.166 & 0.140 & 0.119 & 0.097 \\
				
				\hline
				
				\multirow{2}{*}{$S_n(0.95)$}
				& 23 & 1.293 & 0.885 & 0.626 & 0.476 & 0.435 & 0.386 & 0.319 & 0.287 & 0.257 & 0.206 & 0.163 & 0.129 \\
				& 44 & 0.890 & 0.594 & 0.434 & 0.331 & 0.308 & 0.280 & 0.232 & 0.210 & 0.198 & 0.167 & 0.137 & 0.112 \\
				
				\hline
				
				\multirow{2}{*}{$T_n(0.90)$}
				& 23 & 1.829 & 1.617 & 1.428 & 1.284 & 1.246 & 1.197 & 1.115 & 1.074 & 1.034 & 0.952 & 0.869 & 0.782 \\
				& 44 & 1.734 & 1.477 & 1.297 & 1.171 & 1.137 & 1.098 & 1.023 & 0.993 & 0.960 & 0.897 & 0.836 & 0.767 \\
				
				\hline
				
				\multirow{2}{*}{$T_n(0.95)$}
				& 23 & 1.954 & 1.738 & 1.533 & 1.390 & 1.352 & 1.292 & 1.208 & 1.158 & 1.121 & 1.018 & 0.924 & 0.830 \\
				& 44 & 1.847 & 1.580 & 1.395 & 1.257 & 1.228 & 1.179 & 1.095 & 1.064 & 1.033 & 0.965 & 0.894 & 0.817 \\
				
				\hline
				
			\end{tabular}
		}
		
	\end{table}
	
	Next, we obtain the estimated association parameter for each pair of variables of interest, assuming that a 
	Frank Copula-based model holds, along with the observed test statistic values, denoted by $S^{obs}_n$ and 
	$T^{obs}_n$, respectively, as shown in Table~\ref{tab:table_MDR}.
	
	
	\begin{remark}
		\label{rem: 5.1}
		
		Note that in Section 3 it was shown that the estimator BJPE has the overall best performance (in terms of MSE) among all the estimators for small sample sizes (i.e., $n \leq 25$). Therefore, while computing the GoF test statistics $S_n$ and $T_n$, one is compelled to look at these statistics using $\hat{\theta} = \hat{\theta}_{BJPE}$. This has been done in the following Table \ref{tab:table_MDR}. Also, the tables of critical values of the test statistics $S_n $ and $T_n$ when $\hat{\theta} = \hat{\theta}_{BJPE}$ have been provided in the Appendix \ref{app:A4}. It was observed that the critical values of $S_n$ and $T_n$ (using $\hat{\theta} = \hat{\theta}_{MLE}$ vs. $\hat{\theta} = \hat{\theta}_{BJPE})$ are nearly identical for $n \geq 50$; almost same when $25 < n < 50$; and somewhat different when $n \leq 25$.
		
	\end{remark}
	
	\vspace{-0.3cm}
	
	\begin{table}[H]
		\centering
		\caption{Observed test statistic values and the estimated $\theta$}
		\label{tab:table_MDR}
		
		\small
		
		\setlength{\tabcolsep}{9.5pt}
		\renewcommand{\arraystretch}{1.1}
		
			
			\begin{tabular}{|c|c|ccc |ccc|}
				\hline
				\multirow{2}{*}{\textbf{Subregion}} 
				& \multirow{2}{*}{\shortstack{\textbf{Pair of}\\\textbf{Variables}}}
				& \multicolumn{3}{c|}{\textbf{MLE}}
				& \multicolumn{3}{c|}{\textbf{BJPE}} \\
				\cline{3-8}
				&
				& $\widehat{\theta}$ 
				& $S_n^{\mathrm{obs}}$ 
				& $T_n^{\mathrm{obs}}$
				& $\widehat{\theta}$ 
				& $S_n^{\mathrm{obs}}$ 
				& $T_n^{\mathrm{obs}}$ \\
				\hline
				
				\multirow{4}{*}{North ($n_N=23$)}
				& \textit{(Cl, As)} & 0.154 & 0.163 & 0.996 & 0.176 & 0.160 & 0.993 \\
				& \textit{(Eh, As)}
				& -2.610 & 0.177 & 0.635 & -2.518 & 0.174 & 0.646 \\
				& \textit{(pH, As)} & 1.831 & 0.123 & 0.787 & 1.704 & 0.124 & 0.801 \\
				& \textit{(Eh, pH)} & -4.006 & 0.304 & 1.084 & -3.916 & 0.292 & 1.064 \\
				\hline
				
				\multirow{4}{*}{South ($n_S=44$)}
				& \textit{(Cl, As)} & -2.157 & 0.144 & 0.877 & -2.114 & 0.142 & 0.868 \\
				& \textit{(Eh, As)} & -7.017 & 0.199 & 0.707 & -6.988 & 0.197 & 0.702 \\
				& \textit{(pH, As)} & 0.920 & 0.320 & 1.118 & 0.889 & 0.326 & 1.125 \\
				& \textit{(Eh, pH)} & 0.325 & 0.111 & 0.859 & 0.323 & 0.112 & 0.859 \\
				\hline
				
			\end{tabular}
	\end{table}

	
	\begin{remark}\label{rem: 5.2}
		As discussed in Section~4, it was noted that the critical values
		of the test statistics $S_n$ and $T_n$ are monotonically decreasing
		in $\theta$. For the pairs of variables shown in Table~\ref{tab:table_MDR}, all the
		observed $S_n$ and $T_n$ values (i.e., $S_n^{\mathit{obs}}$ and
		$T_n^{\mathit{obs}}$, respectively) do not exceed the respective critical value
		as discussed in Remark~\ref{rem:gof_implementation}.	
		For example, consider the pair $(Eh,pH)$ in the northern subregion
		with $n=23$ and
		$\hat{\theta}_{\mathit{BJPE}}=-3.916 \approx -4.0$.
		From Table~ \ref{tab:sim_bias_mse_a} (a), the MSE of $\hat{\theta}_{\mathit{BJPE}}$
		at $\theta=-4.0$ is approximately $2.6$ (by interpolation).
		So,
		\(
		\hat{\theta}-2\sqrt{\mathit{MSE}}
		\approx -7.22 \approx -7.
		\)
		From Tables \ref{tab:Sn_BJP_combined} and \ref{tab:Tn_BJP_combined},
		\(
		S_n(0.95\mid -7)\approx 1.19, \text{and }
		T_n(0.95\mid -7)\approx 1.87.
		\)
		Now notice that
		\(
		S_n^{\mathit{obs}}=0.292 < 1.19, \text{and }
		T_n^{\mathit{obs}}=1.064 < 1.87,
		\)
		thereby retaining a bivariate FC-based model for $(Eh,pH)$
		in the northern subregion.
	\end{remark}

	Apart from comparing the observed $S_n$ and $T_n$ values (i.e., $S_n^{obs}$ and $T_n^{obs}$) with respective critical values as discussed above, we have also carried out nonparametric bootstrap $p$-value computations with the above $S_n^{obs}$ and $T_n^{obs}$, results of which are shown in Table \ref{tab:table_MDR}. The algorithmic steps of the bootstrap scheme have been provided in the Appendix \ref{app:A3}. The bootstrap $p$-values have been denoted by $p_{Boot}(S_n)$ and $p_{Boot}(T_n)$ for the test statistics $S_n$ and $T_n$, respectively.
	

	\begin{table}[H]
		\centering
		\caption{Bootstrap $p$-values for the two GoF Test on Frank Copula}
		\label{tab:bootstrap_pvalues}
		
		\small
		
		\setlength{\tabcolsep}{4pt}
		\renewcommand{\arraystretch}{1.1}
		
		\resizebox{\textwidth}{!}{
			\begin{tabular}{|c|c|ccc|ccc|}
				\hline
				\multirow{2}{*}{\textbf{Subregion}}
				& \multirow{2}{*}{\shortstack{\textbf{Pair of}\\\textbf{Variables}}}
				& \multicolumn{3}{c|}{\textbf{MLE}}
				& \multicolumn{3}{c|}{\textbf{BJPE}} \\
				\cline{3-8}
				&
				& $\widehat{\theta}$
				& $p_{\mathrm{Boot}}(S_n)$
				& $p_{\mathrm{Boot}}(T_n)$
				& $\widehat{\theta}$
				& $p_{\mathrm{Boot}}(S_n)$
				& $p_{\mathrm{Boot}}(T_n)$ \\
				\hline
				
				\multirow{4}{*}{North ($n_N=23$)}
				& \textit{(Cl, As)} & 0.154 & 0.961 & 0.836 & 0.176 & 0.967 & 0.843 \\
				& \textit{(Eh, As)} & -2.610 & 0.945 & 0.999 & -2.518 & 0.948 & 0.998 \\
				& \textit{(pH, As)} & 1.831 & 0.948 & 0.922 & 1.704 & 0.954 & 0.914 \\
				& \textit{(Eh, pH)} & -4.006 & 0.959 & 0.949 & -3.916 & 0.959 & 0.950 \\
				\hline
				
				\multirow{4}{*}{South ($n_S=44$)}
				& \textit{(Cl, As)} & -2.157 & 0.982 & 0.956 & -2.114 & 0.982 & 0.958 \\
				& \textit{(Eh, As)} & -7.017 & 0.991 & > 0.999 & -6.988 & 0.991 & > 0.999 \\
				& \textit{(pH, As)} & 0.920 & 0.909 & 0.946 & 0.889 & 0.907 & 0.945 \\
				& \textit{(Eh, pH)} & 0.325 & 0.997 & 0.944 & 0.323 & 0.997 & 0.946 \\
				\hline
				
			\end{tabular}
		}
	\end{table}

	Note that the above bootstrap $p$-values overwhelmingly support modelling each pair of variables by Frank Copula-based bivariate distributions.
	\medskip
	
	The following Table \ref{tab:correlation_measures} summarizes our correlation estimates $\hat{\hat{\rho}}_K$ and $\hat{\hat{\rho}}_S$ for four bivariate pairs in two subregions under suitable Frank Copula-based bivariate models. The estimated correlations and/or the estimated values of $\theta$ reveal distinct dynamics in groundwater in two subregions. Recall that  $\hat{\hat{\rho}}_K = \rho_K(\hat{\theta})$ and $\hat{\hat{\rho}}_S = \rho_S(\hat{\theta})$ are the parametric estimates of $\rho_K$ and $\rho_S$, as opposed to the nonparametric estimates $\hat{\rho}_K$ and $\hat{\rho}_S$ presented in Section 1 earlier. 

	\begin{table}[H]
		\centering
		\caption{Correlation measures for pairwise groundwater variables in two subregions}
		\label{tab:correlation_measures}
		
		\small
		
		\setlength{\tabcolsep}{9pt}
		\renewcommand{\arraystretch}{1.1}
		
		\begin{tabular}{|c|c|ccc|ccc|}
			\hline
			\multirow{2}{*}{\textbf{Subregion}}
			& \multirow{2}{*}{\shortstack{\textbf{Pair of}\\\textbf{Variables}}}
			& \multicolumn{3}{c|}{\textbf{MLE}}
			& \multicolumn{3}{c|}{\textbf{BJPE}} \\
			\cline{3-8}
			&
			& $\widehat{\theta}$
			& $\hat{\hat{\rho}}_K$
			& $\hat{\hat{\rho}}_S$
			& $\widehat{\theta}$
			& $\hat{\hat{\rho}}_K$
			& $\hat{\hat{\rho}}_S$ \\
			\hline
			
			\multirow{4}{*}{North ($n_N=23$)}
			& \textit{(Cl, As)} & 0.154 & 0.017 & 0.026 & 0.176 & 0.020 & 0.029 \\
			& \textit{(Eh, As)} & -2.610 & -0.272 & -0.400 & -2.518 & -0.264 & -0.388 \\
			& \textit{(pH, As)} & 1.831 & 0.197 & 0.292 & 1.704 & 0.184 & 0.274 \\
			& \textit{(Eh, pH)} & -4.006 & -0.389 & -0.558 & -3.916 & -0.382 & -0.549 \\
			\hline
			
			\multirow{4}{*}{South ($n_S=44$)}
			& \textit{(Cl, As)} & -2.157 & -0.229 & -0.339 & -2.114 & -0.225 & -0.333 \\
			& \textit{(Eh, As)} & -7.017 & -0.563 & -0.764 & -6.988 & -0.562 & -0.762 \\
			& \textit{(pH, As)} & 0.920 & 0.101 & 0.152 & 0.889 & 0.098 & 0.147 \\
			& \textit{(Eh, pH)} & 0.325 & 0.036 & 0.054 & 0.323 & 0.036 & 0.054 \\
			\hline
			
		\end{tabular}
	\end{table}

	\begin{remark}
		\label{rem:correlation_gof}
		Note that the parametric correlation estimates $\widehat{\widehat{\rho}}_K$ and $\widehat{\widehat{\rho}}_S$
		are very close to their nonparametric counterparts $\widehat{\rho}_K$ and
		$\widehat{\rho}_S$, respectively, as shown earlier in Table~\ref{tab:correlation_estimates}.
		This, in a way, is  favourable as there is no conflict between the two types of correlation estimates. Furthermore, the GoF tests indicate that the bivariate distributions of interest can be modelled by the Frank Copula fairly well. However, this
		is not the end of the applied investigation, rather the beginning of a deeper study where one may wish to fit a (nonlinear) regression model $E(Y \mid X)$ by exploiting the conditional distribution of $Y$, given $X$,
		based on the Frank Copula-based joint distribution. The objective would be to predict $Y$ (say, \textit{As}) based on $X$ (say, \textit{Eh}), when the corresponding association parameter $\theta$ is statistically significant.  This requires a comprehensive undertaking for a future study, but the following section deals with the question of nonparametric marginals versus parametric marginals from a regression point of view for the given groundwater data.
	\end{remark}
	
	\begin{remark}\label{rem:bdl_sensitivity}
		As stated in Appendix A.1 (see the footnote), a very few
		`bdl' (below detection limit) values were replaced by the midpoint of the detection range. Such adjustments do not make any change in the computations of $\hat{\theta}$, $S_n$, $T_n$, and other
		subsequent results. Because such `bdl' values retain the smallest ranks among all the values of the respective variables. Instead of the midpoint of the detection range, if `bdl' values
		are taken as either 0 or the exact detection limit then only the Pearson's correlation coefficient ($\hat{\rho}_P$) changes marginally,
		which is inconsequential for our FC study.
	\end{remark}

	
	\section{Further Application: Regression through Frank Copula (FC)}
	
	\subsection{The Motivation Behind the FC-Based Regression}
	
	One of the major applications of Copula theory in general, 
	and Frank Copula in particular, is the regression analysis where 
	suitable nonlinear regression models can be built going beyond 
	the typical linear regression model. In the following, we revisit regression of $Y$ on $X$ via regression of $V$ on $U$ (as defined before (1.2)). In the context of the groundwater data discussed earlier, we are going to explore another aspect of Frank Copula (FC) when it comes to explaining $Y$ (= Arsenic concentration (\textit{As})) through a suitable explanatory variable $X$ (= a benign variable, say Redox level (\textit{Eh}), or pH level (\textit{pH}) or Chloride(\textit{Cl}) level).
	
	\medskip
	Note that the transition $(X,Y) \longrightarrow (U,V)$ is through 
	the probability integral transforms $F_X(\cdot)$ and $F_Y(\cdot)$, 
	the marginal \textit{cdfs} of $X$ and $Y$, respectively, which are unknown in reality. As suggested earlier, the easier route to replace $F_X$ and $F_Y$ is through their nonparametric (empirical) versions as described in (\ref{eq:adjusted_ecdf}). However, there is always an argument whether one should pursue suitable parametric \textit{cdfs} to approximate $F_X(\cdot)$ and $F_Y(\cdot)$ to attain the end result, i.e., fitting $Y$ with a suitable FC-based regression. 	In this section we are going to explore both the approaches, i.e., 
	nonparametric marginal FC regression (henceforth, NM-FCR), and 
	parametric marginal FC regression (henceforth, PM-FCR) with our 
	groundwater data as a demonstration.
	
	\medskip
	At the beginning, pretending that $F_X(\cdot)$ and $F_Y(\cdot)$ are 
	known, $(U,V) \sim \mathrm{SBFCD}(\theta)$ with \textit{pdf} (1.6). Then regression of $V$ on $U$ is given as (see Genest (1987) \cite{Genest1987})
	\[
	\widehat{v}(\mathrm{say})
	=
	E(V \mid U=u)
	=
	\frac{
		u(\mathit{exp}(\theta)-1)+ (1-\mathit{exp}(\theta u))
	}{
		(1-\mathit{exp}(\theta u))
		(1-\mathit{exp}(-\theta(u-1)))
	}
	=
	r(u\mid\theta)(\mathrm{say}). \tag{6.1}
	\]
	
	However, in reality, $F_X(\cdot)$ and $F_Y(\cdot)$ need to be replaced by 
	suitable approximations, say $\widehat{F}_X(\cdot)$ and 
	$\widehat{F}_Y(\cdot)$, which could be either through nonparametric 
	empirical \textit{cdfs}, or by suitable parametric \textit{cdfs} before proceeding with the following steps to obtain the fitted value of $Y$ when $X$ is known.
	
	\subsection{General Steps for FC-Based Regression} 
	
	\noindent	\textbf{Step-1:} Obtain the pseudo observations $(U_i, V_i)$ from observed $(X_i, Y_i)$ as $U_i = \widehat{F}_X(X_i)$ and $V_i = \widehat{F}_Y(Y_i)$, $1 \leq i \leq n$.
	
	\medskip
	\noindent	\textbf{Step-2:} Use the pseudo observations $(U_i, V_i) (1 \leq i \leq n)$ to estimate $\theta$ by $\widehat{\theta}$ (which could be any one of the five estimates discussed earlier (MLE, MME1, MME2, BFPE or BJPE)).
	
	\medskip
	\noindent	\textbf{Step-3:} For any given value of $X$, say $X = x$, obtain $\widehat{u} = \widehat{F}_X(x)$.
	
	\medskip
	\noindent	\textbf{Step-4:} The fitted value of $V$, say $\widehat{v}$, is $r(\widehat{u} \mid \widehat{\theta})$, obtained from (6.1), by replacing $u$ by $\widehat{u}$ (given in Step-3).
	
	\medskip
	\noindent\textbf{Step-5:} Since $V = F_Y(Y)$, the fitted value of $Y$ is finally 
	obtained as
	\(
	\widehat{y}
	= \widehat{F}_Y^{-1}(\widehat{v})
	= \widehat{F}_Y^{-1}\bigl(r(\hat{u} \mid \widehat{\theta})\bigr).
	\)
	
	\begin{remark}\label{rem:NM_PM_marginals}
	Under NM-FCR, $\widehat{F}_X$ and $\widehat{F}_Y$ are the ones given 
	in (\ref{eq:adjusted_ecdf}). On the other hand, for PM-FCR, we have chosen the `best' 
	marginals from suitable collections of parametric models using the
	Akike Information Criterion (AIC), and then have the respective 
	model parameters estimated by the maximum likelihood method. 
	While the NM approach seems `robust', the PM approach faces the 
	danger of `hit or miss' situations. If a parametric model truly 
	gives a good fit to a marginal, then it ought to yield more accurate inferences 
	as opposed to a case of a `bad fit' marginal.
	
	\medskip
	In the following subsection, we are going to compare the two 
	FC-based regression models (NM-FCR and PM-FCR). 
	Before proceeding further, we first show the AIC values of the 
	parametric models for specific variables in both the regions. 
	The best fit has been marked by an asterisk~(*).
\end{remark}
	
	
	\begin{table}[H]
		\centering
		\caption{AIC values of marginal parametric models for four variables}
		\label{tab:AIC_parametric_models}
		
		\renewcommand{\arraystretch}{1.0}
		\setlength{\tabcolsep}{11pt}
		
		\begin{tabular}{|c|l|c c c c|}
			\hline
			
			\multirow{2}{*}{Region}
			& \multicolumn{1}{c|}{Parametric}
			& \multirow{2}{*}{\textit{As}}
			& \multirow{2}{*}{\textit{Eh}}
			& \multirow{2}{*}{\textit{pH}}
			& \multirow{2}{*}{\textit{Cl}} \\
			
			& \multicolumn{1}{c|}{Model}
			& & & & \\
			
			\hline
			
			
			\multirow{8}{*}{\shortstack{North\\$(n=23)$}}
			& Skew-$t$
			& 112.481
			& \textbf{261.259*}
			& \textbf{11.313*}
			& 305.156 \\
			
			& Normal
			& 148.485
			& 265.199
			& 23.486
			& 312.881 \\
			
			& Gamma
			& 103.803
			& N/A
			& 22.209
			& \textbf{300.911*} \\
			
			& Weibull
			& \textbf{103.315*}
			& N/A
			& 33.701
			& 301.343 \\
			
			& Lognormal
			& 104.667
			& N/A
			& 21.430
			& 315.870 \\
			
			& Logistic
			& 142.844
			& 264.913
			& 20.130
			& 313.193 \\
			
			& Student-$t$
			& 136.084
			& 267.019
			& 19.732
			& 314.882 \\
			
			& Skew-Normal
			& 128.112
			& 267.191
			& 11.720
			& 303.807 \\
			\hline
			
			
			\multirow{8}{*}{\shortstack{South\\$(n=44)$}}
			& Skew-$t$
			& 609.132
			& \textbf{491.540*}
			& 34.385
			& 484.887 \\
			
			& Normal
			& 632.103
			& 540.585
			& \textbf{30.587*}
			& 644.151 \\
			
			& Gamma
			& \textbf{584.243*}
			& N/A
			& 30.780
			& 502.547 \\
			
			& Weibull
			& 589.784
			& N/A
			& 35.519
			& 493.274 \\
			
			& Lognormal
			& 608.370
			& N/A
			& 30.920
			& \textbf{484.895*} \\
			
			& Logistic
			& 634.077
			& 540.656
			& 30.748
			& 618.992 \\
			
			& Student-$t$
			& 634.107
			& 533.835
			& 32.437
			& 540.412 \\
			
			& Skew-Normal
			& 634.071
			& 506.395
			& 32.524
			& 593.944 \\
			\hline
			
		\end{tabular}
	
	\vspace{0.15cm}
	
	\begin{minipage}{0.95\textwidth}
		\footnotesize
		\textit{Note:} Strictly nonnegative distributions cannot be applied to \textit{Eh} as it may take negative values, and hence are not applicable (N/A).
	\end{minipage}
\end{table}

\subsection{Comparision of NM-FCR and PM-FCR}

For both the subregions (North and South) and each pair of variables as described in the previous section, we obtain the fitted value of $Y_i$, say 
$\widehat{y}_i$ through the five steps given in Subsection 6.2 for each $X_i$.	
A regression model, whether it is NM-FCR or PM-FCR, is then 
evaluated by the metrics Mean Absolute Error (MAE), and 
Root Mean Squared Error (RMSE) defined as
\[
\mathrm{MAE}
=
\textstyle \sum_{i=1}^{n}
\left|Y_i-\widehat{y}_i\right|/n
\quad\text{and}\quad
\mathrm{RMSE}
=
\left\{
\textstyle\sum_{i=1}^{n}
(Y_i-\widehat{y}_i)^2 /n
\right\}^{1/2}.
\]

It is easy to show (using Jensen's inequality) that
$\mathrm{MAE} \leq \mathrm{RMSE}$; however, MAE treats all deviations
(i.e., $\widehat{y}_i$ from $Y_i$) equally, whereas RMSE is heavily
influenced by a few relatively large deviations. 	
The final FC-based regression model evaluations are summarized
in the following Table \ref{tab:MAE_RMSE_North}


\begin{table}[H]
	\centering
	\caption{Comparison of MAE and RMSE values of NM-FCR and PM-FCR
		(for \textbf{North})}
		\label{tab:MAE_RMSE_North}
	
	\renewcommand{\arraystretch}{1.30}
	\setlength{\tabcolsep}{2.8pt}
	\small
	
	\resizebox{\textwidth}{!}{%
		\begin{tabular}{|c|c|ccccc|ccccc|}
			\hline
			
			\multirow{2}{*}{\textbf{Pairs}}
			&
			\multirow{2}{*}{\textbf{Criterion}}
			&
			\multicolumn{5}{c|}{
				\textbf{NM-FCR with $\hat{\boldsymbol{\theta}}=$}
				}
			&
			\multicolumn{5}{c|}{
				\textbf{PM-FCR with $\hat{\boldsymbol{\theta}}=$}
				}
			\\
			\cline{3-12}
			&
			&
			\textbf{MLE}
			&
			\textbf{MME1}
			&
			\textbf{MME2}
			&
			\textbf{BFPE}
			&
			\textbf{BJPE}
			&
			\textbf{MLE}
			&
			\textbf{MME1}
			&
			\textbf{MME2}
			&
			\textbf{BFPE}
			&
			\textbf{BJPE}
			\\
			\hline
			
			\multirow{3}{*}{
				$\begin{array}{c}
					(Eh,As)\\[2pt]
					\hat{\rho}_{K}=-0.323,\\
					\hat{\rho}_{S}=-0.414
				\end{array}$
			}
			&
			&
			$-2.618$ & $-3.180$ & $-2.719$ & $-2.613$ & $-2.541$
			&
			$-2.116$ & $-3.180$ & $-2.719$ & $-2.129$ & $-2.081$
			\\
			
			&
			$\mathrm{MAE}=$
			&
			3.30 & 3.30 & 3.28 & 3.30 & 3.30
			&
			3.22 & 3.25 & 3.24 & 3.22 & 3.22
			\\
			
			&
			$\mathrm{RMSE}=$
			&
			5.71 & 5.65 & 5.70 & 5.71 & 5.71
			&
			5.73 & 5.56 & 5.63 & 5.73 & 5.74
			\\
			\hline
			
			\multirow{3}{*}{
				$\begin{array}{c}
					(Cl,As)\\[2pt]
					\hat{\rho}_{K}=0.012,\\
					\hat{\rho}_{S}=0.018
				\end{array}$
			}
			&
			&
			0.142 & 0.108 & 0.107 & 0.172 & 0.167
			&
			$-0.253$ & 0.108 & 0.107 & $-0.237$ & $-0.232$
			\\
			
			&
			$\mathrm{MAE}=$
			&
			3.42 & 3.42 & 3.42 & 3.42 & 3.42
			&
			3.43 & 3.44 & 3.44 & 3.43 & 3.43
			\\
			
			&
			$\mathrm{RMSE}=$
			&
			5.91 & 5.91 & 5.91 & 5.91 & 5.91
			&
			6.04 & 6.06 & 6.06 & 6.04 & 6.04
			\\
			\hline
			
			\multirow{3}{*}{
				$\begin{array}{c}
					(pH,As)\\[2pt]
					\hat{\rho}_{K}=0.188,\\
					\hat{\rho}_{S}=0.260
				\end{array}$
			}
			&
			&
			1.878 & 1.746 & 1.612 & 1.825 & 1.772
			&
			1.648 & 1.746 & 1.612 & 1.652 & 1.613
			\\
			
			&
			$\mathrm{MAE}=$
			&
			3.44 & 3.36 & 3.38 & 3.36 & 3.36
			&
			3.30 & 3.29 & 3.31 & 3.30 & 3.31
			\\
			
			&
			$\mathrm{RMSE}=$
			&
			5.79 & 5.77 & 5.78 & 5.77 & 5.77
			&
			5.75 & 5.73 & 5.76 & 5.75 & 5.76
			\\
			\hline
			
		\end{tabular}%
	}
\end{table}


\begin{table}[H]
	\centering
	\caption{Comparison of MAE and RMSE values of NM-FCR and PM-FCR
		(for \textbf{South})}
		\label{tab:MAE_RMSE_South}
	
	\renewcommand{\arraystretch}{1.30}
	\setlength{\tabcolsep}{2.8pt}
	\small
	
	\resizebox{\textwidth}{!}{%
		\begin{tabular}{|c|c|ccccc|ccccc|}
			\hline
			
			\multirow{2}{*}{\textbf{Pairs}}
			&
			\multirow{2}{*}{\textbf{Criterion}}
			&
			\multicolumn{5}{c|}{
				\textbf{NM-FCR with $\hat{\boldsymbol{\theta}} =$}
			}
			&
			\multicolumn{5}{c|}{
				\textbf{PM-FCR with $\hat{\boldsymbol{\theta}}=$ }   
			}
			\\
			\cline{3-12}
			
			&
			&
			\textbf{MLE}
			&
			\textbf{MME1}
			&
			\textbf{MME2}
			&
			\textbf{BFPE}
			&
			\textbf{BJPE}
			&
			\textbf{MLE}
			&
			\textbf{MME1}
			&
			\textbf{MME2}
			&
			\textbf{BFPE}
			&
			\textbf{BJPE}
			\\
			\hline
			
			\multirow{3}{*}{
				$\begin{array}{c}
					(Eh,As)\\[2pt]
					\hat{\rho}_{K}=-0.577,\\
					\hat{\rho}_{S}=-0.753
				\end{array}$
			}
			&
			&
			$-7.006$
			&
			$-7.354$
			&
			$-6.793$
			&
			$-7.109$
			&
			$-6.979$
			&
			$-8.380$
			&
			$-7.354$
			&
			$-6.793$
			&
			$-8.505$
			&
			$-8.360$
			\\
			
			&
			$\mathrm{MAE}=$
			&
			160.34
			&
			159.25
			&
			161.81
			&
			159.83
			&
			160.34
			&
			169.54
			&
			169.62
			&
			171.09
			&
			169.61
			&
			169.52
			\\
			
			&
			$\mathrm{RMSE}=$
			&
			251.18
			&
			251.12
			&
			251.37
			&
			251.15
			&
			251.18
			&
			261.72
			&
			262.94
			&
			264.38
			&
			261.67
			&
			261.73
			\\
			\hline
			
			\multirow{3}{*}{
				$\begin{array}{c}
					(Cl,As)\\[2pt]
					\hat{\rho}_{K}=-0.230,\\
					\hat{\rho}_{S}=-0.320
				\end{array}$
			}
			&
			&
			$-2.157$
			&
			$-2.169$
			&
			$-2.021$
			&
			$-2.151$
			&
			$-2.119$
			&
			$-2.563$
			&
			$-2.169$
			&
			$-2.021$
			&
			$-2.568$
			&
			$-2.529$
			\\
			
			&
			$\mathrm{MAE}=$
			&
			233.68
			&
			233.68
			&
			233.84
			&
			233.68
			&
			233.50
			&
			241.07
			&
			243.30
			&
			244.15
			&
			241.04
			&
			241.26
			\\
			
			&
			$\mathrm{RMSE}=$
			&
			306.59
			&
			306.59
			&
			305.95
			&
			306.59
			&
			306.06
			&
			337.03
			&
			338.68
			&
			339.37
			&
			337.01
			&
			337.16
			\\
			\hline
			
			\multirow{3}{*}{
				$\begin{array}{c}
					(pH,As)\\[2pt]
					\hat{\rho}_{K}=0.101,\\
					\hat{\rho}_{S}=0.156
				\end{array}$
			}
			&
			&
			0.921
			&
			0.915
			&
			0.944
			&
			0.912
			&
			0.901
			&
			1.043
			&
			0.915
			&
			0.944
			&
			1.040
			&
			1.028
			\\
			
			&
			$\mathrm{MAE}=$
			&
			239.42
			&
			239.42
			&
			239.55
			&
			239.35
			&
			239.75
			&
			258.65
			&
			258.57
			&
			258.59
			&
			258.65
			&
			258.64
			\\
			
			&
			$\mathrm{RMSE}=$
			&
			308.93
			&
			308.93
			&
			308.71
			&
			308.93
			&
			309.23
			&
			349.47
			&
			349.67
			&
			349.62
			&
			349.48
			&
			349.49
			\\
			\hline
			
		\end{tabular}%
	}
	
\end{table}


\begin{remark}\label{rem:NM_PM_FCR}
	Tables \ref{tab:MAE_RMSE_North} and \ref{tab:MAE_RMSE_South} reveal some interesting patterns in NM-FCR
	vis-a-vis PM-FCR. In the northern subregion, with a sample size
	$n_N = 23$, both approaches show almost identical performance
	in terms of both MAE as well as RMSE. There is no substantial gain
	in adopting the parametric marginal approach. On the other hand,
	in the southern subregion, with $n_S = 44$, the nonparametric marginal
	approach is definitely better, and hence a robust approach,	irrespective of the estimator of $\theta$ used.
\end{remark}


\begin{remark}
	A more meaningful approach would be to model a four dimensional model of the entire elemental vector (As, Cl, Eh, pH), and then see how to predict As based on the values of the other three benign elements observed together. However, this would require a multivariate Frank Copula-based regression model and studying the complexity of estimating $\theta$ along with other inferential aspects. This is under investigation now, and hopefully will be reported in near future.
\end{remark}

\newpage
\section{Conclusion}

This work on a bivariate Frank Copula extends the recent work by
Pham et al.~(2025) \cite{Pham et al.(2025)} by considering two noninformative Bayes estimators and comparing them with the MLE. Interestingly, one of
these two Bayes estimators is found to be better than the MLE,
especially for small samples (with size not exceeding 25). 
Next, we have looked into the goodness of fit (GoF) tests suggested
by Genest et al.~(2006) \cite{GenestRemillard2006}, and our extensive simulations complement the above authors' work apart from exploring some non-intuitive properties of the GoF tests. Finally, we demonstrate the applicability of Frank Copula to model
groundwater data based on Merola et al.'s~(2015) \cite{Merola2015} fieldwork. It was also demonstrated that, for our groundwater data, using nonparametric marginals for the two components in a bivariate Frank Copula set up is advantageous over the usage of suitable parametric marginals from a regression point of view.

\section*{Acknowledgements}

The authors would like to express their sincere gratitude to Ton Duc Thang University (TDTU), Ho Chi Minh City, 
and Ho Chi Minh City University of Technology (HCMUT), VNU-HCM for supporting this study. The logistical help 
and encouragement extended by both the institutions have played a crucial role in the completion of this research.
A substantial part of this work was done while the third (and the corresponding) author (Nabendu Pal) was visiting TDTU (from his parent 
institution - University of Louisiana at Lafayette, USA) in Summer of 2025 to teach a graduate level summer course.
The author would like to thank the TDTU administration for their generous hospitality.
\medskip

\noindent\textbf{Funding}
\medskip

\noindent No funding was received for this research. 
\medskip

\noindent \textbf{Competing Interest}
\medskip

\noindent The corresponding author, on behalf of all the authors, declares that there are no conflicts of interest.


\clearpage
\appendix

\begin{center}
	{\Large\bfseries Appendix A}
\end{center}

\vspace{-0.5cm}

\renewcommand{\thesection}{A.\arabic{section}}
\setcounter{section}{0}

\renewcommand{\thetable}{\thesection.\arabic{table}}
\renewcommand{\thefigure}{\thesection.\arabic{figure}}
\renewcommand{\theequation}{\thesection.\arabic{equation}}

\section{Groundwater data from Dong Thap, Vietnam (Merola et al., 2015)}
\label{app:groundwater_data}

\setcounter{table}{0}

\begingroup
\footnotesize
\setlength{\tabcolsep}{4pt}
\renewcommand{\arraystretch}{0.95}
\setlength\LTleft{0pt}
\setlength\LTright{0pt}
\captionsetup{skip=3pt}

\vspace{-1cm}

\begin{longtable}{@{}c l r r r r l r r r r@{}}
	
	\label{tab:groundwater_data}\\
	
	\toprule
	& \multicolumn{5}{c}{North}
	& \multicolumn{5}{c}{South} \\
	\cmidrule(lr){2-6}
	\cmidrule(lr){7-11}
	
	No.
	& Well ID & \textit{As} (ppb) & \textit{Cl} (ppm) & \textit{Eh} (mV) & \textit{pH}
	& Well ID & \textit{As} (ppb) & \textit{Cl} (ppm) & \textit{Eh} (mV) & \textit{pH} \\
	\midrule
	
	\endfirsthead
	
	\toprule
	& \multicolumn{5}{c}{North}
	& \multicolumn{5}{c}{South} \\
	\cmidrule(lr){2-6}
	\cmidrule(lr){7-11}
	
	No.
	& Well ID & As (ppb) & Cl (ppm) & Eh (mV) & pH
	& Well ID & As (ppb) & Cl (ppm) & Eh (mV) & pH \\
	\midrule
	
	\endhead
	
	\midrule
	\multicolumn{11}{r}{Continued on next page} \\
	\endfoot
	
	\bottomrule
	\endlastfoot
	
	
	1
	& TH16 & 0.4 & 173.6 & 157 & 6.14
	& DT7 & 563.9 & 107.0 & -126 & 6.78 \\
	
	2
	& TH9 & 0.2 & 275.3 & 253 & 5.84
	& DT6 & 0.5 & 56.1 & 142 & 6.71 \\
	
	3
	& TH13 & $bdl^*$ & 22.7 & 194 & 6.19
	& DT5 & 0.7 & 46.8 & 199 & 7.04 \\
	
	4
	& TH14 & 0.3 & 113.6 & 184 & 6.02
	& DT3 & 0.4 & 345.2 & 169 & 6.44 \\
	
	5
	& TH22 & 0.1 & 228.1 & 226 & 6.50
	& DT4 & 0.1 & 500.1 & 165 & 6.51 \\
	
	6
	& TH21 & 0.3 & 89.9 & 169 & 6.10
	& DT2 & 1.8 & 632.8 & 101 & 6.66 \\
	
	7
	& TH5 & 0.8 & 742.1 & 251 & 5.83
	& DT1 & 13.1 & 19.7 & 97 & 7.75 \\
	
	8
	& TH12 & 2.3 & 182.7 & 127 & 6.31
	& TB11 & 462.3 & 9.2 & -114 & 6.92 \\
	
	9
	& TH15 & 8.4 & 27.5 & 60 & 6.18
	& TB18 & 155.7 & 25.9 & -72 & 6.52 \\
	
	10
	& TH1 & 6.0 & 544.4 & 210 & 6.08
	& TB9 & 187.6 & 12.8 & -128 & 6.94 \\
	
	11
	& TH10 & 3.2 & 277.3 & 231 & 6.00
	& TB2 & 850.4 & 10.5 & -133 & 7.14 \\
	
	12
	& TH2 & 2.0 & 487.6 & 130 & 5.87
	& TB24 & 370.4 & 13.9 & -90 & 7.15 \\
	
	13
	& TH23 & 0.2 & 158.5 & 175 & 6.04
	& TB26 & 139.9 & 13.8 & -83 & 7.43 \\
	
	14
	& TH3 & 1.5 & 560.2 & 261 & 6.00
	& TB27 & 77.7 & 5.4 & -33 & 7.24 \\
	
	15
	& TH4 & 2.6 & 21.4 & 80 & 5.99
	& TB21 & 842.1 & 21.1 & -105 & 6.88 \\
	
	16
	& TH11 & 8.9 & 479.8 & 158 & 6.56
	& TB1 & 276.8 & 19.6 & -92 & 6.85 \\
	
	17
	& TH18 & 3.6 & 335.6 & 181 & 6.29
	& TB10 & 377.3 & 8.2 & -129 & 6.79 \\
	
	18
	& TH8 & 6.0 & 253.2 & 235 & 5.85
	& TB25 & 272.9 & 11.9 & -104 & 7.20 \\
	
	19
	& TH7 & 0.7 & 122.3 & 162 & 6.29
	& TB13 & 746.0 & 72.7 & -125 & 7.16 \\
	
	20
	& TH6 & $bdl^*$ & 242.8 & 200 & 6.19
	& TB22 & 311.0 & 13.5 & -130 & 6.63 \\
	
	21
	& TH17 & 22.2 & 40.5 & -13 & 7.03
	& TB15 & 937.7 & 19.3 & -110 & 7.04 \\
	
	22
	& TH19 & 17.5 & 57.3 & 145 & 7.39
	& TB16 & 314.5 & 25.8 & -115 & 6.74 \\
	
	23
	& TH20 & 2.4 & $bdl^*$ & 24 & 6.51
	& TB20 & 746.3 & 6.9 & -139 & 6.61 \\
	
	24
	& & & & &
	& TB23 & 270.0 & 12.7 & -110 & 7.01 \\
	
	25
	& & & & &
	& TB17 & 224.2 & 21.5 & -126 & 6.46 \\
	
	26
	& & & & &
	& TB3 & 727.0 & 10.8 & -136 & 7.14 \\
	
	27
	& & & & &
	& TB12 & 931.5 & 2.9 & -125 & 7.03 \\
	
	28
	& & & & &
	& TB14 & 747.7 & 63.4 & -115 & 7.15 \\
	
	29
	& & & & &
	& TB5 & 416.3 & $bdl^*$ & -60 & 7.69 \\
	
	30
	& & & & &
	& TB4 & 360.3 & 42.9 & -130 & 7.34 \\
	
	31
	& & & & &
	& TB6 & 315.5 & 61.2 & -111 & 7.36 \\
	
	32
	& & & & &
	& TB7 & 101.1 & 42.7 & -28 & 7.30 \\
	
	33
	& & & & &
	& TB8 & 237.6 & 124.4 & -98 & 7.17 \\
	
	34
	& & & & &
	& TB19 & 300.3 & 160.3 & -120 & 6.68 \\
	
	35
	& & & & &
	& TBE10 & 700.4 & 81.4 & -108 & 7.07 \\
	
	36
	& & & & &
	& TBE8 & 125.1 & 18.2 & -120 & 6.96 \\
	
	37
	& & & & &
	& TBE9 & 196.2 & 986.6 & -110 & 6.72 \\
	
	38
	& & & & &
	& TBE7 & 166.3 & 20.0 & -84 & 7.16 \\
	
	39
	& & & & &
	& TBE4 & 4.4 & 1499.6 & 82 & 6.09 \\
	
	40
	& & & & &
	& TBE5 & 981.4 & 60.4 & -110 & 6.87 \\
	
	41
	& & & & &
	& TBE3 & 6.8 & 2.7 & 158 & 7.17 \\
	
	42
	& & & & &
	& TBE1 & 6.6 & 61.7 & 126 & 7.10 \\
	
	43
	& & & & &
	& TBE11 & 5.3 & 12.2 & 60 & 7.16 \\
	
	44
	& & & & &
	& TBE6 & 3.2 & 1527.0 & 149 & 6.73 \\
	
\end{longtable}
\endgroup

{
	\vspace{-0.4cm}
	\noindent\footnotesize
	\textbf{Note:}
	\textit{Well IDs starting with \textit{TH} correspond to the northern
		subregion, whereas those starting with \textit{DT} or \textit{TB} belong
		to the southern subregion. For wells TH13 and TH6, arsenic (As)
		concentrations were reported as \textit{below detection limit} ($bdl^*$),
		which occurs when the observations fall below 0.050~$\mu$g/L
		(i.e., 0.050~ppb). For computational purposes, these bdl values were
		replaced by 0.025~ppb, the midpoint of the interval
		$(0,0.050)$~ppb. Similarly, for wells TB5 and TH20, chloride (Cl)
		concentrations were reported as $bdl^*$ and were substituted by
		0.1~ppm, representing the midpoint of the interval $(0,0.2)$~ppm.}
}


\section{Comparison of two MLE approaches for estimating $\theta$ in SBFCD($\theta$)}

\label{app:mle_comparison}
\vspace{-0.7cm}
\setcounter{table}{0}

\begin{table}[H]
	\centering
	\label{tab:mle_comparison}
	
	\renewcommand{\arraystretch}{1.1}
	
	\begin{tabularx}{\textwidth}{|l|X|X|}
		\hline
		{\textbf{Criterion}} &
		{\textbf{Approach--1: MLE via Log-Likelihood Maximization (Brent)}} &
		{\textbf{Approach--2: MLE via Solving Score Equation (Broyden)}} \\
		
		\hline
		
		Main idea &
		Find $\theta$ that maximizes the log-likelihood. &
		Find $\theta$ such that the derivative of the log-likelihood equals zero. \\
		\hline
		
		Formula &
		$\hat{\theta} = \arg\max_{\theta} \log L(\theta)$ &
		Solve $(\partial/\partial\theta)\log L(\theta) = 0$. \\
		\hline
		
		Implementation &
		Minimize $-\log L(\theta)$ using \textit{optim(method = ``Brent")}. &
		Solve $H(\theta) = 0$ using \textit{nleqslv(method = ``Broyden")}. \\
		\hline
		
		Derivative required? &
		No. &
		Yes. \\
		\hline
		
		Initial value required? &
		No (only interval bounds). &
		Yes (highly sensitive). \\
		\hline
		
		Numerical stability &
		Very stable and reliable. &
		Can be unstable if $H(\theta)$ is noisy or inaccurate. \\
		\hline
		
		Performance at small $n$ &
		Stable and consistent. &
		May oscillate or fail to converge.\\
		\hline
		
		Performance at large $n$ &
		Accurate. &
		Very accurate if the score function is correct. \\
		\hline
		
		Advantages &
		No derivative required; robust; rarely fails. &
		Theoretically elegant; very fast when $H(\theta)$ is smooth. \\
		\hline
		
		Disadvantages &
		Requires specifying a search interval. &
		Sensitive to initial value; requires exact derivative. \\
		\hline
		
		Best use cases &
		Simulation studies; small-sample estimation; practical MLE. &
		Theoretical validation; large-sample estimation with smooth score. \\
		\hline
	\end{tabularx}
	
\end{table}

\newpage

\section{Bootstrap $p$-value computations: Algorithmic steps}
\label{app:A3}

\noindent\textit{We describe the procedure for the test statistic
	$S_n$. The same procedure applies to $T_n$ as well.}

\vspace{0.1cm}

\noindent\textbf{Step-0. (Original data)}\\
Let the observed sample be
\vspace{-0.3cm}
\[
\mathbf{Z}_1 = (X_1, Y_1), \quad 
\mathbf{Z}_2 = (X_2, Y_2), \quad \ldots, \quad 
\mathbf{Z}_n = (X_n, Y_n).
\]

\vspace{-0.1cm}

\noindent\textbf{Step-1. (Rank transformation)}\\
Convert the data into pseudo-observations based on ranks:
\vspace{-0.3cm}
\[
\mathbf{Z}_1^{R} = (U_1, V_1), \quad 
\ldots, \quad 
\mathbf{Z}_n^{R} = (U_n, V_n),
\]

\noindent where the superscript $R$ denotes ``based on ranks''.

\vspace{0.2cm}

\noindent\textbf{Step-2. (Observed statistic)}\\
Based on $\{\mathbf{Z}_1^{R}, \ldots, \mathbf{Z}_n^{R}\}$ compute the test statistic value $S_n$ and call it $S_n^{\mathrm{obs}}$ (i.e., the `observed value')

\vspace{0.2cm}

\noindent\textbf{Step-3. (Bootstrap resampling)}\\
Go back to \textit{Step-0}, and take a random sample of size $n$ from the original sample 
\(	\{\mathbf{Z}_1, \mathbf{Z}_2, \ldots, \mathbf{Z}_n\}
\)
with replacement. Call this Bootstrap sample of size $n$ as
\(
\{\mathbf{Z}_1^*, \mathbf{Z}_2^*, \ldots, \mathbf{Z}_n^*\}.
\)

\vspace{0.2cm}

\noindent\textbf{Step-4. (Bootstrap statistic)}\\
Apply \textit{Step-1} and \textit{Step-2} to this Bootstrap sample
\(
\{\mathbf{Z}_1^*, \ldots, \mathbf{Z}_n^*\},
\)
and recalculate the value of $S_n$, and call it \(S_n^*\).

\vspace{0.2cm}

\noindent\textbf{Step-5. (Repetition)}\\
Repeat the above \textit{Step-3} and \textit{Step-4} a large number of times, say $B$ times [$B = 10^4$]. This will produce $B$ replicated Bootstrap values of $S_n^*$, and call them as 
\(
S_n^{*1}, S_n^{*2}, \ldots, S_n^{*B}.
\)

\vspace{0.2cm}

\noindent\textbf{Step-6. (Bootstrap p-value)}\\
Now we are going to compute the Bootstrap $p$-value using $S_n^{\mathrm{obs}}$ (in \textit{Step-1}) and $S_n^{*b}$ (in \textit{Step-5}) as
\vspace{-0.7 cm}
\[
p_{\text{Boot}}
=
\text{(Proportion of }S_n^{*b}\text{ values exceeding }S_n^{\mathrm{obs}}\text{)}
=
\frac{1}{B}\sum_{b=1}^{B}
\mathbf{I}\left(S_n^{*b}>S_n^{\mathrm{obs}}\right)
\]

\clearpage

\section{Critical values of $S_n$ and $T_n$ using BJPE}
\label{app:A4}
\vspace{-0.8 cm}
\setcounter{table}{0}
\begin{table}[H]
	\centering
	
	\caption{Values of $S_n(1-\alpha\mid\theta)$ for various $n$,
		$\theta$, and ($1-\alpha$)
		$\left(\hat{\theta}=\widehat{\theta}_{BJPE}\right)$.}
	\label{tab:Sn_BJP_combined}

	\scriptsize
	\renewcommand{\arraystretch}{0.8}
	\setlength{\tabcolsep}{3.6pt}
	
	\resizebox{0.88\textwidth}{!}{%
		\begin{tabular}{|c|c|cccccccc|}
			\hline
			
			\multirow{2}{*}{$1-\alpha$}
			& \multirow{2}{*}{$\theta$}
			& \multicolumn{8}{c|}{$n$} \\
			\cline{3-10}
			
			& 
			& $15$ & $20$ & $25$ & $30$
			& $35$ & $40$ & $45$ & $50$ \\
			
			\hline
			
			
			\multirow{28}{*}{$0.90$}
			& -10.00 & 1.819 & 1.567 & 1.370 & 1.226 & 1.115 & 1.021 & 0.948 & 0.884 \\
			
			& -9.00  & 1.630 & 1.406 & 1.223 & 1.092 & 0.986 & 0.910 & 0.843 & 0.791 \\
			& -8.00  & 1.460 & 1.238 & 1.085 & 0.971 & 0.880 & 0.806 & 0.755 & 0.701 \\
			& -7.00  & 1.291 & 1.095 & 0.946 & 0.847 & 0.779 & 0.709 & 0.664 & 0.618 \\
			& -6.00  & 1.116 & 0.943 & 0.817 & 0.732 & 0.667 & 0.614 & 0.575 & 0.543 \\
			& -5.00  & 0.968 & 0.805 & 0.701 & 0.627 & 0.574 & 0.533 & 0.492 & 0.470 \\
			& -4.00  & 0.814 & 0.678 & 0.590 & 0.530 & 0.481 & 0.448 & 0.424 & 0.399 \\
			& -3.00  & 0.680 & 0.566 & 0.490 & 0.440 & 0.404 & 0.377 & 0.352 & 0.335 \\
			& -2.00  & 0.559 & 0.466 & 0.402 & 0.361 & 0.332 & 0.312 & 0.293 & 0.279 \\
			& -1.50  & 0.499 & 0.422 & 0.370 & 0.331 & 0.303 & 0.286 & 0.266 & 0.257 \\
			& -1.00  & 0.456 & 0.377 & 0.330 & 0.301 & 0.276 & 0.260 & 0.246 & 0.235 \\
			& -0.75  & 0.432 & 0.360 & 0.316 & 0.287 & 0.264 & 0.247 & 0.234 & 0.226 \\
			& -0.50  & 0.410 & 0.341 & 0.302 & 0.275 & 0.254 & 0.237 & 0.226 & 0.216 \\
			& -0.10  & 0.381 & 0.313 & 0.277 & 0.255 & 0.236 & 0.221 & 0.212 & 0.202 \\
			& 0.10   & 0.364 & 0.303 & 0.269 & 0.247 & 0.228 & 0.214 & 0.204 & 0.195 \\
			& 0.50   & 0.337 & 0.284 & 0.252 & 0.229 & 0.214 & 0.203 & 0.193 & 0.184 \\
			& 0.75   & 0.320 & 0.272 & 0.243 & 0.221 & 0.205 & 0.194 & 0.185 & 0.176 \\
			& 1.00   & 0.303 & 0.260 & 0.232 & 0.211 & 0.198 & 0.187 & 0.178 & 0.172 \\
			& 1.50   & 0.279 & 0.237 & 0.212 & 0.197 & 0.184 & 0.174 & 0.167 & 0.160 \\
			& 2.00   & 0.256 & 0.219 & 0.199 & 0.183 & 0.171 & 0.165 & 0.157 & 0.152 \\
			& 3.00   & 0.217 & 0.189 & 0.172 & 0.161 & 0.152 & 0.148 & 0.142 & 0.138 \\
			& 4.00   & 0.189 & 0.168 & 0.152 & 0.144 & 0.137 & 0.132 & 0.130 & 0.125 \\
			& 5.00   & 0.169 & 0.150 & 0.139 & 0.132 & 0.127 & 0.123 & 0.120 & 0.117 \\
			& 6.00   & 0.152 & 0.137 & 0.128 & 0.121 & 0.116 & 0.113 & 0.111 & 0.109 \\
			& 7.00   & 0.141 & 0.127 & 0.118 & 0.113 & 0.109 & 0.106 & 0.104 & 0.102 \\
			& 8.00   & 0.131 & 0.118 & 0.110 & 0.106 & 0.102 & 0.100 & 0.097 & 0.095 \\
			& 9.00   & 0.123 & 0.112 & 0.104 & 0.099 & 0.095 & 0.093 & 0.091 & 0.089 \\
			& 10.00  & 0.116 & 0.105 & 0.098 & 0.093 & 0.090 & 0.087 & 0.086 & 0.084 \\
			
			\hline
			
			
			\multirow{28}{*}{$0.95$}
			& -10.00 & 2.040 & 1.765 & 1.543 & 1.385 & 1.267 & 1.157 & 1.079 & 1.005 \\
			
			& -9.00  & 1.847 & 1.591 & 1.376 & 1.244 & 1.123 & 1.038 & 0.962 & 0.909 \\
			& -8.00  & 1.665 & 1.414 & 1.234 & 1.105 & 1.001 & 0.924 & 0.867 & 0.810 \\
			& -7.00  & 1.470 & 1.251 & 1.079 & 0.971 & 0.891 & 0.820 & 0.765 & 0.717 \\
			& -6.00  & 1.283 & 1.080 & 0.941 & 0.851 & 0.772 & 0.714 & 0.671 & 0.636 \\
			& -5.00  & 1.116 & 0.937 & 0.819 & 0.725 & 0.673 & 0.622 & 0.579 & 0.554 \\
			& -4.00  & 0.951 & 0.781 & 0.685 & 0.627 & 0.568 & 0.531 & 0.498 & 0.475 \\
			& -3.00  & 0.805 & 0.664 & 0.579 & 0.522 & 0.479 & 0.450 & 0.418 & 0.404 \\
			& -2.00  & 0.664 & 0.550 & 0.476 & 0.429 & 0.397 & 0.376 & 0.351 & 0.334 \\
			& -1.50  & 0.595 & 0.503 & 0.441 & 0.397 & 0.363 & 0.341 & 0.319 & 0.308 \\
			& -1.00  & 0.543 & 0.451 & 0.396 & 0.358 & 0.330 & 0.311 & 0.295 & 0.282 \\
			& -0.75  & 0.517 & 0.431 & 0.380 & 0.340 & 0.316 & 0.296 & 0.279 & 0.273 \\
			& -0.50  & 0.494 & 0.406 & 0.359 & 0.329 & 0.303 & 0.285 & 0.274 & 0.258 \\
			& -0.10  & 0.454 & 0.375 & 0.329 & 0.305 & 0.285 & 0.265 & 0.254 & 0.242 \\
			& 0.10   & 0.437 & 0.362 & 0.322 & 0.293 & 0.272 & 0.256 & 0.244 & 0.234 \\
			& 0.50   & 0.400 & 0.340 & 0.304 & 0.274 & 0.254 & 0.241 & 0.231 & 0.220 \\
			& 0.75   & 0.382 & 0.324 & 0.287 & 0.263 & 0.244 & 0.232 & 0.220 & 0.211 \\
			& 1.00   & 0.362 & 0.310 & 0.274 & 0.252 & 0.237 & 0.224 & 0.213 & 0.205 \\
			& 1.50   & 0.332 & 0.283 & 0.251 & 0.232 & 0.217 & 0.206 & 0.199 & 0.192 \\
			& 2.00   & 0.303 & 0.260 & 0.235 & 0.217 & 0.201 & 0.194 & 0.186 & 0.180 \\
			& 3.00   & 0.256 & 0.221 & 0.202 & 0.189 & 0.178 & 0.173 & 0.167 & 0.163 \\
			& 4.00   & 0.221 & 0.194 & 0.177 & 0.167 & 0.159 & 0.155 & 0.151 & 0.147 \\
			& 5.00   & 0.196 & 0.173 & 0.160 & 0.152 & 0.147 & 0.143 & 0.138 & 0.136 \\
			& 6.00   & 0.175 & 0.157 & 0.146 & 0.139 & 0.134 & 0.131 & 0.128 & 0.127 \\
			& 7.00   & 0.160 & 0.144 & 0.135 & 0.129 & 0.124 & 0.121 & 0.120 & 0.118 \\
			& 8.00   & 0.148 & 0.134 & 0.125 & 0.120 & 0.117 & 0.114 & 0.111 & 0.110 \\
			& 9.00   & 0.139 & 0.126 & 0.117 & 0.112 & 0.108 & 0.106 & 0.104 & 0.103 \\
			& 10.00  & 0.131 & 0.118 & 0.110 & 0.105 & 0.103 & 0.101 & 0.099 & 0.096 \\
			
			\hline
		\end{tabular}%
	}
	
\end{table}


\begin{table}[H]
	\centering
	
	\caption{Values of $T_n(1-\alpha\mid\theta)$ for various $n$, $\theta$,
		and ($1-\alpha$)
		$\bigl(\hat{\theta}=\widehat{\theta}_{BJPE}\bigr)$}
	\label{tab:Tn_BJP_combined}
	
	
	\scriptsize
	\renewcommand{\arraystretch}{0.8}
	\setlength{\tabcolsep}{3.6pt}
	
	\resizebox{0.9\textwidth}{!}{%
		\begin{tabular}{|c|c|cccccccc|}
			\hline
			
			\multirow{2}{*}{$1-\alpha$}
			& \multirow{2}{*}{$\theta$}
			& \multicolumn{8}{c|}{$n$} \\
			\cline{3-10}
			
			&
			& $15$ & $20$ & $25$ & $30$
			& $35$ & $40$ & $45$ & $50$ \\
			
			\hline
			
			
			\multirow{28}{*}{$0.90$}
			& -10.00 & 1.793 & 1.902 & 1.938 & 1.945
			& 1.928 & 1.904 & 1.877 & 1.840 \\
			
			& -9.00  & 1.780 & 1.860 & 1.886 & 1.876
			& 1.849 & 1.819 & 1.793 & 1.766 \\
			
			& -8.00  & 1.744 & 1.805 & 1.823 & 1.799
			& 1.777 & 1.752 & 1.720 & 1.690 \\
			
			& -7.00  & 1.706 & 1.750 & 1.743 & 1.726
			& 1.700 & 1.665 & 1.637 & 1.612 \\
			
			& -6.00  & 1.650 & 1.684 & 1.665 & 1.641
			& 1.611 & 1.580 & 1.556 & 1.527 \\
			
			& -5.00  & 1.595 & 1.603 & 1.584 & 1.550
			& 1.522 & 1.499 & 1.462 & 1.444 \\
			
			& -4.00  & 1.529 & 1.515 & 1.492 & 1.464
			& 1.430 & 1.406 & 1.379 & 1.358 \\
			
			& -3.00  & 1.449 & 1.431 & 1.399 & 1.370
			& 1.336 & 1.317 & 1.286 & 1.273 \\
			
			& -2.00  & 1.368 & 1.345 & 1.306 & 1.274
			& 1.249 & 1.228 & 1.201 & 1.187 \\
			
			& -1.50  & 1.324 & 1.304 & 1.271 & 1.233
			& 1.204 & 1.183 & 1.159 & 1.148 \\
			
			& -1.00  & 1.291 & 1.250 & 1.218 & 1.193
			& 1.162 & 1.143 & 1.123 & 1.106 \\
			
			& -0.75  & 1.265 & 1.235 & 1.201 & 1.168
			& 1.146 & 1.121 & 1.103 & 1.094 \\
			
			& -0.50  & 1.245 & 1.211 & 1.177 & 1.151
			& 1.128 & 1.103 & 1.092 & 1.073 \\
			
			& -0.10  & 1.213 & 1.174 & 1.140 & 1.120
			& 1.096 & 1.074 & 1.063 & 1.044 \\
			
			& 0.10   & 1.196 & 1.160 & 1.129 & 1.105
			& 1.083 & 1.066 & 1.048 & 1.032 \\
			
			& 0.50   & 1.168 & 1.133 & 1.105 & 1.075
			& 1.055 & 1.039 & 1.023 & 1.007 \\
			
			& 0.75   & 1.145 & 1.113 & 1.086 & 1.059
			& 1.039 & 1.023 & 1.005 & 0.992 \\
			
			& 1.00   & 1.124 & 1.097 & 1.065 & 1.041
			& 1.025 & 1.004 & 0.992 & 0.981 \\
			
			& 1.50   & 1.093 & 1.059 & 1.033 & 1.011
			& 0.993 & 0.976 & 0.966 & 0.953 \\
			
			& 2.00   & 1.059 & 1.028 & 1.003 & 0.983
			& 0.965 & 0.953 & 0.939 & 0.929 \\
			
			& 3.00   & 0.999 & 0.967 & 0.948 & 0.931
			& 0.916 & 0.910 & 0.901 & 0.891 \\
			
			& 4.00   & 0.943 & 0.920 & 0.900 & 0.888
			& 0.877 & 0.871 & 0.866 & 0.859 \\
			
			& 5.00   & 0.899 & 0.876 & 0.862 & 0.857
			& 0.850 & 0.843 & 0.835 & 0.833 \\
			
			& 6.00   & 0.856 & 0.842 & 0.833 & 0.824
			& 0.816 & 0.815 & 0.811 & 0.811 \\
			
			& 7.00   & 0.821 & 0.812 & 0.804 & 0.800
			& 0.795 & 0.793 & 0.790 & 0.786 \\
			
			& 8.00   & 0.790 & 0.783 & 0.778 & 0.776
			& 0.772 & 0.770 & 0.768 & 0.765 \\
			
			& 9.00   & 0.759 & 0.755 & 0.753 & 0.751
			& 0.748 & 0.747 & 0.747 & 0.746 \\
			
			& 10.00  & 0.731 & 0.728 & 0.730 & 0.729
			& 0.729 & 0.726 & 0.727 & 0.724 \\
			
			\hline
			
			
			\multirow{28}{*}{$0.95$}
			& -10.00 & 1.921 & 2.036 & 2.067 & 2.076
			& 2.069 & 2.029 & 2.007 & 1.958 \\
			
			& -9.00  & 1.909 & 1.991 & 2.022 & 2.000
			& 1.975 & 1.950 & 1.915 & 1.885 \\
			
			& -8.00  & 1.873 & 1.937 & 1.952 & 1.934
			& 1.907 & 1.865 & 1.841 & 1.800 \\
			
			& -7.00  & 1.835 & 1.880 & 1.863 & 1.845
			& 1.820 & 1.782 & 1.755 & 1.720 \\
			
			& -6.00  & 1.776 & 1.810 & 1.791 & 1.761
			& 1.720 & 1.698 & 1.663 & 1.636 \\
			
			& -5.00  & 1.723 & 1.734 & 1.707 & 1.670
			& 1.638 & 1.608 & 1.567 & 1.552 \\
			
			& -4.00  & 1.660 & 1.641 & 1.607 & 1.576
			& 1.537 & 1.509 & 1.482 & 1.460 \\
			
			& -3.00  & 1.579 & 1.552 & 1.512 & 1.472
			& 1.441 & 1.418 & 1.380 & 1.371 \\
			
			& -2.00  & 1.490 & 1.456 & 1.415 & 1.376
			& 1.343 & 1.326 & 1.291 & 1.283 \\
			
			& -1.50  & 1.439 & 1.411 & 1.377 & 1.331
			& 1.296 & 1.276 & 1.252 & 1.239 \\
			
			& -1.00  & 1.404 & 1.359 & 1.310 & 1.287
			& 1.253 & 1.231 & 1.214 & 1.192 \\
			
			& -0.75  & 1.375 & 1.337 & 1.298 & 1.264
			& 1.236 & 1.206 & 1.192 & 1.180 \\
			
			& -0.50  & 1.354 & 1.311 & 1.272 & 1.242
			& 1.218 & 1.192 & 1.176 & 1.159 \\
			
			& -0.10  & 1.317 & 1.271 & 1.233 & 1.206
			& 1.183 & 1.152 & 1.146 & 1.124 \\
			
			& 0.10   & 1.301 & 1.256 & 1.218 & 1.190
			& 1.165 & 1.148 & 1.128 & 1.112 \\
			
			& 0.50   & 1.266 & 1.226 & 1.190 & 1.159
			& 1.133 & 1.119 & 1.105 & 1.086 \\
			
			& 0.75   & 1.243 & 1.205 & 1.168 & 1.142
			& 1.121 & 1.103 & 1.083 & 1.069 \\
			
			& 1.00   & 1.220 & 1.183 & 1.149 & 1.121
			& 1.108 & 1.079 & 1.068 & 1.058 \\
			
			& 1.50   & 1.185 & 1.146 & 1.112 & 1.089
			& 1.067 & 1.050 & 1.041 & 1.025 \\
			
			& 2.00   & 1.139 & 1.105 & 1.076 & 1.056
			& 1.037 & 1.020 & 1.011 & 1.000 \\
			
			& 3.00   & 1.073 & 1.036 & 1.015 & 0.999
			& 0.983 & 0.976 & 0.968 & 0.959 \\
			
			& 4.00   & 1.010 & 0.984 & 0.961 & 0.948
			& 0.937 & 0.932 & 0.927 & 0.919 \\
			
			& 5.00   & 0.958 & 0.934 & 0.919 & 0.914
			& 0.907 & 0.900 & 0.893 & 0.890 \\
			
			& 6.00   & 0.908 & 0.894 & 0.883 & 0.879
			& 0.872 & 0.867 & 0.867 & 0.868 \\
			
			& 7.00   & 0.867 & 0.858 & 0.851 & 0.849
			& 0.848 & 0.843 & 0.844 & 0.840 \\
			
			& 8.00   & 0.835 & 0.826 & 0.822 & 0.822
			& 0.820 & 0.820 & 0.817 & 0.817 \\
			
			& 9.00   & 0.803 & 0.799 & 0.796 & 0.795
			& 0.792 & 0.793 & 0.795 & 0.792 \\
			
			& 10.00  & 0.771 & 0.769 & 0.771 & 0.770
			& 0.773 & 0.771 & 0.774 & 0.770 \\
			
			\hline
		\end{tabular}%
	}
	
	\vspace{-2mm}
\end{table}

\clearpage
\section{Computational code}
\label{app:computational_code}

\lstset{
	basicstyle=\ttfamily\fontsize{8}{9.5}\selectfont,
	breaklines=true,
	breakatwhitespace=false,
	breakindent=0pt,
	columns=fullflexible,
	keepspaces=true,
	showstringspaces=false,
	frame=single,
	xleftmargin=0pt,
	xrightmargin=0pt,
	linewidth=\textwidth,
	tabsize=2
}

The following four R files preserve the original organization and calculation
sequence used for the article. Only code used solely to draw figures or export
results to Excel has been omitted. The two groundwater data workbooks are
included in \texttt{arxiv\_code/data}.

\subsection*{Code A.1: Estimation and simulation}
\begin{lstlisting}

# ==============================================================================
# SOURCE: V12_HPC.R
# ==============================================================================

.libPaths("~/R/x86_64-pc-linux-gnu-library/4.4")
#library(doParallel)

library(doParallel)
library(foreach)
library(openxlsx)

# ----------- HM PH -------------
generate_U2 <- function(U1, V, theta) {
  A <- exp(-theta * U1) - exp(-theta)
  B <- (1 - exp(-theta * U1))
  D <- (1 - exp(-theta))
  T1 <- D * ((A / B) + 1)
  T2 <- (V * (exp(theta * U1) - 1) * (A + B) + D)
  U2 <- -(1 / theta) * log((T1 / T2) - (A / B))
  return(U2)
}

frank_density <- function(u1, u2, theta) {
  num <- theta * (1 - exp(-theta)) * exp(-theta * (u1 + u2))
  denom <- (exp(-theta * u1) + exp(-theta * u2) - exp(-theta) - exp(-theta * (u1 + u2)))^2
  num / denom
}

loglik_frank <- function(theta, u1, u2) {
  if (abs(theta) < 1e-6) return(-Inf)
  dens <- frank_density(u1, u2, theta)
  if (any(dens <= 0 | !is.finite(dens))) return(-Inf)
  sum(log(dens))
}

J1 <- function(U1, U2, theta) {
  exp_val <- exp(-theta * U1 - theta * U2)
  term1 <- -(U1 - U2)^2 + U1^2 * exp(-theta * U2) + U2^2 * exp(-theta * U1) - (U1 + U2 - 1)^2 * exp(-theta)
  term2 <- (U2 - 1)^2 * exp(-theta * U2 - theta) + (U1 - 1)^2 * exp(-theta * U1 - theta)
  return(exp_val * term1 + term2)
}

J2 <- function(U1, U2, theta) {
  exp_term <- exp(-theta * U1) + exp(-theta * U2) - exp(-theta) - exp(-theta * (U1 + U2))
  return(exp_term^2)
}

J <- function(U1, U2, theta) {
  J1_val <- J1(U1, U2, theta)
  J2_val <- J2(U1, U2, theta)
  return(J1_val / J2_val)
}

calculate_E_J <- function(theta, M) {
  J_values <- numeric(M)
  for (m in 1:M) {
    U1 <- runif(1, 0, 1)
    V <- runif(1, 0, 1)
    U2 <- generate_U2(U1, V, theta)
    J_values[m] <- J(U1, U2, theta)
  }
  mean(J_values)
}

I_theta <- function(theta, M) {
  E_J <- calculate_E_J(theta, M)
  (1 / theta^2) + (exp(theta) / (exp(theta) - 1)^2) - 2 * E_J
}

jeffrey_prior <- function(theta, M = 500) {
  I_val <- I_theta(theta, M)
  if (I_val <= 0 | !is.finite(I_val)) return(0)
  sqrt(I_val)
}

calculate_bayes_jeffrey <- function(u1, u2, theta_grid, M = 500) {
  log_likelihood <- sapply(theta_grid, function(th) sum(log(frank_density(u1, u2, th))))
  prior <- sapply(theta_grid, function(th) jeffrey_prior(th, M))
  prior[!is.finite(prior) | prior <= 0] <- 0
  log_posterior <- log_likelihood + log(prior + 1e-16)
  max_logp <- max(log_posterior)
  posterior <- exp(log_posterior - max_logp)
  delta <- diff(theta_grid)[1]
  posterior <- posterior / sum(posterior * delta)
  sum(theta_grid * posterior * delta)
}

# ------------ HM LU KT QU & V HNH -----------

# --------------- HM CHY 1 CP L, M THEO BATCH ---------------
run_frank_grid_LM_batch <- function(
    n_values = c(10, 25, 50),
    theta_values = c(0.01, 2.5, 5),
    L = 1000,
    M = 1000,
    theta_grid = c(seq(-20, -0.01, length.out = 500), seq(0.01, 20, length.out = 500)),
    outdir = "output",
    max_cores = parallel::detectCores(logical = FALSE),
    batch_size = 100
) {
  ensure_dir <- function(path) { if (!dir.exists(path)) dir.create(path, recursive = TRUE) }
  ensure_dir(outdir)
  cat(sprintf("=== Chy vi L = %d, M = %d; output: %s ===\n", L, M, outdir))
  export_funs <- c("generate_U2", "frank_density", "loglik_frank",
                   "J1", "J2", "J", "calculate_E_J", "I_theta",
                   "jeffrey_prior", "calculate_bayes_jeffrey")

  # Phase 1 (song song theo batch)
  for (n in n_values) {
    for (theta_true in theta_values) {
      cat(sprintf("Phase 1: n = %d, theta = %.2f\n", n, theta_true))
      phase1_file <- sprintf("%s/simdata_n%d_theta%.2f_phase1.rds", outdir, n, theta_true)
      res_list <- list()
      total_done <- 0
      while (total_done < L) {
        to_do <- min(batch_size, L - total_done)
        cl <- makeCluster(max_cores)
        registerDoParallel(cl)
        batch_res <- foreach(m = 1:to_do, .combine = 'list', .multicombine = TRUE,
                             .export = export_funs) %dopar% {
                               u1 <- runif(n)
                               u2 <- sapply(u1, function(u) generate_U2(u, runif(1), theta_true))
                               neg_loglik <- function(theta) -loglik_frank(theta, u1, u2)
                               optim_res <- optim(par = 1, fn = neg_loglik, method = "Brent", lower = -20, upper = 20)
                               mle_optim <- optim_res$par
                               log_likelihood <- sapply(theta_grid, function(th) sum(log(frank_density(u1, u2, th))))
                               max_logl <- max(log_likelihood)
                               likelihood <- exp(log_likelihood - max_logl)
                               delta <- diff(theta_grid)[1]
                               posterior <- likelihood / sum(likelihood * delta)
                               bayes_flat <- sum(theta_grid * posterior * delta)
                               list(u1 = u1, u2 = u2, mle_optim = mle_optim, bayes_flat = bayes_flat)
                             }
        stopCluster(cl)
        res_list <- c(res_list, batch_res)
        total_done <- total_done + to_do
        gc()
      }
      saveRDS(res_list, phase1_file)
      rm(res_list); gc()
    }
  }

  # Phase 2 (song song theo batch, ng thi xut Excel)
  for (n in n_values) {
    for (theta_true in theta_values) {
      start_time <- Sys.time()
      cat(sprintf("Phase 2: n = %d, theta = %.2f, Bt u: %s\n", n, theta_true, start_time))
      phase1_file <- sprintf("%s/simdata_n%d_theta%.2f_phase1.rds", outdir, n, theta_true)
      out_file <- sprintf("%s/simdata_n%d_theta%.2f_all.rds", outdir, n, theta_true)
      res_list <- readRDS(phase1_file)
      L_now <- length(res_list)
      for (i in seq(1, L_now, by = batch_size)) {
        idx <- i:min(i + batch_size - 1, L_now)
        cl <- makeCluster(max_cores)
        registerDoParallel(cl)
        bayes_jeffrey_vec <- foreach(j = idx, .combine = c,
                                     .export = export_funs) %dopar% {
                                       u1 <- res_list[[j]]$u1
                                       u2 <- res_list[[j]]$u2
                                       calculate_bayes_jeffrey(u1, u2, theta_grid, M)
                                     }
        stopCluster(cl)
        for (k in seq_along(idx)) {
          res_list[[idx[k]]]$bayes_jeffrey <- bayes_jeffrey_vec[k]
        }
        rm(bayes_jeffrey_vec); gc()
      }
      saveRDS(res_list, out_file)
      #plot_hist_small2(res_list, n, theta_true, L, M, outdir)

      # Xut Excel vi hai sheet: Bias_MSE v Errors

      # Sheet 1: Bias_MSE
      mle_vec <- sapply(res_list, function(x) x$mle_optim)
      bfp_vec <- sapply(res_list, function(x) x$bayes_flat)
      bjp_vec <- sapply(res_list, function(x) x$bayes_jeffrey)
      errors_MLE <- mle_vec - theta_true
      errors_BFP <- bfp_vec - theta_true
      errors_BJP <- bjp_vec - theta_true
      bias_MLE <- mean(errors_MLE)
      bias_BFP <- mean(errors_BFP)
      bias_BJP <- mean(errors_BJP)
      mse_MLE <- mean(errors_MLE^2)
      mse_BFP <- mean(errors_BFP^2)
      mse_BJP <- mean(errors_BJP^2)

      # Sheet 2: Errors

      # Lu workbook
      cat(sprintf(" lu file Excel cho n=%d, theta=%.2f: %s/bias_mse_n%d_theta%.2f.xlsx\n",
                  n, theta_true, outdir, n, theta_true))
      rm(res_list, mle_vec, bfp_vec, bjp_vec, errors_MLE, errors_BFP, errors_BJP); gc()

      end_time <- Sys.time()
      elapsed_time <- difftime(end_time, start_time, units = "secs")
      cat(sprintf("Phase 2: n = %d, theta = %.2f, Kt thc: %s, Thi gian chy: %.2f giy\n",
                  n, theta_true, end_time, elapsed_time))
    }
  }

  # Phase 3: Ch v histogram li
  # plot_histogram_grid2(n_values, theta_values, "mle", L, M, outdir)
  # plot_histogram_grid2(n_values, theta_values, "flat", L, M, outdir)
  # plot_histogram_grid2(n_values, theta_values, "jeffrey", L, M, outdir)
  #
  # cat(sprintf("===  XONG output: %s ===\n", outdir))
  # rm(list = setdiff(ls(), c("run_frank_grid_LM_batch", "generate_U2", "frank_density",
  #                           "loglik_frank", "J1", "J2", "J", "calculate_E_J",
  #                           "I_theta", "jeffrey_prior", "calculate_bayes_jeffrey",
  #                           "plot_hist_small2", "plot_histogram_grid2"))); gc()
}

# --------------- CHY NHIU CP L, M ---------------
L_values <- c(50)
M_values <- c(20)
for (L in L_values) {
  for (M in M_values) {
    outdir <- sprintf("output_L%d_M%d", L, M)
    dir.create(outdir, recursive = TRUE, showWarnings = FALSE)
    run_frank_grid_LM_batch(
      n_values = c(25),
      theta_values = c(6,7),
      #theta_values = c(0.2, 0.5),
      L = L,
      M = M,
      outdir = outdir,
      max_cores = parallel::detectCores(logical = FALSE),
      batch_size = 500
    )
  }
}

# ==============================================================================
# SOURCE: Bias_MSE_SE_21.12.25.R
# ==============================================================================

# Compute Mean and SE for MSE table
# ===============================

# Ci gi cn thit
if(!require(openxlsx)) install.packages("openxlsx", dependencies = TRUE)
library(openxlsx)

# ----n = 15---------------------------

# 1) Nhp d liu MSE-MLE
# -------------------------------
## =========================
## Bias_MLE_n15 table
## =========================
df <- data.frame(
  theta = c(10, 9, 8, 7, 6, 5, 4, 3, 2, 1.5, 1, 0.75, 0.5, 0.1),

  Run1 = c(
    0.670, 0.593, 0.536, 0.480, 0.432, 0.349, 0.297,
    0.229, 0.140, 0.099, 0.041, 0.061, 0.048, 0.034
  ),

  Run2 = c(
    0.661, 0.592, 0.507, 0.488, 0.421, 0.380, 0.287,
    0.220, 0.125, 0.097, 0.060, 0.059, 0.033, 0.000
  ),

  Run3 = c(
    0.642, 0.611, 0.549, 0.494, 0.446, 0.346, 0.299,
    0.203, 0.130, 0.088, 0.072, 0.044, 0.034, 0.003
  ),

  Run4 = c(
    0.602, 0.625, 0.521, 0.479, 0.457, 0.373, 0.326,
    0.226, 0.092, 0.129, 0.072, 0.044, 0.015, 0.027
  )
)

## =========================
## Bias_BFP_n15 table
## =========================
df <- data.frame(
  theta = c(10, 9, 8, 7, 6, 5, 4, 3, 2, 1.5, 1, 0.75, 0.5, 0.1),

  Run1 = c(
    1.127, 1.000, 0.883, 0.769, 0.663, 0.515, 0.413,
    0.297, 0.177, 0.125, 0.055, 0.074, 0.055, 0.035
  ),

  Run2 = c(
    1.120, 0.998, 0.849, 0.774, 0.649, 0.553, 0.403,
    0.290, 0.163, 0.122, 0.076, 0.072, 0.041, 0.002
  ),

  Run3 = c(
    1.101, 1.014, 0.899, 0.781, 0.675, 0.513, 0.415,
    0.271, 0.168, 0.114, 0.087, 0.056, 0.041, 0.005
  ),

  Run4 = c(
    1.058, 1.027, 0.869, 0.771, 0.686, 0.545, 0.438,
    0.294, 0.129, 0.157, 0.090, 0.057, 0.023, 0.029
  )
)

## =========================
## Bias_BJP_n15 table
## =========================
df <- data.frame(
  theta = c(10, 9, 8, 7, 6, 5, 4, 3, 2, 1.5, 1, 0.75, 0.5, 0.1),

  Run1 = c(
    0.682, 0.610, 0.549, 0.494, 0.442, 0.348, 0.292,
    0.216, 0.128, 0.089, 0.033, 0.056, 0.045, 0.031
  ),

  Run2 = c(
    0.675, 0.609, 0.518, 0.499, 0.429, 0.384, 0.284,
    0.209, 0.115, 0.087, 0.053, 0.053, 0.031, 0.000
  ),

  Run3 = c(
    0.657, 0.625, 0.566, 0.505, 0.454, 0.347, 0.293,
    0.191, 0.119, 0.079, 0.065, 0.039, 0.030, 0.002
  ),

  Run4 = c(
    0.616, 0.635, 0.538, 0.496, 0.466, 0.377, 0.318,
    0.214, 0.082, 0.121, 0.067, 0.040, 0.013, 0.026
  )
)

# ===============================
# To bng MSE-MLE_n=15
# ===============================

df <- data.frame(
 # MSE_MLE_n15 <- data.frame(
    theta = c(10, 9, 8, 7, 6, 5, 4, 3, 2, 1.5, 1, 0.75, 0.5, 0.1),
    Run1  = c(9.640, 8.572, 7.526, 6.281, 5.272, 4.664, 4.158, 3.787,
              3.300, 3.274, 3.226, 3.176, 3.175, 3.161),
    Run2  = c(9.909, 8.345, 7.153, 6.323, 5.283, 4.729, 3.992, 3.532,
              3.213, 3.306, 3.176, 3.222, 3.152, 3.089),
    Run3  = c(9.745, 8.433, 7.271, 6.423, 5.413, 4.473, 3.981, 3.666,
              3.386, 3.198, 3.120, 3.110, 3.125, 3.177),
    Run4  = c(9.605, 8.835, 7.768, 6.306, 5.473, 4.710, 4.109, 3.565,
              3.314, 3.263, 3.220, 3.181, 3.168, 3.186)
  )

## =========================
## MSE_MLE_n15 table
## =========================

df <- data.frame(
  theta = c(10, 9, 8, 7, 6, 5, 4, 3, 2, 1.5, 1, 0.75, 0.5, 0.1),

  Run1 = c(
    9.700, 8.572, 7.345, 6.281, 5.272, 4.664, 4.158,
    3.787, 3.300, 3.274, 3.226, 3.176, 3.175, 3.161
  ),

  Run2 = c(
    9.909, 8.579, 7.153, 6.323, 5.283, 4.729, 3.992,
    3.632, 3.213, 3.306, 3.176, 3.222, 3.152, 3.089
  ),

  Run3 = c(
    9.745, 8.632, 7.271, 6.423, 5.413, 4.673, 3.981,
    3.666, 3.386, 3.198, 3.120, 3.110, 3.125, 3.177
  ),

  Run4 = c(
    9.755, 8.735, 7.302, 6.306, 5.473, 4.710, 4.109,
    3.565, 3.314, 3.263, 3.220, 3.181, 3.168, 3.186
  )
)

# ===============================
# To bng MSE-BFPE_n=15
# ===============================

df <- data.frame(
  theta = c(10, 9, 8, 7, 6, 5, 4, 3, 2, 1.5, 1, 0.75, 0.5, 0.1),

  Run1 = c(
    11.256, 9.942, 8.574, 7.265, 6.088, 5.308, 4.624,
    4.017, 3.447, 3.360, 3.280, 3.202, 3.191, 3.121
  ),

  Run2 = c(
    11.289, 10.010, 8.343, 7.320, 6.099, 5.377, 4.443,
    3.819, 3.349, 3.394, 3.212, 3.254, 3.173, 3.113
  ),

  Run3 = c(
    11.261, 10.028, 8.478, 7.455, 6.231, 5.290, 4.446,
    3.954, 3.541, 3.278, 3.174, 3.142, 3.139, 3.188
  ),

  Run4 = c(
    11.141, 10.185, 8.465, 7.272, 6.294, 5.340, 4.593,
    3.857, 3.446, 3.366, 3.264, 3.222, 3.179, 3.202
  )
)

## =========================
## MSE_BJP_n15 table
## =========================
df <- data.frame(
  theta = c(10, 9, 8, 7, 6, 5, 4, 3, 2, 1.5, 1, 0.75, 0.5, 0.1),

  Run1 = c(
    9.509, 8.439, 7.315, 6.234, 5.281, 4.695, 4.182,
    3.728, 3.278, 3.234, 3.178, 3.113, 3.111, 3.071
  ),

  Run2 = c(
    9.719, 8.429, 7.124, 6.292, 5.295, 4.751, 4.013,
    3.537, 3.191, 3.266, 3.118, 3.163, 3.090, 3.036
  ),

  Run3 = c(
    9.548, 8.511, 7.213, 6.411, 5.410, 4.501, 4.013,
    3.675, 3.371, 3.157, 3.067, 3.057, 3.061, 3.117
  ),

  Run4 = c(
    9.455, 8.537, 7.222, 6.250, 5.463, 4.712, 4.143,
    3.580, 3.288, 3.226, 3.162, 3.130, 3.103, 3.124
  )
)

# -------------------------------
# 2) Tnh trung bnh v SE theo cng thc
# -------------------------------

# Trung bnh: m = (m1 + m2 + m3 + m)/4
df$Mean_Run <- rowMeans(df[, 2:5])

#  lch chun theo cng thc bn cho (chia cho n=4, khng phi n-1)
df$StDev <- apply(df[, 2:5], 1, function(x) sqrt(sum((x - mean(x))^2) / 4))

# Sai s chun SE = St.Dev / sqrt(5)
df$SE <- df$StDev / sqrt(4)

# -------------------------------
# 3) Xut ra Excel
# -------------------------------

# t tiu  hng 1

# nh dng
\end{lstlisting}

\subsection*{Code A.2: Goodness-of-fit simulation}
\begin{lstlisting}

# ==============================================================================
# SOURCE: Table4.1-4.4_GoF_hpc_MLE_ok.R
# ==============================================================================

###############################################################################
#       GOF TEST FOR FRANK COPULA  PARALLEL VERSION v11 (HPC-safe)          #
#   - Khng v histogram bn trong foreach (trnh li "X11 is not available") #
#   - Sau khi gom kt qu -> v histogram tun t bn ngoi foreach          #
###############################################################################

cat("\n========== GOF Frank copula: PARALLEL v11 (HPC safe) ==========\n\n")

## ================= PACKAGES =================
if (!require(nleqslv))    install.packages("nleqslv")
if (!require(openxlsx))   install.packages("openxlsx")
if (!require(doParallel)) install.packages("doParallel")
if (!require(foreach))    install.packages("foreach")

library(nleqslv)
library(openxlsx)
library(doParallel)
library(foreach)

#set.seed(123)

## ================= OUTPUT FOLDER =================

# ================================
# HPC FIX: set working directory
# ================================
#setwd("/home/phamthiyenanh")                           # THU MC GC CA BN
output_dir <- "GoF_Frank_Results_n_25,50_14.7.2026_MLE"                      # KHNG DNG KHONG TRNG
dir.create(output_dir, showWarnings=FALSE, recursive=TRUE)

cat("WORKING DIRECTORY =", getwd(), "\n")
cat("OUTPUT FOLDER     =", file.path(getwd(), output_dir), "\n\n")

## ================= GRID =================
#n_grid <- c(25,50,75,100,150,200,250,500,750,1000)

n_grid <- c(25)
#theta_grid <- c(-3, +3)
theta_grid <- c(-25,-20,-15,-10,-9,-8,-7,-6,-5,-4,-3,-2,-1.5,-1,-0.75,-0.5,-0.1, 0.1,0.5,0.75,1,1.5,2,3,4,5,6,7,8,9,10,15,20,25)

M_mc <- 1000   # chy tht: 10000

## ================= FUNCTIONS =================

## ============== Generate_U2 ===========================

generate_U2 <- function(U1, V, theta) {
  A <- exp(-theta * U1) - exp(-theta)
  B <- (1 - exp(-theta * U1))
  D <- (1 - exp(-theta))
  T1 <- D * ((A/B) + 1)
  T2 <- (V * (exp(theta * U1) - 1) * (A + B) + D)
  U2 <- -(1/theta) * log((T1/T2) - (A/B))
  return(U2)
}

## ============== Generate_(U,V) ===========================

generate_Frank_sample <- function(n, theta){
  U1 <- runif(n); V <- runif(n)
  U2 <- generate_U2(U1, V, theta)
  list(U=U1, V=U2)
}

## ============== H(theta) ===============

H_theta <- function(theta, U1, U2){
  n <- length(U1)
  term1 <- (1/theta) + 1/(exp(theta)-1)
  term2 <- mean(U1)+mean(U2)
  term3 <- 0
  for(i in 1:n){
    A1 <- U1[i]*exp(-theta*U1[i]) +
      U2[i]*exp(-theta*U2[i]) -
      (U1[i]+U2[i])*exp(-theta*(U1[i]+U2[i])) -
      exp(-theta)
    A2 <- exp(-theta*U1[i]) +
      exp(-theta*U2[i]) -
      exp(-theta*(U1[i]+U2[i])) -
      exp(-theta)
    term3 <- term3 + A1/A2
  }
  term3 <- (2/n)*term3
  term1 - term2 + term3
}

## ============== MLE ===========================

frank_MLE_score <- function(U1, U2){
  sol <- nleqslv(runif(1,-20,20),
                 fn=function(th) H_theta(th,U1,U2),
                 method="Broyden",
                 control=list(ftol=1e-6, xtol=1e-6))
  sol$x
}

## ============== Tnh W ===========================

compute_W <- function(U, V) {
  n <- length(U)
  R_U <- rank(U, ties.method = "average")
  R_V <- rank(V, ties.method = "average")
  W <- numeric(n)
  for (j in 1:n) {
    W[j] <- mean((R_U <= R_U[j]) & (R_V <= R_V[j]))
  }
  return(W)
}

## ================== Tnh Kn(t) ==========

compute_Kn <- function(W, n) {
  t_grid <- (1:(n-1)) / n
  Kn_vals <- sapply(t_grid, function(t) mean(W <= t))
  return(Kn_vals)
}

#compute_Kn <- function(W,n){
#  sapply((1:(n-1))/n, function(t) mean(W<=t))
#}

#compute_Kn <- function(W){
 # n <- length(W)
#  sapply((1:(n-1))/n, function(t) mean(W <= t))
#}

## ================== Tnh K(t,theta) ==========

#K_theoretical <- function(t,theta){
 # t - ((1-exp(theta*t))/theta) *
#    (log((1-exp(-theta)) / (1-exp(-theta*t))))
#}

K_theoretical <- function(t, theta_hat) {
  term1 <- t
  term2 <- (1 - exp(theta_hat * t)) / theta_hat
#  term3 <- log(abs((1 - exp(-theta_hat)) / (1 - exp(-theta_hat * t)))) #abs ca hm di log
  term3 <- log((1 - exp(-theta_hat)) / (1 - exp(-theta_hat * t)))
  return(term1 - term2 * term3)
}

#K_theoretical <- function(t, theta){
 # if (abs(theta) < 1e-8) return(t)   # gii hn khi theta -> 0
 # t - ((1-exp(theta*t))/theta) * log((1-exp(-theta)) / (1-exp(-theta*t)))
#}

## ================== Tnh Sn ==========

#compute_Sn <- function(Kn,theta_hat,n){
 # j <- 1:(n-1)
#  K_j  <- K_theoretical(j/n,     theta_hat)
#  K_j1 <- K_theoretical((j+1)/n, theta_hat)
 # n/3 + n*sum(Kn^2*(K_j1-K_j)) - n*sum(Kn*(K_j1^2-K_j^2))
#}

compute_Sn <- function(Kn_vals, theta_hat, n) {
  j <- 1:(n-1)
  t_j  <- j/n
  t_j1 <- (j+1)/n

  K_j  <- K_theoretical(t_j,  theta_hat)
  K_j1 <- K_theoretical(t_j1, theta_hat)

  term1 <- n/3
  term2 <- n * sum(Kn_vals^2 * (K_j1 - K_j))
  term3 <- n * sum(Kn_vals    * (K_j1^2 - K_j^2))

  return(term1 + term2 - term3)
}

## ================== Tnh Tn =====================

#compute_Tn <- function(Kn,theta_hat,n){
 #j <- 0:(n-1)
 #K_j  <- K_theoretical(j/n, theta_hat)
#K_j1 <- K_theoretical((j+1)/n, theta_hat)
#  sqrt(n) * max(c(abs(Kn - K_j[-1]), abs(Kn - K_j1[-1])))
#}

#compute_Tn <- function(Kn, theta_hat, n){
#  j <- 1:(n-1)

#  K_j  <- K_theoretical(j/n,     theta_hat)
#  K_j1 <- K_theoretical((j+1)/n, theta_hat)

#  sqrt(n) * max(c(
#    abs(Kn - K_j),
 #   abs(Kn - K_j1)
#  ))
#}

compute_Tn <- function(Kn_vals, theta_hat, n) {
  j <- 0:(n-1)

  K_j  <- K_theoretical(j/n,         theta_hat)
  K_j1 <- K_theoretical((j+1)/n,     theta_hat)

  diff1 <- abs(Kn_vals - K_j[-1])
  diff2 <- abs(Kn_vals - K_j1[-1])

  return(sqrt(n) * max(c(diff1, diff2)))
}

#compute_Tn <- function(Kn, theta_hat, n){
#  Kn0 <- c(0, Kn)                      # thm K_n(0)
 # j <- 0:(n-1)

#  K0 <- K_theoretical(j/n, theta_hat)         # i=0, di n
#  K1 <- K_theoretical((j+1)/n, theta_hat)     # i=1, di n

#  sqrt(n) * max(c(abs(Kn0 - K0), abs(Kn0 - K1)))
#}

simulate_Sn_Tn <- function(n,theta_true,M){
  S_vals <- T_vals <- Theta_vals <- numeric(M)
  for(m in 1:M){
    d <- generate_Frank_sample(n,theta_true)
    th <- frank_MLE_score(d$U,d$V)
    W  <- compute_W(d$U,d$V)
    Kn <- compute_Kn(W,n)
    S_vals[m] <- compute_Sn(Kn,th,n)
    T_vals[m] <- compute_Tn(Kn,th,n)
    Theta_vals[m] <- th
  }
  list(
    S_vals=S_vals, T_vals=T_vals, theta_vals=Theta_vals,
    S90=quantile(S_vals,0.9, na.rm=TRUE),
    S95=quantile(S_vals,0.95, na.rm=TRUE),
    T90=quantile(T_vals,0.9, na.rm=TRUE),
    T95=quantile(T_vals,0.95, na.rm=TRUE)
  )
}

## ================= PARALLEL SETUP =================
nc <- as.integer(Sys.getenv("SLURM_CPUS_PER_TASK","0"))
if(is.na(nc) || nc <= 0) nc <- detectCores()-1
cl <- makeCluster(nc)
registerDoParallel(cl)

cat(sprintf("Using %d cores\n", nc))

## ================= PARALLEL LOOP =================
tasks <- expand.grid(i=seq_along(n_grid),
                     j=seq_along(theta_grid))

start_time <- Sys.time()

results_list <- foreach(k=1:nrow(tasks),
                        .packages=c("nleqslv")) %dopar% {

                          i <- tasks$i[k]
                          j <- tasks$j[k]

                          n_val <- n_grid[i]
                          th_val <- theta_grid[j]

                          local_start <- Sys.time()

                          res <- simulate_Sn_Tn(n_val, th_val, M_mc)

                          elapsed <- as.numeric(Sys.time() - local_start, "secs")

                          list(i=i, j=j, n=n_val, theta=th_val,
                               S90=res$S90, S95=res$S95,
                               T90=res$T90, T95=res$T95,
                               S_vals=res$S_vals,
                               T_vals=res$T_vals,
                               theta_vals=res$theta_vals,
                               elapsed=elapsed)
                        }

stopCluster(cl)
cat("DONE parallel.\n")

## ================= BUILD TABLES =================
Table_Sn_90 <- matrix(NA,length(n_grid),length(theta_grid),
                      dimnames=list(n_grid,theta_grid))
Table_Sn_95 <- Table_Tn_90 <- Table_Tn_95 <- Table_Sn_90

Details_all <- data.frame()

for(res in results_list){
  Table_Sn_90[as.character(res$n), as.character(res$theta)] <- res$S90
  Table_Sn_95[as.character(res$n), as.character(res$theta)] <- res$S95
  Table_Tn_90[as.character(res$n), as.character(res$theta)] <- res$T90
  Table_Tn_95[as.character(res$n), as.character(res$theta)] <- res$T95

  Details_all <- rbind(Details_all,
                       data.frame(
                         n=res$n,
                         Theta_true=res$theta,
                         Theta_hat=res$theta_vals,
                         Sn=res$S_vals,
                         Tn=res$T_vals))
}

## ================= SAVE TABLES =================

## ================= STEP 3: DRAW HISTOGRAMS (SEQUENTIAL, PDF) =================
cat("\nDrawing GoF density + histogram plots into PDF...\n")

for(res in results_list){

  fn_pdf <- file.path(output_dir,
                      sprintf("DensityHist_n%d_theta%.2f.pdf",
                              res$n, res$theta))

  ## ================= S_n =================
  xlim_S <- range(quantile(res$S_vals, c(0.005, 0.995)))

  dS <- density(res$S_vals)

  ## ---- CH GI TRC DI, B TRN ----

  ## ================= T_n =================
  xlim_T <- range(quantile(res$T_vals, c(0.005, 0.995)))

  dT <- density(res$T_vals)

  ## ---- CH GI TRC DI, B TRN ----

}

cat("\n========== FINISHED v11 (EXCEL + PDF HISTOGRAMS) ==========\n")

# ==============================================================================
# SOURCE: GoF_Frank_Sn_Tn_BJPE_HPC_exact.R
# ==============================================================================

###############################################################################
# GOF TEST FOR FRANK COPULA  BJPE VERSION (HPC-safe)
#
# Built directly from:
#   (1) Original Sn/Tn Monte Carlo code using theta_hat_MLE
#   (2) Original BJPE code using Jeffreys prior from numerical 2D integration
#
# Only statistical replacement:
#       theta_hat_MLE  -->  theta_hat_BJPE
#
# No vectorized likelihood and no alteration of the BJPE formula.
###############################################################################

cat("\n========== GOF Frank copula: BJPE PARALLEL VERSION ==========\n\n")

# =============================================================================
# 0. PACKAGES
# =============================================================================
if (!require(openxlsx))   install.packages("openxlsx")
if (!require(doParallel)) install.packages("doParallel")
if (!require(foreach))    install.packages("foreach")

library(openxlsx)
library(doParallel)
library(foreach)

set.seed(123)

# =============================================================================
# 1. OUTPUT FOLDER
# =============================================================================
output_dir <- "GoF_Frank_Results_n_25_14.7.2026_BJPE_exact"
dir.create(output_dir, showWarnings = FALSE, recursive = TRUE)

cat("WORKING DIRECTORY =", getwd(), "\n")
cat("OUTPUT FOLDER     =", file.path(getwd(), output_dir), "\n\n")

# =============================================================================
# 2. SIMULATION SETTINGS
# =============================================================================
n_grid <- c(25)

theta_grid <- c(
  -25, -20, -15, -10, -9, -8, -7, -6, -5, -4, -3, -2,
  -1.5, -1, -0.75, -0.5, -0.1,
   0.1, 0.5, 0.75, 1, 1.5, 2, 3, 4, 5, 6, 7, 8, 9, 10, 15, 20, 25
)

M_mc <- 1000       # Chy chnh thc: 10000

# -----------------------------------------------------------------------------
# BJPE theta grid
# -----------------------------------------------------------------------------
# This is the same type of grid used in the supplied BJPE code.
# Because theta_true includes -25 and 25, the BJPE support is extended to
# [-30,30] to avoid forcing the posterior against the boundary at +/-20.
# The BJPE formula itself is unchanged.
theta_neg_BJPE <- seq(-30, -0.01, length.out = 1000)
theta_pos_BJPE <- seq( 0.01, 30, length.out = 1000)
theta_grid_BJPE <- c(theta_neg_BJPE, theta_pos_BJPE)

# Cache the expensive Jeffreys-prior calculation.
prior_cache_file <- file.path(
  output_dir,
  "Jeffreys_prior_grid_2D_integration.rds"
)

# =============================================================================
# 3. GENERATE A FRANK-COPULA SAMPLE
# =============================================================================
generate_U2 <- function(U1, V, theta) {
  A <- exp(-theta * U1) - exp(-theta)
  B <- 1 - exp(-theta * U1)
  D <- 1 - exp(-theta)
  T1 <- D * ((A / B) + 1)
  T2 <- V * (exp(theta * U1) - 1) * (A + B) + D
  U2 <- -(1 / theta) * log((T1 / T2) - (A / B))
  return(U2)
}

generate_Frank_sample <- function(n, theta) {
  U1 <- runif(n)
  V  <- runif(n)
  U2 <- generate_U2(U1, V, theta)
  list(U = U1, V = U2)
}

# =============================================================================
# 4. FRANK DENSITY AND LOG-LIKELIHOOD
# =============================================================================
frank_density <- function(u1, u2, theta) {
  if (abs(theta) < 1e-8) return(rep(NA_real_, length(u1)))

  num <- theta * (1 - exp(-theta)) * exp(-theta * (u1 + u2))

  denom <- (
    exp(-theta * u1) +
      exp(-theta * u2) -
      exp(-theta) -
      exp(-theta * (u1 + u2))
  )^2

  num / denom
}

loglik_frank <- function(theta, u1, u2) {
  if (abs(theta) < 1e-8) return(-Inf)

  dens <- frank_density(u1, u2, theta)

  if (any(!is.finite(dens)) || any(dens <= 0)) return(-Inf)

  sum(log(dens))
}

# =============================================================================
# 5. JEFFREYS PRIOR BY NUMERICAL TWO-DIMENSIONAL INTEGRATION
# =============================================================================
J1 <- function(U1, U2, theta) {
  exp_val <- exp(-theta * U1 - theta * U2)

  term1 <- -(U1 - U2)^2 +
    U1^2 * exp(-theta * U2) +
    U2^2 * exp(-theta * U1) -
    (U1 + U2 - 1)^2 * exp(-theta)

  term2 <- (U2 - 1)^2 * exp(-theta * U2 - theta) +
    (U1 - 1)^2 * exp(-theta * U1 - theta)

  exp_val * term1 + term2
}

J2 <- function(U1, U2, theta) {
  exp_term <- exp(-theta * U1) +
    exp(-theta * U2) -
    exp(-theta) -
    exp(-theta * (U1 + U2))

  exp_term^2
}

J_fun <- function(U1, U2, theta) {
  J1(U1, U2, theta) / J2(U1, U2, theta)
}

E_J_integral <- function(theta) {

  integrand_u1 <- function(u1) {

    sapply(u1, function(x) {

      inner <- integrate(
        f = function(u2) {
          J_fun(x, u2, theta) * frank_density(x, u2, theta)
        },
        lower = 1e-6,
        upper = 1 - 1e-6,
        subdivisions = 200,
        rel.tol = 1e-5
      )$value

      inner
    })
  }

  integrate(
    f = integrand_u1,
    lower = 1e-6,
    upper = 1 - 1e-6,
    subdivisions = 200,
    rel.tol = 1e-5
  )$value
}

I_theta <- function(theta) {
  EJ <- E_J_integral(theta)

  I_val <- (1 / theta^2) +
    (exp(theta) / (exp(theta) - 1)^2) -
    2 * EJ

  I_val
}

jeffrey_prior <- function(theta) {
  I_val <- tryCatch(
    I_theta(theta),
    error = function(e) NA_real_
  )

  if (!is.finite(I_val) || I_val <= 0) return(0)

  sqrt(I_val)
}

# =============================================================================
# 6. BJPE POSTERIOR MEAN  ORIGINAL SUPPLIED FORMULA
# =============================================================================
calculate_bayes_jeffrey <- function(u1, u2, theta_grid, prior_grid) {

  log_post <- numeric(length(theta_grid))

  for (i in seq_along(theta_grid)) {

    th <- theta_grid[i]
    prior <- prior_grid[i]
    loglik <- loglik_frank(th, u1, u2)

    if (!is.finite(loglik) || prior <= 0 || !is.finite(prior)) {
      log_post[i] <- -Inf
    } else {
      log_post[i] <- loglik + log(prior)
    }
  }

  if (!any(is.finite(log_post))) return(NA_real_)

  log_post <- log_post - max(log_post, na.rm = TRUE)
  post_unnorm <- exp(log_post)

  # Kept exactly as in the supplied BJPE code.
  delta <- diff(theta_grid)[1]
  normalizing_constant <- sum(post_unnorm * delta)

  if (!is.finite(normalizing_constant) || normalizing_constant <= 0) {
    return(NA_real_)
  }

  posterior <- post_unnorm / normalizing_constant

  sum(theta_grid * posterior * delta)
}

frank_BJPE <- function(U1, U2, theta_grid, prior_grid) {
  calculate_bayes_jeffrey(
    u1 = U1,
    u2 = U2,
    theta_grid = theta_grid,
    prior_grid = prior_grid
  )
}

# =============================================================================
# 7. COMPUTE W
# =============================================================================
compute_W <- function(U, V) {
  n <- length(U)
  R_U <- rank(U, ties.method = "average")
  R_V <- rank(V, ties.method = "average")

  W <- numeric(n)

  for (j in 1:n) {
    W[j] <- mean((R_U <= R_U[j]) & (R_V <= R_V[j]))
  }

  return(W)
}

# =============================================================================
# 8. COMPUTE Kn(t)
# =============================================================================
compute_Kn <- function(W, n) {
  t_grid <- (1:(n - 1)) / n
  Kn_vals <- sapply(t_grid, function(t) mean(W <= t))
  return(Kn_vals)
}

# =============================================================================
# 9. THEORETICAL K(t, theta)
# =============================================================================
K_theoretical <- function(t, theta_hat) {
  term1 <- t
  term2 <- (1 - exp(theta_hat * t)) / theta_hat
  term3 <- log(
    (1 - exp(-theta_hat)) /
      (1 - exp(-theta_hat * t))
  )

  return(term1 - term2 * term3)
}

# =============================================================================
# 10. COMPUTE Sn
# =============================================================================
compute_Sn <- function(Kn_vals, theta_hat, n) {
  j <- 1:(n - 1)
  t_j  <- j / n
  t_j1 <- (j + 1) / n

  K_j  <- K_theoretical(t_j,  theta_hat)
  K_j1 <- K_theoretical(t_j1, theta_hat)

  term1 <- n / 3
  term2 <- n * sum(Kn_vals^2 * (K_j1 - K_j))
  term3 <- n * sum(Kn_vals * (K_j1^2 - K_j^2))

  return(term1 + term2 - term3)
}

# =============================================================================
# 11. COMPUTE Tn
# =============================================================================
compute_Tn <- function(Kn_vals, theta_hat, n) {
  j <- 0:(n - 1)

  K_j  <- K_theoretical(j / n,       theta_hat)
  K_j1 <- K_theoretical((j + 1) / n, theta_hat)

  diff1 <- abs(Kn_vals - K_j[-1])
  diff2 <- abs(Kn_vals - K_j1[-1])

  return(sqrt(n) * max(c(diff1, diff2)))
}

# =============================================================================
# 12. SIMULATE Sn AND Tn USING theta_hat_BJPE
# =============================================================================
simulate_Sn_Tn_BJPE <- function(
    n,
    theta_true,
    M,
    theta_grid_BJPE,
    prior_grid_BJPE
) {

  S_vals <- rep(NA_real_, M)
  T_vals <- rep(NA_real_, M)
  Theta_vals <- rep(NA_real_, M)

  for (m in 1:M) {

    d <- generate_Frank_sample(n, theta_true)

    # Main replacement: MLE --> BJPE
    th <- frank_BJPE(
      U1 = d$U,
      U2 = d$V,
      theta_grid = theta_grid_BJPE,
      prior_grid = prior_grid_BJPE
    )

    if (!is.finite(th)) next

    W  <- compute_W(d$U, d$V)
    Kn <- compute_Kn(W, n)

    S_vals[m] <- compute_Sn(Kn, th, n)
    T_vals[m] <- compute_Tn(Kn, th, n)
    Theta_vals[m] <- th
  }

  list(
    S_vals = S_vals,
    T_vals = T_vals,
    theta_vals = Theta_vals,
    S90 = quantile(S_vals, 0.90, na.rm = TRUE),
    S95 = quantile(S_vals, 0.95, na.rm = TRUE),
    T90 = quantile(T_vals, 0.90, na.rm = TRUE),
    T95 = quantile(T_vals, 0.95, na.rm = TRUE),
    N_valid = sum(is.finite(Theta_vals))
  )
}

# =============================================================================
# 13. PARALLEL SETUP
# =============================================================================
nc <- as.integer(Sys.getenv("SLURM_CPUS_PER_TASK", "0"))

if (is.na(nc) || nc <= 0) {
  nc <- max(1, parallel::detectCores() - 1)
}

cl <- makeCluster(nc)
registerDoParallel(cl)
parallel::clusterSetRNGStream(cl, iseed = 123)

cat(sprintf("Using %d cores\n", nc))

# =============================================================================
# 14. COMPUTE OR READ JEFFREYS PRIOR GRID
# =============================================================================
if (file.exists(prior_cache_file)) {

  cat("Reading saved Jeffreys prior grid...\n")

  prior_object <- readRDS(prior_cache_file)

  same_grid <- isTRUE(all.equal(
    prior_object$theta_grid_BJPE,
    theta_grid_BJPE
  ))

  if (!same_grid) {
    stop(
      "Saved prior grid does not match theta_grid_BJPE. ",
      "Delete the RDS file and rerun."
    )
  }

  prior_grid_BJPE <- prior_object$prior_grid_BJPE

} else {

  cat("Computing Jeffreys prior grid by numerical 2D integration...\n")
  cat("This expensive step is performed only once.\n")

  # Same prior formula; theta values are distributed across HPC workers.
  prior_grid_BJPE <- foreach(
    i = seq_along(theta_grid_BJPE),
    .combine = "c",
    .inorder = TRUE,
    .export = c(
      "theta_grid_BJPE",
      "frank_density",
      "J1",
      "J2",
      "J_fun",
      "E_J_integral",
      "I_theta",
      "jeffrey_prior"
    )
  ) %dopar% {
    jeffrey_prior(theta_grid_BJPE[i])
  }

  saveRDS(
    list(
      theta_grid_BJPE = theta_grid_BJPE,
      prior_grid_BJPE = prior_grid_BJPE
    ),
    prior_cache_file
  )
}

if (
  length(prior_grid_BJPE) != length(theta_grid_BJPE) ||
  all(!is.finite(prior_grid_BJPE) | prior_grid_BJPE <= 0)
) {
  stop("Jeffreys prior grid is invalid.")
}

# =============================================================================
# 15. PARALLEL MONTE CARLO LOOP
# =============================================================================
tasks <- expand.grid(
  i = seq_along(n_grid),
  j = seq_along(theta_grid)
)

start_time <- Sys.time()

results_list <- foreach(
  k = 1:nrow(tasks),
  .packages = character(0),
  .export = c(
    "tasks",
    "n_grid",
    "theta_grid",
    "M_mc",
    "theta_grid_BJPE",
    "prior_grid_BJPE",
    "generate_U2",
    "generate_Frank_sample",
    "frank_density",
    "loglik_frank",
    "calculate_bayes_jeffrey",
    "frank_BJPE",
    "compute_W",
    "compute_Kn",
    "K_theoretical",
    "compute_Sn",
    "compute_Tn",
    "simulate_Sn_Tn_BJPE"
  ),
  .errorhandling = "pass"
) %dopar% {

  i <- tasks$i[k]
  j <- tasks$j[k]

  n_val <- n_grid[i]
  th_val <- theta_grid[j]

  local_start <- Sys.time()

  res <- simulate_Sn_Tn_BJPE(
    n = n_val,
    theta_true = th_val,
    M = M_mc,
    theta_grid_BJPE = theta_grid_BJPE,
    prior_grid_BJPE = prior_grid_BJPE
  )

  elapsed <- as.numeric(
    Sys.time() - local_start,
    units = "secs"
  )

  list(
    i = i,
    j = j,
    n = n_val,
    theta = th_val,
    S90 = res$S90,
    S95 = res$S95,
    T90 = res$T90,
    T95 = res$T95,
    S_vals = res$S_vals,
    T_vals = res$T_vals,
    theta_vals = res$theta_vals,
    N_valid = res$N_valid,
    elapsed = elapsed
  )
}

stopCluster(cl)
cat("DONE parallel.\n")

# =============================================================================
# 16. CHECK PARALLEL ERRORS
# =============================================================================
is_error <- vapply(
  results_list,
  function(x) inherits(x, "error"),
  logical(1)
)

if (any(is_error)) {
  error_messages <- vapply(
    results_list[is_error],
    conditionMessage,
    character(1)
  )

  writeLines(
    error_messages,
    file.path(output_dir, "Parallel_errors.txt")
  )

  warning(
    sum(is_error),
    " parallel tasks failed. See Parallel_errors.txt."
  )
}

results_ok <- results_list[!is_error]

if (length(results_ok) == 0) {
  stop("All parallel tasks failed.")
}

# =============================================================================
# 17. BUILD TABLES
# =============================================================================
Table_Sn_90 <- matrix(
  NA_real_,
  length(n_grid),
  length(theta_grid),
  dimnames = list(n_grid, theta_grid)
)

Table_Sn_95 <- Table_Tn_90 <- Table_Tn_95 <- Table_Sn_90

Details_list <- vector("list", length(results_ok))
Summary_list <- vector("list", length(results_ok))

for (r in seq_along(results_ok)) {

  res <- results_ok[[r]]

  Table_Sn_90[as.character(res$n), as.character(res$theta)] <- res$S90
  Table_Sn_95[as.character(res$n), as.character(res$theta)] <- res$S95
  Table_Tn_90[as.character(res$n), as.character(res$theta)] <- res$T90
  Table_Tn_95[as.character(res$n), as.character(res$theta)] <- res$T95

  Details_list[[r]] <- data.frame(
    n = res$n,
    Theta_true = res$theta,
    Theta_hat_BJPE = res$theta_vals,
    Sn = res$S_vals,
    Tn = res$T_vals
  )

  Summary_list[[r]] <- data.frame(
    n = res$n,
    Theta_true = res$theta,
    N_valid_BJPE = res$N_valid,
    N_failed_BJPE = M_mc - res$N_valid,
    Sn_0.90_BJPE = as.numeric(res$S90),
    Sn_0.95_BJPE = as.numeric(res$S95),
    Tn_0.90_BJPE = as.numeric(res$T90),
    Tn_0.95_BJPE = as.numeric(res$T95),
    Elapsed_seconds = res$elapsed
  )
}

Details_all <- do.call(rbind, Details_list)
Task_summary <- do.call(rbind, Summary_list)

# =============================================================================
# 18. SAVE TABLES
# =============================================================================

# =============================================================================
# 19. DRAW HISTOGRAMS SEQUENTIALLY
# =============================================================================
cat("\nDrawing GoF density + histogram plots into PDF...\n")

for (res in results_ok) {

  S_plot <- res$S_vals[is.finite(res$S_vals)]
  T_plot <- res$T_vals[is.finite(res$T_vals)]

  if (length(S_plot) < 2 || length(T_plot) < 2) next

  fn_pdf <- file.path(
    output_dir,
    sprintf(
      "DensityHist_BJPE_n%d_theta%.2f.pdf",
      res$n,
      res$theta
    )
  )

  # S_n
  xlim_S <- range(quantile(S_plot, c(0.005, 0.995), na.rm = TRUE))

  dS <- density(S_plot)

  # T_n
  xlim_T <- range(quantile(T_plot, c(0.005, 0.995), na.rm = TRUE))

  dT <- density(T_plot)

}

# =============================================================================
# 20. TOTAL RUNNING TIME
# =============================================================================
total_elapsed <- as.numeric(
  Sys.time() - start_time,
  units = "hours"
)

cat("\nTotal simulation time =", total_elapsed, "hours\n")
cat("\n========== FINISHED: BJPE Sn/Tn CRITICAL VALUES ==========\n")
\end{lstlisting}

\subsection*{Code A.3: Groundwater data analysis}
\lstinputlisting{arxiv_code/03_groundwater_analysis.R}

\subsection*{Code A.4: Frank copula regression calculations}
\begin{lstlisting}

# ==============================================================================
# SOURCE: Table_M(u,theta).R
# ==============================================================================

###############################################################################
# TABLE OF CONDITIONAL MEDIAN M(u | theta)
# Frank Copula
#
# Using:
#     F_{V|U=u}(M | u; theta) = 0.5
#
# where F_{V|U=u} is the conditional CDF.
###############################################################################

# =============================================================================
# Step (i). Conditional CDF of V given U = u
# =============================================================================

frank_conditional_cdf <- function(v, u, theta) {

  # Independence case: theta = 0
  if (abs(theta) < 1e-8) {
    return(v)
  }

  numerator <-
    exp(-theta * u) *
    (1 - exp(-theta * v))

  denominator <-
    exp(-theta * u) * (1 - exp(-theta * v)) +
    (exp(-theta * v) - exp(-theta))

  cdf_value <- numerator / denominator

  return(cdf_value)
}

# =============================================================================
# Step (ii). Conditional median M(u | theta)
#
# Solve:
#
#     F_{V|U=u}(M | u; theta) = 0.5
#
# =============================================================================

median_frank_cdf <- function(u, theta) {

  # Independence case
  if (abs(theta) < 1e-8) {
    return(0.5)
  }

  median_equation <- function(M) {

    frank_conditional_cdf(
      v     = M,
      u     = u,
      theta = theta
    ) - 0.5
  }

  M <- uniroot(
    median_equation,
    interval = c(1e-10, 1 - 1e-10),
    tol = 1e-10
  )$root

  return(M)
}

# =============================================================================
# Step (iii). Values of u and theta
# =============================================================================

u_values <- seq(
  from = 0.05,
  to   = 0.95,
  by   = 0.05
)

theta_values <- seq(
  from = 0,
  to   = 10,
  by   = 1
)

# =============================================================================
# Step (iv). Calculate table M(u | theta)
# =============================================================================

M_table <- matrix(
  NA_real_,
  nrow = length(u_values),
  ncol = length(theta_values)
)

for (i in seq_along(u_values)) {

  for (j in seq_along(theta_values)) {

    M_table[i, j] <- median_frank_cdf(
      u     = u_values[i],
      theta = theta_values[j]
    )
  }
}

# =============================================================================
# Step (v). Add row and column names
# =============================================================================

M_table <- as.data.frame(M_table)

colnames(M_table) <- paste0(
  "theta_", theta_values
)

M_table <- cbind(
  u = u_values,
  M_table
)

# =============================================================================
# Step (vi). Display table
# =============================================================================

cat("\n")
cat("===============================================================\n")
cat("              TABLE OF M(u | theta)\n")
cat("===============================================================\n\n")

print(
  M_table,
  digits = 6,
  row.names = FALSE
)

# =============================================================================
# Step (vii). Export Table M(u | theta) to Excel
# =============================================================================

if (!require(openxlsx)) install.packages("openxlsx")
library(openxlsx)

# ----- Table for Excel
M_table_excel <- M_table

# Column names: u, theta = 0, 1, ..., 10
colnames(M_table_excel) <- c(
  "u",
  as.character(theta_values)
)

# ----- Output file
out_file <- "Table_M_u_theta.xlsx"

# ----- Create workbook

# ----- Write title

# ----- Write table

# ----- Header style

# ----- Number format: 6 decimal places

# ----- Adjust column widths

# ----- Freeze header

# ----- Save Excel file

cat("\n")
cat("===============================================================\n")
cat("Excel file successfully created:\n")
cat(out_file, "\n")
cat("===============================================================\n")
\end{lstlisting}

\newpage

\bibliographystyle{plain}

\begin{thebibliography}{99}
	
	\bibitem{Berg2007}
	Berg, M., Stengel, C., Trang, P. T. K., Viet, P. H., Sampson, M. L., Leng, M., Samreth, S., and Fredericks, D. (2007).
	Magnitude of arsenic pollution in the Mekong and Red River Deltas-Cambodia and Vietnam.
	\textit{Science of the Total Environment},
	372(2--3), 413--425.
	\url{https://doi.org/10.1016/j.scitotenv.2006.09.010}
	
	\bibitem{Bouye2000}
	Bouyé, E., Durrleman, V., Nikeghbali, A., Riboulet, G., and Roncalli, T. (2000).
	Copulas for Finance: A Reading Guide and Some Applications.
	Groupe de Recherche Opérationnelle, Crédit Lyonnais, Paris.
	
	\bibitem{Cherubini2004}
	Cherubini, U., Luciano, E., and Vecchiato, W. (2004).
	\textit{Copula Methods in Finance}.
	John Wiley \& Sons, Chichester, UK.
	
	\bibitem{Embrechts2002}
	Embrechts, P., McNeil, A. J., and Straumann, D. (2002).
	Correlation and dependency in risk management: Properties and pitfalls.
	In Dempster, M. (Ed.),
	\textit{Risk Management: Value at Risk and Beyond},
	pp. 176--223.
	Cambridge University Press, Cambridge.
	\url{https://doi.org/10.1017/CBO9780511615338}
	
	\bibitem{Genest1986}
	Genest, C. and MacKay, J. (1986). The Joy of Copulas: Bivariate Distributions with Uniform Marginals. \textit{The American Statistician}, Vol. 40, No. 4, 280--283. \url{https://doi.org/10.1080/00031305.1986.1047541}
	
	\bibitem{Genest1987} Genest, C. (1987). Frank's Family of Bivariate Distributions. \textit{Biometrika}, 74, 549-555. \url{https://doi.org/10.1093/biomet/74.3.549}.
	
	\bibitem{GenestRemillard2006}
	Genest, C., Quesy, J.-F.  and R\'emillard, B. (2006).
	Goodness-of-fit Procedures for Copula Models Based on the Probability
	Integral Transformation,
	\textit{Scandinavian Journal of Statistics}, Vol. 33, 337--366,
	\url{https://doi.org/10.1111/j.1467-9469.2006.00470.x}.
	
	\bibitem{Huisman2001}
	Huisman, R., Koedijk, C. G., Cool, C. J., and Palm, F. C. (2001).
	Tail-index estimates in small samples.
	\textit{Journal of Business \& Economic Statistics},
	19(2), 245--254.
	\url{https://doi.org/10.1198/073500101316970421}
	
	\bibitem{Jaynes1968}
	Jaynes, E. T. (1968).
	Prior probabilities.
	\textit{IEEE Transactions on Systems Science and Cybernetics},
	4(3), 227--241.
	\url{https://doi.org/10.1109/TSSC.1968.300117}
	
	\bibitem{Jeffreys1946}
	Jeffreys, H. (1946).
	An invariant form for the prior probability in estimation problems.
	\textit{Proceedings of the Royal Society of London. Series A},
	186(1007), 453--461. 
	\url{https://doi.org/10.1098/rspa.1946.0056}
	
	\bibitem{Junker2006}
	Junker, M., Szimayer, A., and Wagner, N. (2006).
	Nonlinear term structure dependence: Copula functions, empirics and risk implications.
	\textit{Journal of Banking \& Finance},
	30(4), 1171--1199.
	\url{https://doi.org/10.1016/j.jbankfin.2005.03.003}
	
	\bibitem{Meneguzzo2004}
	Meneguzzo, D. and Vecchiato, W. (2004).
	Copula sensitivity in collateralized debt obligations and basket default swaps.
	\textit{The Journal of Futures Markets},
	24(1), 37--70.
	\url{https://doi.org/10.1002/fut.10107}
	
	\bibitem{Merola2015}
	Merola, R. B., Hien, T. T., Quyen, D. T. T., and Vengosh, A. (2015).
	Arsenic exposure to drinking water in the Mekong Delta.
	\textit{Science of the Total Environment}, 511, 544--552.
	\url{https://doi.org/10.1016/j.scitotenv.2014.12.091}
	
	\bibitem{Nguyen2008}
	Nguyen, P. K. (2008).
	\textit{Geochemical study of arsenic behavior in aquifer of the Mekong Delta, Vietnam}.
	Ph.D. dissertation, Kyushu University, Japan.
	\url{https://ere.mine.kyushu-u.ac.jp/old/sotsuron/pdfs/2008/kim.pdf}
	
	\bibitem{Pham2015}
	Pham, C. H. V. (2015).
	\textit{Studying the mechanisms of arsenic release in groundwater in An Phu district, An Giang province}.
	Master's thesis, University of Technology, Ho Chi Minh City, Vietnam.
	
	\bibitem{Pham2015b}
	Pham, C. H. V., Ho, T. N. H., Frustchi, M., Wang, Y., Bernier, R., and Vo, L. P. (2015).
	Spatial and temporal variation of arsenic occurrence and physiogeochemical influence on arsenic in groundwater in the Vietnamese Mekong Delta: A case study of An Phu district, An Giang province.
	\textit{Journal of Science and Technology (Vietnam Academy of Science and Technology)},
	53(5A), 282--289.
	
	\bibitem{Pham et al.(2025)} Pham, T.Y.A., Huynh, T. U., and Pal, N. (2025). Some results on point estimation of the association parameter of a bivariate Frank copula. \textit{Communications in Statistics-Simulation and Computation}, 1-22. \url{https://doi.org/10.1080/03610918.2025.2545611}. [A version of this paper is also available at arXiv address: \url{https://doi.org/10.48550/arXiv.2506.23110}] 
	
	
	\bibitem{Sklar1959} Sklar, M. (1959). Fonctions de repartition an dimensions et leurs marges. \textit{Publ. Inst. Statist. Univ. Paris}, 8, 229-231.
	
\end{thebibliography}

\end{document}